# K Giants in Baade's Window. I. Velocity and Line–strength Measurements

D. M. Terndrup[1,2]

Department of Astronomy, The Ohio State University,
174 W. 18th Ave., Columbus, OH 43210
Electronic mail: terndrup@baade.mps.ohio–state.edu

Elaine M. Sadler

School of Physics, University of Sydney, NSW 2006, Australia
Electronic mail: ems@physics.usyd.edu.au

R. Michael Rich[2,3]

Department of Astronomy, Pupin Laboratories, Columbia University,
538 West 120th Street, New York, NY 10027
Electronic mail: rmr@cuphyd.columbia.edu

## ABSTRACT

This is the first in a series of papers in which we analyze medium–resolution spectra of over 400 K and M giants in Baade's Window. Our sample was selected from the proper motion study of Spaenhauer *et al.* [AJ, 103, 297 (1992)]. We have measured radial velocities for most of the sample, as well as line–strength indices on the system of Faber *et al.* [ApJS, 57, 711 (1985)]. We analyze the random and systematic errors in velocities and line strengths, and show that the bright ($V < 16.0$) stars in our sample are predominantly foreground disk stars along the line–of–sight toward Baade's Window. We find that most of the bulge K giants have stronger Mg absorption at a given color than do stars in the solar neighborhood. If the K giants in our sample are moderately old, we suggest that on average they may have [Mg/Fe] $\approx$ +0.3, consistent with the results of recent high–resolution spectroscopy in Baade's Window.

*Subject headings:*

---

## 1. Introduction

Ever since Baade (1944) formulated the concept of stellar populations, astronomers have recognized that the nuclear bulge is a key to our understanding of the formation and evolution of the Galaxy. Research on the bulge has been particularly active in the past fifteen years, and recent reviews by Frogel (1988), Rich (1992), and King (1993) summarize much of this work.

Several important aspects of the bulge population, however, remain poorly understood. The first is how metallicity controls the late stages of stellar evolution. The K giants in Baade's Window (hereafter BW) have $\langle [\mathrm{Fe/H}] \rangle \approx 0$ with a wide range in abundances ranging from $[\mathrm{Fe/H}] \approx -1.0$ to $+0.5$ (Whitford & Rich 1983; Rich 1988; Tyson 1991; Geisler & Friel 1992; McWilliam & Rich 1994). The RR Lyraes, in contrast, have a narrow distribution of metallicities with a mean value near $[\mathrm{Fe/H}] = -1$ (Walker & Terndrup 1991), thus indicating that only the most metal–poor stars in BW have a significant probability of becoming horizontal–branch stars at moderately high temperatures (see also Renzini 1994). In addition, the bulge M giants have significantly stronger molecular absorption at a given color than do cool stars in the solar neighborhood. This has been interpreted to indicate that the M giants are evolutionary descendants of the metal–rich K giants only (e.g., Terndrup *et al.* 1990; Sharples *et al.* 1990; Terndrup 1993; Morrison *et al.* 1993), though it may be that the M giants have enhanced relative abundances of Ti resulting in strong TiO bands (McWilliam & Rich 1994).

Determining the metallicity distribution in the bulge is critical for setting limits on the age (or distribution of ages) of bulge stars, currently poorly known. Observations of the turnoff in BW (Holtzman *et al.* 1993) and at larger galactocentric distances (Terndrup 1988) yield a mean age of $\approx 10$ Gyr for a mean distance to the bulge of $R_0 = 8$ kpc. Lee (1992), however, has used the properties of the bulge RR Lyraes to argue that the metal–poor stars (at least) are 1–2 Gyr older than globular clusters of similar metallicity, making them the oldest stars in the Galaxy.

In this regard, it is also important to determine the abundances of the alpha–capture and iron–peak elements. According to recent models of stellar evolution, the effect of enhanced alpha–element abundances on stellar evolution is similar to that produced by an equivalent increase in overall metallicity (Pinsonneault, private communication). The consequences of selective enrichment of the light elements, however, may be considerably different at low and high metallicities because the mass of turnoff stars is a strong function of the metal abundance for $[\mathrm{Fe/H}] > 0$ (e.g., VandenBerg & Laskarides 1987; Greggio & Renzini 1990). There is at present some evidence, based on spectra of a small number of BW stars (McWilliam & Rich 1994) and of nearby disk stars on eccentric orbits which take them in towards the bulge (Barbuy & Grenon 1990), that the stars in the bulge may have enhanced abundances of alpha–capture elements.

Aside from their importance as tracers of the chemical evolution history, bulge stars are potentially valuable as templates for the study of the integrated light of ellipticals and the bulges of other spirals. As demonstrated some time ago (Whitford 1978), the energy distribution and line strengths of the integrated spectrum of the bulge quantitatively resemble those of other bulges and ellipticals. Straightforward integration of the empirical luminosities, colors, and absorption strengths of bulge K and M giants provides a reasonable match to the observations of ellipticals and the bulges of other spirals (Frogel & Whitford 1987; Terndrup *et al.* 1990). In the bulge we have the hope of exploring in individual stars the effects of gravity and metallicity, which are extremely difficult to sort out in integrated light (Worthey *et al.* 1992; Silva & Elston 1994; Worthey 1994).

Bulge stars are also valuable probes of the mass distribution in the inner Galaxy, as shown by recent studies of the bulge's light distribution and gas motions in the inner disk (Binney *et al.* 1991; Blitz & Spergel 1991; Dwek *et al.* 1994; Stanek *et al.* 1994, Weiland *et al.* 1994), the spatial distribution of Mira variables (Whitelock & Cathpole 1992), and the radial velocities and proper motions of a small sample of K giants in BW (Zhao *et al.* 1994). Detailed constraints on the shape of the inner Galaxy cannot currently be made because there is a paucity of stellar data, particularly on the three–dimensional motions of stars in the bulge and inner disk. Most studies have concentrated on the distribution of line–of–sight radial velocities (Rich 1990, Sharples *et al.* 1990, Blum *et al.* 1994, 1995), which can be well matched by axisymmetric models of the bulge (Kent 1992, Kuijken 1994).

Motivated by these issues, we have conducted an extensive survey of K giants in the bulge. We have obtained radial velocities and metallicity/gravity indicators for nearly 400 stars in BW, which is located at $(\ell, b) = (1°, -3.9°)$, and 180 stars in another field at $(\ell, b) = (0°, -8°)$. The sample of stars in BW was chosen from a recent proper–motion survey to generate the first extensive sample of the three–dimensional stellar orbits in the bulge. Large samples were also chosen to map out the metallicity distribution and metallicity gradient in the bulge with significantly better statistical accuracy than in previous studies.

In this paper, which is the first in a series, we present photometry and spectroscopy of K and M giants in the Baade's Window field of the bulge and analyze the sample as a whole, with emphasis on the strengths of Mg and Fe absorption. Future papers will discuss the metallicity distribution of the individual stars in our sample, the metallicity gradient in the bulge, and the kinematics of the K giants in BW.

This paper is organized as follows: In Sec. 2, we describe the photometric and spectroscopic observations which constitute our database, as well as basic reductions and analysis techniques. We proceed in Secs. 3 and 4 to discuss the measurements of radial velocities and line–strength measures from the spectra, and analyze the random and systematic errors in our measurements. We derive the extinction toward the BW in Sec. 5, while in Sec. 6 we do a simple analysis of the combined database of velocities, line strengths, and proper motions to identify regions of the color–magnitude diagram that are dominated by bulge stars; we also discuss biases in the selection of stars in the proper–motion survey. Finally, we conclude (Sec. 7) with a comparison of the line strengths to those in the solar neighborhood and globular clusters.

## 2. Observations and Data Reduction

### 2.1. Selection of Stars and Observational Strategy

We chose to observe the sample of 427 BW stars with relative proper motions measured by Spaenhauer *et al.* (1992, hereafter SJW), both because the proper motion data are valuable for kinematic studies and because, being color–selected, the SJW sample is biased towards bulge stars rather than members of the foreground disk. The SJW stars are distributed throughout the annulus delineated by Arp (1965), which has diameters of $4'$ and $8'$ centered on the globular cluster NGC 6522 in BW.

Arp (1965) suggested that most of the bluer stars in BW belong to the disk, while redder stars are dominated by the bulge population (plus some red giants in the inner disk). This has been confirmed by later studies (Terndrup 1988; Paczyński *et al.* 1994). To minimize disk contamination, SJW first selected stars redder than $B - V = 1.4$ from photographic photometry (Arp 1965; van den Bergh 1971). They also confined their measurements to stars whose images were relatively uncrowded.

We constructed observing lists for the target stars from a list of plate $(x, y)$ coordinates kindly supplied by B. Jones in advance of the publication of the SJW paper. We computed a coordinate transformation using eleven M stars in common with Blanco *et al.* (1984). The residuals in this transformation were 0.1 arcsec rms in both right ascension and declination. This coordinate transformation was then applied to the remaining SJW stars.

Spectra were obtained in a coordinated program with the 3.9 m Anglo–Australian Telescope (AAT) and the 4 m telescope at Cerro Tololo (CTIO). In all, we observed 401 of the SJW stars. Most of the remaining 26 stars were too close to a brighter star for us to get an uncontaminated spectrum, though a few were missed for other reasons (the star was too faint, poorly centered, was assigned to a broken fiber for the CTIO observations, or a fiber fell out of the AAT plugged–plate system during the observation).

We observed 313 stars at the AAT and 227 at CTIO; 139 stars were observed with both telescopes. This large overlap was deliberate: since we used a photon–counting detector on a Cassegrain spectrograph at the AAT and a CCD on a bench–mounted spectrograph at CTIO, we wanted to compare the data from the two systems to check that there were no systematic differences in the final measurements. All reductions were performed independently, and then the two data sets were combined.

### 2.2. AAT spectroscopy

The AAT observations were made during the 1988, 1989 and 1990 observing seasons. Two different fiber systems were used interchangeably to feed the RGO spectrograph: AUTOFIB at the main Cassegrain focus and FOCAP at the auxiliary focus. It was possible to switch quickly from one to the other by moving a mirror. The detector was the Image Photon Counting System (IPCS).

AUTOFIB (Parry & Sharples 1988) is a robot positioner which allows 64 fibers to be placed independently within a $40'$ field. It can reconfigure the fibers quickly, but requires a minimum separation of $33''$ between targets. FOCAP is a manual plugged–plate system with a $12'$ field at auxiliary focus. It allows the fibers to be packed more closely, but takes longer to reconfigure. Both AUTOFIB and FOCAP used Polymicro–FHP fibers with a diameter of 320 $\mu$m (corresponding to $2.1''$ on the sky). Since the BW field is only a few arcminutes across and is very crowded, we found it more efficient to use FOCAP to observe the fainter stars in our sample and AUTOFIB for the brighter ones.

With both FOCAP and AUTOFIB, 8 to 10 fibers in each field were allocated to sky spectra, using "sky" positions chosen from CCD frames to be free of bright stars. As many as possible of the remaining fibers were allocated to program stars. The total exposure time in each configuration was typically 30–60 minutes with AUTOFIB and 2–3 hours with FOCAP. With the 600V grating, the IPCS spectra covered the region 3750–5700 Å at a resolution of 4.6 Å FWHM (2.0 Å/pixel).

Data reduction was done at the AAO with the FIGARO software package. To extract the individual star and sky spectra, we used twilight sky frames taken at the start and end of each night as templates. Because spectra near the ends of the slit had significant curvature from geometrical distortions in the IPCS image tube, we used a 7th–order polynomial to trace each spectrum. The spectra were then extracted using an aperture 5 pixels wide (corresponding to 360 $\mu$m at the slit, slightly larger than the 320 $\mu$m fibers). Scattered light was also removed by this process. No flat–fielding was necessary for the IPCS spectra.

Next, we corrected for differences in the relative throughput of individual fibers (the "response function"). This is affected both by differences in fiber transmission and by vignetting along the spectrograph slit. Again, we used the twilight sky frames for this calibration. Night–to–night variations in the response function of each fiber bundle were typically 2–3%.

For the wavelength calibration, we used spectra of an argon arc lamp observed each time the fibers were reconfigured (and once an hour during long integrations). After extracting the individual arc spectra, we fitted each with a fifth–order polynomial. Residuals in this fit were typically $\pm 0.2$ Å rms. The fit was then used to transform each program spectrum to a linear wavelength scale covering the region 3860–5650 Å. We then subtracted a "mean sky" spectrum from each program spectrum to give a final, sky–subtracted spectrum for each star.

## 2.3. Spectroscopy at CTIO

We used the CTIO 4 m telescope and ARGUS fiber positioning system (Ingerson 1988, Lutz *et al.* 1990) in the 1989 and 1990 observing seasons. The ARGUS system can independently position 24 pairs of fibers across a $30'$ field, and feeds the light from the fibers to a bench–mounted spectrograph located in an isolated room below the observing floor. Each fiber pair consists of a "star" fiber placed at the position of the primary target, and a "sky" fiber mounted approximately $15''$ radially back from the star fiber. The positions of the star fibers were checked at the telescope with a periscope arrangement, whereby television images of the target star and the fiber, backlit from the spectrograph room, were aligned. In 1989 only 12 fiber pairs were available, and in both years it was sometimes the case that 1–2 fibers per run were broken or their positioners inoperable. Exposure times were typically 1800–2400 s.

For both year's observations we used the blue–optimized air Schmidt camera and the KPGL2 grating, blazed at 4400 Å, in first order. The detector was a thick GEC (#12 in CTIO's designation), with $384 \times 576$ pixels of size 22 $\mu m$, arranged so that the long dimension was in the spectral direction. This chip is coated with a fluorescent dye, giving an efficiency of $\sim 15\%$ below 5000 Å. No order separating filter was used. The spectral range for all observations was 3980–5650 Å ($\pm 50$ Å, depending on the individual nightly setups) at 2.85 Å pixel $^{-1}$. The fibers were 100 $\mu m$ in diameter, corresponding to $1.8''$ on the sky, and projected onto 2.1 pixels FWHM over most of the spectral range. The focus was slightly poorer (2.5 pixels) at the red end. The velocity resolution was therefore approximately 350 km s$^{-1}$ per resolution element in the central portion of the spectrum. The fibers had poor transmission below 4200 Å.

The first steps in the reduction were to correct for overscan level and bias structure. The latter was measured by constructing nightly averages of 25–50 zero–s frames. Laboratory tests on the GEC chip did not reveal problems with the charge transfer efficiency at low light levels, so we did not apply a preflash to the exposures.

In 1989, we obtained one–dimensional flat–fields by placing the fibers in a center of the ARGUS field and illuminating them with a quartz lamp reflecting off a white screen on the inside of the telescope dome. This produced an image which was illuminated only by the fibers and was dark elsewhere. The resulting "dome quartz flats" — one for each star and sky fiber — were extracted as described below, and were normalized to have an average intensity of unity via division by a low–order spline fit. This produced spectra which represented the pixel–to–pixel variations in the chip's sensitivity, as averaged across the spatial extent of each fiber.

In 1990, we constructed a two–dimensional flat field by placing a diffusing glass filter at the end of the fiber bundle on the spectrograph, then illuminating the fibers by pointing the telescope at the afternoon sky. The glass filter was sufficiently thick that variations in intensity in the spatial direction (from the point–spread function of the images of the fibers) and in the wavelength direction (from absorption lines in the solar and terrestrial atmospheres) were smoothed out. From the white appearance of the diffusing filter, we termed this the "milk flat." This flat was normalized to a mean of 1.0 by fitting spline functions along each row and column, and was divided into all the data frames before the individual spectra were extracted, thus correcting all pixels for sensitivity variations whether or not they were illuminated by the fibers.

The first step in the processing was to correct for scattered light which was evident as a smooth pattern of illumination between the parts of the detector which were exposed through the fibers. We fit a second–order polynomial surface to the scattered light, and subtracted this surface from all images.

To begin the extraction of the star and sky spectra, we first traced the location of each spectrum in the well–exposed spectra of the afternoon sky ("solar spectra"), and used apertures centered on these traces for all the stellar spectra. From the location of the spectra the brightest bulge stars observed on each night, we found that the location of the apertures varied by no more than 0.2 pixel in the spatial direction from the start to the end of the night. (This stability is necessary for the success of one–dimensional flat–field technique used in 1989.) We had to employ relatively narrow apertures (full width 4 pixels) because the wings of the stellar spectra overlapped. With such narrow apertures, we typically retrieved about 70% of the signal from each spectrum on the CCD.

At this stage, the 1990 data were properly flattened and extracted. For the 1989 data, we divided the extracted spectra by the normalized spectra from the dome quartz flats.

For both years' observations, we used the dome quartz flats to determine the relative fiber transmission. The individual spectra from the quartz flats varied in intensity by up to 35% in intensity because of the differences in fiber throughput. We ignored the small (1–3%) fiber–to–fiber variations in the dependence of the throughput on wavelength. We also measured the relative fiber throughput by moving the fibers to a central position and taking solar spectra. Here, we arbitrarily designated one aperture as having unit throughput, and computed multiplicative factors to represent the average transmission of the other fibers with respect to the fiducial one. The scale factors from the dome quartz flats agreed with those from the solar spectra to better than 0.5% rms, so we took the average as the factor by which each spectrum was scaled after extraction from the data frame. After this scaling, all spectra are on the same intensity scale, allowing accurate sky subtraction.

Before further processing, we removed the numerous single pixel events ("cosmic rays") from the spectrum by hand. Their locations were recorded so that later measures of line strength at the sites of the cosmic rays could be flagged.

We obtained a preliminary wavelength calibration for each night by placing the fibers in a central position and illuminating them with light from a He–Ne–Ar lamp which was fed to the fiber bundle by a periscope arrangement. A fourth–order polynomial relating pixel location and wavelength was fitted to the spectra extracted in each aperture and then used to interpolate the spectra to a linear wavelength scale. The wavelength calibration was done for each spectrum individually, since the relation of pixel location and wavelength varied by up to 3 pixels between adjacent fibers.

The wavelength solutions were independent of the telescope position for a range in hour angle of at least $\pm 4.5$ hours, and the arc spectra at different telescope positions were always aligned better than than 0.04 pixels ($\approx 15$ km s$^{-1}$). Thus we were able to use arc spectra taken at the start and end of the night rather than taking arc spectra for each fiber configuration (one of the advantages of using a bench–mounted spectrograph).

The preliminary wavelength solution for each spectrum could not be used for radial velocity work because the illumination by the arc lamp was not identical to that from the sky, so we determined velocity corrections for each aperture from the solar spectra. As with the correction for fiber transmission, one fiber (usually the one with the most observations of velocity standards) was chosen to define the zero of velocity. The solar spectra for all the other apertures were then cross–correlated with this reference in the way as we measured stellar radial velocities (see Sec. 3.2). The resulting velocity corrections were $\leq 10$ km s$^{-1}$ for most fibers, but sometimes as large as 25 km s$^{-1}$. The errors in the velocity corrections, as measured from repeated high signal–to–noise observations of solar spectra, are about 4 km s$^{-1}$.

For each observation of a set of bulge stars, we computed an average sky spectrum to correct for moonlight, telluric emission lines and (in the bulge) the faint underlying light from main–sequence stars. For the 1989 observations, since there were only 12 fiber pairs available, we pointed two star fibers at patches in BW which, based on CCD frames, did not contain resolved stars. Most of the sky fibers (50–75%), which were in fixed positions relative to the star fibers, had count levels indistinguishable from the two sky patches, implying that there was no significant contamination from resolved stars in the sky fibers. We therefore computed a mean sky spectrum from the average of the spectra in the designated sky regions and the sky spectra with the lowest counts. For the 1990 observations, we simply averaged the 50–70% of sky spectra with the lowest counts after correction for the relative transmission of the fibers. The average sky spectrum was then subtracted from each of the stellar spectra.

## 2.4. Photometry

Photometry of the target stars comes from three sources:

Run 1: Images from the CTIO 4 m telescope in 1985 May of a single field to the northwest of NGC 6522. The detector was the Kitt Peak PFCCD camera, and the field was $4.96' \times 2.93'$ (with the longer dimension east/west) at a scale of $0.59''$ pixel$^{-1}$. These data are in $BVI$ (Cousins) and were analyzed by Terndrup (1988).

Run 2: Previously unpublished frames of a field to the northeast of NGC 6522, obtained in 1984 June. These frames used the same setup as run 1, and are in $V$ and $I$.

Run 3: Observations in 1988 May with the CTIO 0.9 m telescope and GEC CCD camera. This data set consists of $VI$ images of seven fields which cover most of Arp's (1965) annulus in BW. The frames have a scale of $0.49''$ pixel$^{-1}$ and a field size $4.11' \times 2.55'$, with the longer dimension east/west. Terndrup & Walker (1994) have analyzed a subset of these frames.

Table 1 summarizes the positions, exposure times and seeing for all the photometry. The fields are designated by $m.n$, where $m$ is the run number (see above) and $n$ refers to the $n$th field from that run.

We first processed the raw images from the three runs with the usual combination of overscan regions, zero–exposure frames, dome and sky flats. We then averaged multiple exposures together, after spatially shifting the images so that the stellar images were aligned. The $v$ and $i$ instrumental magnitudes in each frame were measured with DoPHOT (Schechter $et\ al.$ 1993), which was kindly supplied to us by Paul Schechter. The remaining steps in the reduction were to tie the instrumental magnitudes for the various exposures together, then compute the transformation to the Cousins $VI$ system.

In each field, we designated the shortest exposure in each filter as the "standard" frame. We then computed the instrumental magnitude offset between the standard frames and the longer exposures by computing the mean difference in magnitudes for all stars detected on both frames. In each case the magnitude shift that was derived was within 0.02 mag of that expected from the ratio of exposure times. For stars observed more than once, the $v$ and $i$ magnitudes were averaged after weighting by $1/\sigma$, where $\sigma$ is the error in the instrumental magnitude estimated by DoPHOT. At this point, we had average $v$ and $i$ magnitudes for each field.

There are three reference points for transforming our instrumental photometry to the Cousins system: a photometric sequence by Walker & Mack (1986); Terndrup's (1988) calibrations of the frames designated here as run 1; and photometric standards observed conjointly with the frames in run 3 (Graham 1982). Comparing these three photometric scales allowed us to estimate the systematic errors in the photometry.

For run 3, we made 34 observations of 13 standards on two nights and derived transformation equations of the form:

$$
\begin{aligned}
V &= v + c_1 X + \text{const.}, \\
V - I &= 1.012(v - i) + c_2 X + \text{const.},
\end{aligned}
$$

where $X$ is the airmass. The coefficients $c_1$ and $c_2$ were determined independently for each night's observations. The instrumental magnitudes for the standards were measured on the CCD frames using an circular aperture of diameter $9''$ and a sky annulus of diameters $12''$ and $15''$. The rms residuals were 0.017 mag in $V$ and 0.020 mag in $V - I$.

The key step in bringing the instrumental magnitudes to the Cousins system was to determine the offset between the DoPHOT magnitudes and a set of aperture magnitudes on the bulge frames, which were measured in the same way as the photometric standards. We first identified the brightest 7–15 stars on the shortest–exposure frame in each color, then used DAOPHOT (Stetson 1987) to subtract all the remaining stellar images from those frames. We chose DAOPHOT for this operation because it is easier to do the subtraction than in the fully automatic routines in DoPHOT; we also found that DoPHOT — at least in the way we were using it — systematically overestimated the absolute counts in the crowded bulge frames, producing a calibration which was up to 0.05 mag too bright in $V$ and $I$. We then computed a constant offset to bring the instrumental magnitudes onto the aperture values. The typical error in the offset was 0.032 mag in $V$ and 0.015 mag in $V - I$; these errors represent the accuracy to which the magnitudes on each system could be absolutely calibrated. The color error is smaller, suggesting that the presence or absence of faint nearby stars, rather than photon statistics, dominates errors in photometry of the brightest stars.

We then independently calibrated the photometry from each field in run 3 and compared that calibration to the photometry from run 1 and Walker & Mack (1986) sequence. The mean differences (in the sense of run 3 *minus* Walker/Mack) are $-0.010$ and $-0.038$ mag in $V$ and $V - I$ respectively (the calibration for run 3 is slightly redder). The mean differences for run 3 *minus* run 1 are 0.050 and 0.042 mag in $V$ and $V - I$, respectively (the calibration for run 1 is brighter and bluer). These differences, though not small, are consistent with the errors in the absolute calibration of each frame. An independent measure of the systematic error in the photometry is provided by Udalski *et al.* (1992, 1993), who carried out independent photometry in BW. They compared their sequences both to the Terndrup (1988) calibration and to their own observations of standards. They also performed a re-reduction of some frames from run 1, which one of us (DT) supplied to them. They derive $V$ and $V - I$ zero points that are 0.084 brighter and 0.030 mag redder, respectively, than the Walker/Mack photometry which is close to our adopted scale. They discuss possible sources for the difference, suggesting that part of the problem is the determination of the background brightness in the very crowded BW field. Another source may be that they employed a different method for determining the zero point of the photometry on their frames.

The final step was to average the magnitudes and colors for those stars which were measured more than once on overlapping fields. In this step, we calculated a simple average, not correcting for the offset between the independent photometric calibrations.

Figure 1 displays the photometry for the stars in this survey. The small points show the $V$ and $V - I$ values for each star on the CCD frames from runs 1–3, where multiple observations have been averaged together. The larger filled points show the photometry the SJW stars. The magnitudes and colors, along with errors for each, are compiled in A few of the stars with proper motions do not have photometry because they were just outside the CCD frames we obtained. Figure 2 shows the internal errors estimated by DoPHOT in $V$ and $V - I$ for the spectroscopic targets as a function of $V$; in addition to these, the photometry for the individual CCD frames can have systematic errors which are (above) about 0.04 in both $V$ and $V - I$. Those stars with errors which are larger than average were from crowded images, or near frame edges or defects. The final photometry is listed in Columns 2–5 of Table 2.

## 3. Radial velocity measurements

### 3.1. AAT data

We measured radial velocities for all the BW stars by cross–correlation methods. For the AAT data, several spectra of the radial velocity standard HD 203638 (a K0 giant with $V_h = 22.6$ km s$^{-1}$, Anderson *et al.* 1987) were co-added and used as a template for the K–giant program stars. We prepared a separate template for M stars by co-adding our program spectra of seven M6 giants whose radial velocities had already been measured by Sharples *et al.* 1990.

Each object spectrum was continuum–subtracted, then rebinned to a wavelength scale which was linear in $\log_{10}\lambda$ (and corresponded to 158.3 km s$^{-1}$ channel$^{-1}$) before cross–correlation against the template using the spectral region 4200–5500 Å.

We used the cross–correlation coefficient $R$ to estimate the internal error in each velocity measurement. The relation between $R$ and the velocity error was established by a series of Monte Carlo tests in which the K–giant template spectrum was degraded by adding noise and then cross–correlated against the original undegraded spectrum. Typical internal errors for the AAT program spectra were 15–25 km s$^{-1}$. The total (external) errors derived from a comparison with the CTIO and Rich (1988) velocities are about 30% higher, or 20–30 km s$^{-1}$ (see Sec. 3.3 below).

### 3.2. CTIO Data

We measured radial velocities for the CTIO data with the "fxcor" cross–correlation package in IRAF. Each spectrum was first rebinned from a linear wavelength scale to one in which $\log_{10}\lambda$ is a constant, maintaining the same number of pixels as in the original spectrum. For the CTIO data, this corresponded to 179.3 km s$^{-1}$ pixel$^{-1}$. The spectra were then continuum subtracted, and windowed to run from 4750–5500 Å. Several velocity standards from Mayor *et al.* 1984 were observed each night. The standard–star spectra, which had very high signal–to–noise, were extracted and processed in the same way as those of the bulge stars. As a check on our wavelength solution, we observed the velocity standards through several different fibers. It was always possible to reproduce the velocities of the standards (measured against one another) to 10 km s$^{-1}$ rms.

### 3.3. Comparisons of Velocities

Radial velocities for some of the stars in our sample have also been measured by Rich (1990). Rich's spectra were taken either at "low" resolution (4.5 Å; i.e. similar to our AAT and CTIO spectra) or "high" resolution (2.7 Å). We denote these velocities as $v_r(L)$ and $v_r(H)$ respectively. Rich estimates velocity errors of 34 km s$^{-1}$ (L) and 15 km s$^{-1}$ (H) for the two samples.

A cross–comparison between the four BW velocity samples (AAT, CTIO, Rich H, Rich L) allows us to estimate the typical velocity errors for each sample and to check that the zero points of the velocity scales agree. Figure 3 shows the comparison: the velocities in all four samples agree well.

Table 3 lists the mean difference and standard deviation between each pair of velocity samples. Seven stars have discrepant velocities (velocity difference $> 3\sigma$) between the CTIO and AAT samples. After eliminating these, the zero–point difference between the CTIO and AAT samples is very small, only $-0.3 \pm 3.1$ km s$^{-1}$ (mean error). The Rich high– and low–dispersion samples differ in zero point from the AAT/CTIO scale by $4.4 \pm 4.0$ km s$^{-1}$ and $-8.1 \pm 5.3$ km s$^{-1}$, respectively.

From Table 3, we derive external velocity errors of 26 km s$^{-1}$ (AAT), 25 km s$^{-1}$ (CTIO), 17 km s$^{-1}$ (Rich H) and 49 km s$^{-1}$ (Rich L) for the four data sets.

For the CTIO data, this is very close to the mean internal error of 27 km s$^{-1}$ estimated by the cross–correlation program. For the AAT data, however, the external error is 30% higher than the mean estimated internal error (perhaps because the internal error estimate assumed a perfect match in spectral type between the program and template stars). We therefore multiplied our original AAT estimates by a factor of 1.3 to give the final estimated errors in Table 2.

When several velocities were available for a given star, we computed a weighted mean with $1/\sigma^2$ weighting, where $\sigma$ is the estimated internal error. To bring the Rich velocity scales onto the CTIO/AAT zero–point, we added $-4.4$ km s$^{-1}$ to the $v_r$(H) velocities and 8.3 km s$^{-1}$ to the $v_r$(L) values, and used $\sigma = 17$ and $\sigma = 49$ km s$^{-1}$, respectively. Columns 6–13 of Table 2 display the individual velocities, their errors, and the adopted mean velocity. The average velocity error in the combined data set is 12 km s$^{-1}$.

## 4. Line–strength Measurements

We measured several indices of absorption–line strength defined by Faber *et al.* (1985, hereafter FFBG). These indices have been used extensively by the Lick group and others to study both individual stars and the integrated light of galaxies and star clusters. Here we verify that our measurements are on the same system as FFBG. Sec. 6 discusses the line strengths of bulge stars in more detail.

The indices as defined by FFBG are on an instrumental system, since they were measured from Lick Observatory IDS spectra which were not flux calibrated. Their indices were measured on spectra at a resolution of 9 Å and shifted to zero radial velocity. When observing with other detectors and spectrographs, it is therefore necessary to observe enough of the standard stars listed by FFBG to transform the line–strength indices onto the Lick system. There are two aspects to this transformation: correcting for the different continuum response of the detector relative to the Lick IDS (this affects the Mg$_1$, Mg$_2$ and CN indices, which have widely–separated continuum sidebands), and correcting for differences in spectral resolution (usually only important if the observations have a resolution poorer than 9 Å, when weaker lines like Fe 5270 start to be smeared out). For the BW field, we also need to consider the higher extinction along the line–of–sight, which changes the continuum slope in the blue.

For both the AAT and CTIO data sets, the indices were measured after the stellar spectra were shifted to zero radial velocity. Errors in the indices were computed from photon statistics in the spectrum and the nearby mean sky. Each data set was measured independently and transformed to the Lick scale. The two sets were then compared, checked and merged as described below.

### 4.1. The AAT Data

The IPCS is a photon counting system limited to a maximum count rate of $\sim 1$ Hz, so it was necessary to observe the FFBG standards through neutral density filters and to use long integrations to reach the required S/N. For this reason, only nine FFBG stars were observed at the AAT. They spanned a range $2.10 \leq V - K \leq 3.70$ and $-2.6 \leq$ [Fe/H] $\geq +0.42$, as catalogued by FFBG.

These nine stars were used in a first–pass attempt to transform the AAT data to the FFBG scale. The transformations were then checked by a bootstrap comparison of the AAT program stars with observations of the same stars at CTIO, as described below (§4.3).

Line–strength indices were measured for both standard and program stars using the batch-processing rougine BPW (Rich 1988). For the Fe 5270, Fe 5335, $\langle$Fe$\rangle$, H$\beta$, G 4300 and Mg$b$ indices, no corrections were needed to transform to the FFBG system. For the Mg$_1$, Mg$_2$ and CN molecular indices, the standard–star transformations were as follows:

$$\begin{aligned} \text{Mg}_1(\text{AAT}) + 0.045 &= \text{Mg}_1(\text{FFBG}) \\ \text{Mg}_2(\text{AAT}) + 0.044 &= \text{Mg}_2(\text{FFBG}) \\ \text{CN}(\text{AAT}) + 0.023 &= \text{CN}(\text{FFBG}) \end{aligned}$$

For the BW stars, we also checked the effect of reddening by applying a standard extinction curve to the standard–star spectra and remeasuring the indices. These tests showed that only the CN index was significantly affected. For an extinction $A_V = 1.5$ (typical of the BW field, below), we derive:

$$0.982 \times \text{CN}(\text{AAT}, \text{BW}) + 0.016 = \text{CN}(\text{FFBG})$$

to transform the BW observations to the FFBG scale.

### 4.2. The CTIO Data

With the CCD detector at CTIO, we were able to observe a large number of FFBG standards, most with 3–5 repeat observations. Transformations were calculated separately for the 1989 and 1990 CTIO runs. The FFBG standards observed at CTIO spanned the same range in color and metallicity as those observed at the AAT.

We were able to achieve a satisfactory match to the FFBG system for all indices with transformation equations of the form

$$\text{index(FFBG)} = \text{index(CTIO)} + \text{const.}$$

Since different kinds of fiber were used for the two observing seasons at CTIO, it was necessary to transform each year's observations to the FFBG system separately. The transformation coefficients for each index are listed in Table 4, along with the mean error in the mean for the transformation.

Figure 4 shows the correlation between the FFBG indices and our values for the standard stars observed at CTIO and the AAT. The residuals in the transformation to the FFBG system are listed in Table 5. Columns 2–4, respectively, display the rms residuals in the transformation for the various indices, the error in a single measurement for our observations of the FFBG standards, and the mean error in each index quoted by FFBG. These quantities refer to all measures of the FFBG standards from both CTIO and AAT. The error of a single measurement in our data was computed by the rms scatter in repeat observations of the standards from the CTIO data. The last column of Table 5 displays the expected scatter in the transformation to the FFBG system, calculated by combining in quadrature our internal error and FFBG's typical error.

In general, the scatter in Figure 4 is consistent with the errors in a single measurement for our data and the FFBG stars. For the molecular band indices (CN, $Mg_1$ and $Mg_2$), however, the scatter is up to 50% larger than expected from measurement errors alone. These are the indices which use widely–spaced continuum bands, and require a correction for the difference in continuum slope between our spectra and those used by FFBG. The increased scatter therefore reflects the uncertainty in deriving the transformation equations for each of our data sets.

### 4.3.   The merged data set

We now compare the independently–transformed AAT and CTIO indices of our BW sample to confirm that they are on the same system and to check that the errors in each index are estimated correctly.

Figure 5 compares the index values measured on the AAT and CTIO spectra, while Table 6 lists the rms differences and typical errors. Column 2 of that table lists the mean difference in the sense (CTIO *minus* AAT) for each index; the error in this quantity is the error of the mean. Column 3 shows the rms scatter between the data sets, while columns 3 and 4 show the average error for the AAT and CTIO indices.

The rms differences between the AAT and CTIO values for each index are, in general, close to the value expected from the average errors. The scatter is, however, much larger than that for the bright FFBG standards[4], because the BW stars are much fainter than the FFBG standards.

For faint stars, both photon statistics and background light contribute significantly to the measurement errors. To illustrate this, we compute the relative brightness of a star and the background light for a star of magnitude $V = 16.8$, which is the median $V$ magnitude of the BW sample. The mean surface brightness of the unresolved light in the bulge is $\mu_V = 19.7$ mag arcsec$^{-2}$ (Terndrup 1988). The fibers used at CTIO, which have a diameter of 1.8″, therefore see a background of brightness $\mu_V = 18.7$ mag. Since the light from the star is scrambled by the fibers and therefore fills the aperture seen by the spectrograph, a star with $V = 16.8$ will have a surface brightness at the end of the fiber of $\mu_V = 17.8$. This is only a factor of 2.3 brighter than the sky, even if all the light from the star goes down the fiber. The situation is naturally worse in poor seeing or in moonlight.

As can be seen from Figure 5, the molecular indices, which have relatively small errors compared to the range in values of the indices, correlate well between the independent observations. For CN, most of the scatter is due to the larger errors for the CTIO spectra. For Mg$_1$ and Mg$_2$, the correlation probably does not have a slope of unity. This is almost certainly caused by systematic errors in the location of the mean continuum point in these indices because the CTIO spectra did not have the same instrumental continuum slope as the AAT spectra. The correlations are significant, though less strong, for the Mg$b$, Fe 5270, and Fe 5335 indices. H$\beta$, on the other hand, has very large errors compared to the range of this index. For these indices, however, the scatter is statistically what is expected given the errors of measurement. The large errors for these indices have important ramifications for determining metallicities in the bulge.

We computed a weighted mean and error for the AAT and CTIO indices; these are compiled in Table 6. Because the errors in the CN index in the AAT data are generally much smaller than the CTIO values, the weighted mean is close to the AAT value when both measurements are available. For most indices, the mean difference between the AAT and CTIO values is statistically equal to zero and within the uncertainty of the transformations to the FFBG standards (Table 4), indicating that the preliminary AAT transformation discussed above was adequate. For the Mg$b$ index, the mean difference was small but statistically non–zero, so we transformed the AAT indices to match the CTIO data before combining them. The mean difference in the CN index is also non–zero, but here we decided not to apply any constants to bring the two systems into agreement, as the errors in the CTIO values are so large.

In Table 7 we display the averaged indices Fe 5270, Fe 5335, ⟨Fe⟩, H$\beta$, CN, Mg$_1$, and Mg$_2$ for each star. [5] There are a few stars with radial velocity measurements in Table 2 for which the S/N in our spectra was too low to measure line–strength indices. A few measurements of individual indices are also missing because of cosmic ray hits on the CTIO CCD.

---

[4]The values of the CN index from the CTIO observations have extremely large errors because the fiber transmission shortward of 4400 Å is greatly reduced in the ARGUS instrument, and the spectrum near the CN feature was usually very faint.

## 5. The Extinction Towards Baade's Window

Determining the foreground extinction is crucial to any analysis of the BW population. Table 8 summarizes earlier estimates of the extinction to BW. Some of these come from the color–magnitude diagram of the globular cluster NGC 6522 in BW, while others use the $UBV$ colors of foreground stars. Estimates of $E(B − V)_0$ (where the subscript refers to the extinction for a zero–color star) range from 0.49 to 0.56 mag.

In this section, we estimate the extinction to BW using the observed $H\beta$ line–strength indices for stars in our sample, and also compare the extinction in different parts of BW. Most previous studies have adopted an average value for the extinction in BW, or a value appropriate for the clearer parts of the Window, though Blanco $et$ $al.$ (1984) did explore the variation of extinction using the relation between spectral type and photographic $I$ magnitude. Based on deep photographs of BW from Blanco 1984, they divided their BW field into three zones, denoted by the letters A–C. They showed that the M giants in zones B and C were fainter by $\Delta I \approx 0.3$ mag, corresponding to $\Delta E(B − V)_0 = 0.17$. About 2/3 of our sample is in their region A, with the remainder about equally divided between regions B and C.

Our derivation of the reddening to BW assumes that (i) the strength of the $H\beta$ absorption line depends only on temperature, and is insensitive to gravity and metallicity, and (ii) $H\beta$ has the same dependence on temperature in the BW stars as in the FFBG standards. The first assumption is supported by an extensive analysis of the Lick stellar library carried out by Gorgas $et$ $al.$ (1993, hereafter G93).

FFBG and G93 used $V − K$ as their temperature indicator, so to compare our results to theirs we need to transform our $V − I$ colors to $V − K$. We did this by combining our $V$ magnitudes with $K$–band photometry for a small field in BW from Tiede $et$ $al.$ (1995). Figure 6 shows the relation between $V − I$ and $V − K$; both colors are uncorrected for reddening. The solid line is a low–order polynomial which fits $V − K$ as a function of $V − I$; the scatter about this line is consistent with observational error. From this polynomial, we can generate a "calculated" $V − K$ color for stars with $0.7 \leq V − I \leq 4.0$, which we denote $(V − K)_c$. The terms of the polynomial are

$$(V − K)_c = -0.345 + 2.522(V − I) − 0.131(V − I)^2$$

The top panel of Figure 7 plots $H\beta$ against $(V − K)_c$ for our BW sample. Since the errors in

---

H$\beta$ for individual stars are large, we sorted the values by color and binned them into groups of 40 stars. The error bars shown in Figure 7 are errors in the mean for each bin. The solid line shows the run of H$\beta$ with color as derived by FFBG (the more extensive analysis in G93 gives essentially the same result). Because the continuum bands for the H$\beta$ index are very close together in wavelength, the values of H$\beta$ are independent of reddening, and the reddening vector in Figure 7 is therefore horizontal. The dashed line shows the FFBG relation shifted by $E(V - K) = 1.30 \pm 0.07$.

Figure 7 also shows the estimated extinction towards the Blanco *et al.* (1984) regions A (center panel), and the combined regions B/C (lower panel). Given the sizes of our errors, the extinctions towards regions B and C were statistically the same, so we considered regions B and C together. We derive an extinction for region A of $E(V - K) = 1.23 \pm 0.08$, and a value of $E(V - K) = 1.38 \pm 0.12$ for the other regions. This difference is significant at a level of only $1.5\sigma$. Because this is not a large difference, and because the reddening across the whole BW field is patchy, we decided decided to adopt a single reddening for all BW stars using the value derived for the sample as a whole.

To compare our new values of extinction to previous results, we used the relative extinction in different colors as given by Bessell & Brett (1988)

$$
\begin{aligned}
A_V &= 3.12 E(B - V), \\
A_K &= 0.34 E(B - V).
\end{aligned}
$$

which yields $E(V - K) = 2.78 E(B - V)$. We then computed the color excess in $V - I$ and the color excess for a zero–color star using the relations given by Dean *et al.* (1978) as

$$
\begin{aligned}
E(B - V) &= E(B - V)_0 [1 - 0.08(B - V)_0], \\
E(V - I)/E(B - V) &= 1.25 [1 + 0.06(B - V)_0 + 0.014 E(B - V)].
\end{aligned}
$$

We assumed a mean color of $(B - V)_0 = 1.10$ for the entire sample (Terndrup 1988).

Table 8 lists our derived color excesses $E(B - V)_0$ and $E(V - I)$. The value we find for the full sample, $E(B - V)_0 = 0.51 \pm 0.04$, is close to that derived by Walker & Terndrup (1991) from observations of the globular cluster NGC 6522 in BW. The difference in reddening between regions B/C and region A which we derive is about half that found by Blanco *et al.* (1984), but our sample is confined to a smaller area in BW than theirs, and does not extend as far into areas of higher extinction.

## 6. Bulge and Disk Stars

The earliest photometric studies of the bulge (e.g. Arp 1965, van den Bergh 1971) recognized that the CMD in BW contains foreground disk stars as well as those from the giant branch and main sequence of the bulge population. The foreground stars are seen most clearly as a sequence bluer than the bulge's giant branch, arising from disk turnoff stars on the near side of the bulge some 2–5 kpc from the Sun (Terndrup 1988; Ortolani $et$ $al.$ 1993; Paczyński $et$ $al.$ 1994). There are also disk giants mixed on the CMD with the bulge giants. Several previous studies (e.g. Terndrup 1988; Rich 1988; Blanco & Terndrup 1989) used models of galactic structure to estimate that the disk contributes 10 to 20% of the giants on the CMD in BW.

The CMD of the BW field in Fig. 1 suggests that some stars in the SJW sample may be too bright to be members of the bulge. The brightest part of the giant branch in the bulge, as judged from those stars with $V - I \geq 2.5$, is at $V \approx 15.5$. The bulge giant branch is fainter at bluer colors because the stars are further down the giant branch, and at redder colors because of large bolometric corrections in the $V$ band for cool, metal-rich stars.

Figure 8 plots the heliocentric radial velocity $v_r$ and proper motions $\mu_\ell$ and $\mu_b$ of the SJW sample against $V$ magnitude. (The zero point of the SJW proper motions is unknown, so SJW set $\langle \mu_l \rangle = 0$, $\langle \mu_b \rangle = 0$.) All three panels suggest differences in the distribution of stars brighter and fainter than $V \approx 16$. Fainter stars have a larger radial velocity dispersion, and proper motions distributed evenly about $\mu_l = 0$, $\mu_b = 0$. For stars brighter than $V \approx 16$, the radial velocity dispersion in radial velocities is smaller and the mean $\mu_l$ shifts toward positive values.

Table 9 lists the weighted mean velocities and proper motions, and their dispersions, for bright and faint stars. We computed weighted dispersions according to the formalism in Armandroff & Da Costa (1986) and SJW; setting the weights equal to $1/\sigma^2$, where $\sigma$ is the radial velocity error for our data (Table 2), and the tabulated error in proper motion from SJW. The values of $\sigma(v_r)$ and $\langle \mu_\ell \rangle$ differ between the stars with $V \leq 15.5$ and those with $V > 16.0$ by $2\sigma$ and $5\sigma$, respectively.

The difference between $\langle \mu_\ell \rangle$ for stars brighter and fainter than $V \sim 16$ mag is probably the best evidence that many of the brighter stars belong to the old disk. If most of the SJW stars are genuine members of the bulge at distances close to $R_0$, then their mean proper motion relative to the sun should be roughly equal (and opposite) to the sun's transverse velocity about the Galactic center (i.e. $\sim 250$ km s$^{-1}$). In practice, SJW arbitrarily defined $\langle \mu_\ell \rangle$ to be zero. The non–zero $\langle \mu_\ell \rangle$ observed for the bright stars therefore means that they have a net transverse velocity with respect to the Galactic center, as expected for disk stars.

The three–dimensional motions of the bright stars also match what we expect for members of the disk. Column 2 of Table 10 compares the velocity dispersions in $\ell$ and $b$ for the stars with $V < 16.0$ if we assume that they all lie at a distance of 4 kpc from the sun. Since the line–of–sight to BW is close to the Galactic meridian ($\ell = 1°$), we can equate $\sigma_r = \sigma(v_r)$, $\sigma_\phi = \sigma(\mu_\ell)$ and $\sigma_z = \sigma(\mu_b)$. If the bright stars are $\sim 4$ kpc away, they would have a radial velocity anisotropy of $\sigma_\phi^2/\sigma_r^2 = 0.49 \pm 0.09$, close to the expected value of $1/2$ for a flat rotation curve and the value for the velocity anisotropy of disk stars in the solar neighborhood (cf. Lewis & Freeman 1989).

If the same stars were at the bulge distance of 8 kpc (column 3 of Table 10), the radial velocity anisotropy would be four times larger, i.e. $\sigma_\phi^2/\sigma_r^2 = 2.0 \pm 0.4$, in contrast to the fainter stars in column 4 of Table 10, which have a near–isotropic velocity dispersion, and which we argue are true members of the bulge.

That the brighter stars towards BW have a lower velocity dispersion has been noted before. In their study of the kinematics of late M giants in BW, Sharples *et al.* (1990) found a lower velocity dispersion for the brightest stars in their sample (which they identified with a disk population). For M giants with $I < 11.8$, they found $71^{+20}_{-11}$ km s$^{-1}$, close to our value for the K giants with $V < 16.0$ ($79 \pm 16$ km s$^{-1}$). For their fainter stars, they measured a velocity dispersion of $113^{+6}_{-5}$ km s$^{-1}$, identical within the errors to our value of $110 \pm 10$ km s$^{-1}$ for the fainter K giants.

## 7. Mg and Fe strengths

We now examine the Fe and Mg line strengths in our sample of BW K giants, looking in particular at the behavior of the line strength indices with color and comparing this with the cluster and field stars analyzed extensively by G93.

First, however, it is necessary to discuss the philosophy behind the G93 paper. This work was part of a long–term program to obtain a large spectral library of K and M giants for the interpretation of the integrated spectra of elliptical galaxies and bulges. The library contains field giants and dwarfs, along with subgiants and giants in globular and open clusters spanning a range of metallicities. G93 presented an extensive analysis of their library, and computed "fitting functions" which described the behavior of the FFBG indices as a function of temperature (parameterized by $V - K$), gravity ($\log g$), and metal abundance ($Z \equiv$ [Fe/H]). They showed that the Fe 5270 and Fe 5335 indices were sensitive to abundance but insensitive to gravity, while the Mg indices were highly sensitive to both metallicity and gravity. G93 described a procedure for deriving $Z$ and $\log g$ for individual stars from the Mg and Fe indices. For data with good signal–to–noise, this is accurate to about 0.25 dex in $Z$ and 0.23 dex in $\log g$.

G93 also concluded that only one metallicity parameter was needed to characterize abundance variations among the different populations of stars in the library. They reached this conclusion by demonstrating that the scatter about the (three parameter) fitting functions was only slightly larger than the errors of measurement, i.e. that there was no remaining variation which required the introduction of other parameters, particularly those relating to metallicity.

Here, we want to test whether our BW K giants have the same distribution *in index space* as the stars in the FFBG and G93 libraries. We begin by examining the run of the Mg$_2$ and ⟨Fe⟩ indices as a function of $(V - K)_c$ color, shown in Figure 9. The open points denote stars with $V > 15.75$, and the filled points those with $V < 15.75$ (many of which may be foreground disk stars, as discussed above). show indices for the likely foreground disk stars. Both Mg$_2$ and ⟨Fe⟩ show the expected rise for cooler stars, at least up to $(V - K)_c \approx 3.5$. Redder than this, the Mg$_2$ index rises slightly, but ⟨Fe⟩ is sharply reduced because of strong TiO absorption in the continuum sidebands.

The solid lines in Figure 9 represent loci of constant metallicity for a moderately old population at the distance of the bulge (here $R_0 = 8$ kpc). To calculate them, we used giant branches from the Yale isochrones (Green *et al.* 1987) for an age of $10 \times 10^9$ yr, which is a reasonable estimate for the age of the bulk of bulge stars (Holtzman *et al.* 1993; Terndrup 1988 after an adjustment to $R_0 = 8$ kpc). Then for each metallicity [Fe/H] $= -1.0, -0.5, 0.0,$ and $+0.3$, we computed $(V - K)_c$ from the output $V - I$ color, and derived $\log g$ from

$$\log(g/g_\odot) = \log(M/M_\odot) - \log(L/L_\odot) + 4\log(T/T_\odot),$$

where $M$, $L$, and $T$ are respectively the mass, luminosity, and effective temperature at each point in the isochrone. The helium abundance was set to $Y = 0.228 + 2.7Z$ as in Worthey 1994. The resulting sequences represent a set of model giant branches expressed in a space of color/absorption index.

If the G93 fitting functions were applicable to bulge stars, then the distribution of points about the giant branches in Fig. 9 should be the same for both the $Mg_2$ and $\langle Fe \rangle$ indices. In the $\langle Fe \rangle$ plot, most stars lie between the giant branches for $\langle [Fe/H] \rangle = -1.0$ and $+0.3$, with some above the $= +0.3$ locus. Even with some observational scatter, thus suggests that the most metal–rich of our stars have [Fe/H] $\sim +0.5$, consistent with previous work in BW (Whitford & Rich 1983; Rich 1988; McWilliam & Rich 1994). Most of the $Mg_2$ indices for our sample, however, lie above the model sequence for [Fe/H] $= +0.3$; implying that the stars in BW *have significantly stronger Mg absorption* at a given color than cluster and local field stars, upon which the G93 fitting functions are based. This does not appear to be an effect of mis–estimating the reddening to BW. Since the $Mg_2$ index changes more rapidly with temperature than $\langle Fe \rangle$, we could reconcile the two indices by shifting all the points in Figure 9 redwards in $(V - K)_c$ by about 0.5 mag (i.e. reducing the assumed BW reddening E$(V - K)$ from 1.30 mag to 0.8 mag). This would be equivalent to adopting E$(B - V)_0 = 0.32$ mag, which is inconsistent with *all* the reddening determinations in BW.

Figure 10 shows the Mg–Fe discrepancy in a less model–dependent way by plotting $Mg_2$ against $\langle Fe \rangle$ directly. Such a plot is also unaffected by reddening. The top panel of Figure 10 shows individual values for the stars in our sample, along with the model sequences from Figure 9 (solid lines). These are repeated in the lower panel, which also shows (as filled circles) average values for the BW sample, binned in five groups in $V - I$. The error bars show the standard deviation about the mean indices in each bin, and the starred points in the lower panel represent the FFBG standards which we observed in this study.

There is considerable scatter in the distribution of points in Figure 10 (top panel). The scatter above the model sequences (high values of $\langle Fe \rangle$) is roughly what we expect from the observational errors in our sample, which are 0.7 Å in $\langle Fe \rangle$ and 0.066 mag in $Mg_2$ (Table 6). There is more scatter below the model lines (i.e., towards higher $Mg_2$), and the binned values are shifted away from both the model sequences and the FFBG standards toward higher $Mg_2$ at a given $\langle Fe \rangle$.

What could cause the BW stars to have stronger magnesium absorption at a given color or $\langle$Fe$\rangle$? Since Mg line strength increases with gravity, one possibility is that the K giants in BW are ascending the giant branch with significantly higher gravities than the Yale isochrones predict for a moderately old population (here $10 \times 10^9$ yr). It is unlikely that the gravities in the Yale isochrones are seriously in error, since they have similar values to those in other stellar evolution models (Worthey 1994). Higher gravities would imply that the bulge is significantly younger than $10 \times 10^9$ yr, which would be hard to reconcile with CMDs of the main–sequence turnoff.

If many of the BW stars were significantly closer than 8 kpc, they would have lower luminosities and higher gravities than expected, and this might produce the scatter toward higher values of Mg$_2$ in Figure 10. We note, however, that the distribution of points in Figures 9 and 10 for the bright stars in our sample, which we argued above are located at intermediate distances, is not significantly different from those for the bulk of the sample. This is not to say that bulge and disk stars have identical compositions, particularly in the enhancement of light elements; we do conclude, however, that the scatter towards higher Mg$_2$ index in Figure 10 is not caused by the presence in the sample of stars which are much closer than the bulge.

A third alternative is that some of the stars in our sample have [Mg/Fe] $> 0$. The recent high–resolution study of 11 BW stars by McWilliam & Rich 1994 shows that this is possible. We can quantify the effects of age and [Mg/Fe] on the predicted indices as follows: *on average*, the Mg$_2$ indices are about 50% larger than expected from the G93 fitting functions at a given $V - K$. Differentiating the G93 fitting functions yields the expected change in Mg$_2$ and $\langle$Fe$\rangle$ for any change in $\log g$ or $Z$. For $\langle$Fe$\rangle$, we assume (like G93) that $Z \equiv$ [Fe/H]; but use $Z_{\rm Mg} =$ [Fe/H] + [Mg/Fe] in the fitting function for Mg$_2$. At the median color of our sample, $(V - K)_{c,0} = 2.4$, we find $\Delta \log {\rm Mg}_2 / \Delta \log g \approx 0.35$, and $\Delta \log {\rm Mg}_2 / \Delta Z \approx 0.50$ near $Z \approx 0$.

For each assumed value of [Mg/Fe], then, we can compute a change in $log\,g$ which will increase the value of the Mg$_2$ index by 50%. Higher values of [Mg/Fe] require lower gravities to produce the observed strengths of Mg$_2$. Figure 11 shows the relation between turnoff mass for various values of [Mg/Fe]. The solid line is for a mean age of $t = 12 \times 10^9$ yr and the dashed line is for $8 \times 10^9$ yr. If the high Mg$_2$ strengths of our bulge stars are accounted for by gravity alone (i.e., [Mg/Fe] $= 0$), the bulge giants would have to be ascending the giant branches with very high masses ($\sim 2 - 3 M_\odot$). With even moderate Mg enhancements of a few tenths dex, the bulge stars would then have masses as expected for moderately old populations. Indeed, the sensitivity of the Mg$_2$ index to gravity is so strong that we can set tight limits to the Mg enhancement in the bulge by noting that the turnoff photometry in the bulge (Terndrup 1988; Holtzman *et al.* 1993) requires that most of the bulge stars be older than $5 \times 10^9$ yr. Similarly, the bulge stars cannot be older than a Hubble time. The two dotted lines in Figure 11 delimit the range in turnoff mass for these age limits. if the bulge is moderately old we must have [Mg/Fe] $> 0$. We therefore conclude that even for large uncertainties in the mean age of the bulge stars, the high Mg$_2$ index requires $\langle$[Mg/Fe]$\rangle \approx 0.3$ for the the bulk of stars in BW.

## 8. Discussion

We have presented photometry, radial velocities, and indices of absorption strength for a sample of over 400 stars in Baade's Window. The data were obtained in a coordinated program at CTIO and the AAT, and were independently reduced, so that we could assess the random and systematic errors in our measurements. The two data sets were then merged into a single, final set.

A simple analysis of the kinematics suggests that many of the brighter stars in the SJW sample ($V \leq 16.0$) are foreground objects at intermediate distances. Most previous studies of bulge kinematics which have measured the line–of–sight radial velocity dispersion (e.g., Rich 1990; Tyson & Rich 1991; Minniti et al. 1992; Spaenhauer et al. 1992; Blum et al. 1994; Blum et al. 1995) made the simplifying assumption that most of the stars under study are at the same distance as the Galactic center (an exception is Harding & Morrison 1993). This is only justified if the sample reaches faint enough to sample the density maximum through the bulge (cf. Sharples et al. 1990).

The SJW proper motion sample may have other selection effects as well as the inclusion of bright foreground stars. The relatively sharp cutoff at faint magnitudes (Fig. 1) may bias against the inclusion of distant metal–rich stars (since metal–poor stars are more luminous than metal–rich stars of similar spectral type, and so can be seen to greater distances). Such a bias could affect attempts to compare the kinematics of stars of different metallicity. Differences in the kinematics of metal–rich and metal–poor bulge stars are suggested by recent studies (Rich 1990; Zhao et al. 1994; Minniti et al. 1995), but need to be confirmed with large–scale surveys with a proper treatment of selection bias. In the next paper in this series, we will explore selection effects in our sample with metallicity and distance (Sadler et al. 1995).

By comparing the iron and magnesium indices in the SJW stars and stars in the Lick stellar libraries through the G93 fitting functions, we have found that the bulge stars appear to have enhanced magnesium line strengths. The most likely explanation seems to be that many of the BW stars have [Mg/Fe] $\approx 0.3$. The alternatives (both of which appear unlikely) are that the majority of stars in BW are relatively young or much closer than the Galactic center. We caution the reader that we have not yet derived [Mg/Fe] ratios for individual stars. Instead, we have computed, based on the G93 fitting functions, a magnesium enhancement which would shift the average $Mg_2$ index in the $Mg_2$, $\langle Fe \rangle$ plane away from the sequence for local K giants to the locus of points in BW.

Although our suggestion that many of the BW stars are Mg–enhanced needs to be confirmed by a more careful analysis, there are several arguments which support such a conclusion. The magnesium enhancement we find is consistent with the recent analysis of high–resolution spectra of 11 BW stars by McWilliam & Rich (1994), who derive $\langle [Mg/Fe] \rangle = +0.35 \pm 0.15$ (s.d.) and also note that the abundance of Ti is also elevated (those of Si and Ca are nearly solar in their analysis).

More indirectly, from observations of similar absorption indices on low–resolution spectra, Friel & Janes (1993) have argued that there are systematic differences in the [Mg/Fe] ratios of K giants in open clusters, with the amount of [Mg/Fe] correlated with age. Their interpretation was supported by a theoretical analysis of the sensitivity of Mg indices to metallicity, gravity, and [Mg/Fe] (McQuity *et al.* 1994).

A mild overabundance of the light elements in bulge stars would also help reconcile several different metallicity scales which have been proposed for the stars in BW. The average [Fe/H] from iron lines in BW appears to be near solar (Terndrup 1988; McWilliam & Rich 1994). This is lower than the $\langle[\text{Fe/H}]\rangle \approx 0.3$ dex as estimated from the strengths of CO and TiO absorption in M giants (Frogel & Whitford 1987; Frogel *et al.* 1990; Sharples *et al.* 1990) or from the strengths of Ca and Na lines in the infrared (Terndrup *et al.* 1991). All of these metallicity indicators would yield higher abundances than the iron lines if the $\alpha$–capture elements were enhanced in the bulge.

If many bulge stars are enhanced in magnesium, they may resemble stars in elliptical galaxies and the bulges of other spirals. Worthey (1993, Worthey 1994) has computed an extensive set of models based on the G93 fitting functions, which can be used to predict the feature strengths in the integrated light of elliptical galaxies as a function of age and metallicity. In the first published confrontation between the Worthey models and data on ellipticals, Worthey *et al.* 1992 showed that *no combination of age and metallicity* in the models can describe the relative strengths of the Mg and Fe features in elliptical galaxies and the bulges of other spirals such as M 31. With increasing luminosity, the spectra of ellipticals show enhanced Mg indices compared to Fe as compared to the models. They suggest that, in an average large elliptical, the [Mg/Fe] ratio exceeds that of the the stars in the solar neighborhood by $0.2 - 0.3$ dex.

We wish to thank the staff at Cerro Tololo and the Anglo–Australian Observatory who gave us many hours of superb help in the preparation for and execution of our program. We also thank ATAC and the Cerro Tololo TAC, which generously awarded telescope time for a coordinated program at the two observatories. DT acknowledges support from the National Science Foundation awards AST–9157038 (Presidential Young Investigator Program) and INT–9215844 (US/Australia Cooperative Science Program), funds from which greatly aided the analysis of our data. He also thanks the staff of the Anglo–Australian Observatory and the University of Sydney for their hospitality during his visits to Australia. RMR received support from NASA grant NAGW 2479. We received many valuable comments and advice from Michael Bessell, Tim Bedding, Sandra Faber, Jay Gallagher, Ruth Peterson, Nick Suntzeff, Albert Whitford, and Guy Worthey. Glenn Tiede provided computer programs which generated the giant branches in Figures 9 and 10.

Fig. 1.— Color-magnitude diagram in Baade's Window. The small points show photometry for all stars on the CCD frames from this survey, while the larger points identify the stars with proper motions from Spaenhauer *et al.* (1992).

Fig. 2.— Random photometric errors in the photometry of the Baade's Window stars with proper motions. Shown are (bottom panel) the error in $V$ and (top panel) the error in $V - I$ as computed by DoPHOT. In addition to the random error, there are systematic errors of $\sim 0.04$ mag in $V$ and $V - I$ in the zero-point of the photometry on each CCD frame, as discussed in the text.

Fig. 3.— Comparison of radial velocity samples. Shown are the correlations between the AAT and CTIO radial velocities (this paper) and velocities from Rich's (1988) high- and low-dispersion samples (denoted "H" and "L", respectively.) The solid lines in each panel denote identity. The error bars show average internal errors for each sample as estimated in the text. The filled points in the CTIO/AAT comparison denote stars which have discrepant velocities at the $3\sigma$ level.

Fig. 4.— Comparison between measured and standard indices of absorption strength. Symbols are: (*filled triangles*) AAO data; (*open circles*) CTIO observations in 1989; and ($\times$) CTIO observations in 1990. Units are magnitudes for the $Mg_1$, $Mg_2$ and CN indices, Å for the others.

Fig. 5.— Comparison of the CTIO and AAO measures of absorption strength for bulge stars. The solid lines in each panel denote identity, while the error bars represent the typical error of measurement as described in the text.

Fig. 6.— Correlation between $V - I$ (this paper) and $V - K$ for a small region in BW. The latter colors are from Tiede *et al.* (1995). The solid line shows the polynomial relation we used to derive a computed color $(V - K)_c$ from $V - I$.

Fig. 7.— Determination of the extinction towards BW. Each panel shows (solid line) the relation between H$\beta$ index (FFBG) and $V - K$ for nearby stars, and (points) data for this survey binned in groups of 40 stars. The error bars on the points are errors of the mean. The dashed line shows the FFBG relation shifted in $V - K$ to match the points (the H$\beta$ index is insensitive to reddening). The top panel shows the mean reddening for our entire sample, while the central and lower panels show the values derived from the Blanco et al. (1984) regions A and B/C in Baade's Window.

Fig. 8.— Motions of Baade's Window Stars as a function of $V$ magnitude. Plotted are (bottom to top) the heliocentric radial velocity, proper motion in $\ell$, and proper motion in $b$.

Fig. 9.— Theroetical and observed $Mg_2$ and $\langle Fe \rangle$ indices. The small filled points are for $V \leq 16$ while the open circles are for $V > 16$. The solid lines are calculated sequences for isochrones of age $10 \times 10^9$ yr as described in the text. From top to bottom these sequences are for [Fe/H] = 0.3, 0.0, $-0.5$, and $-1.0$, respectively.

Fig. 10.— $Mg_2$ and $\langle Fe \rangle$ indices. Symbols are as in Figure 9. The upper panel shows the individual stars in our sample along with the calculated sequences from Figure 9. The lower panel shows the mean indices binned in four groups, after sorting by $V - I$. The error bars on the points show the dispersion about the mean values; this dispersion is real and much larger than observational scatter. In both panels, the starred points show the indices for FFBG standards.

Fig. 11.— Possible age-Mg enhancement for the bulge. Shown as solid points with line is the locus of turnoff mass and [Mg/Fe] value which would produce the mean enhancement in the Mg index for bulge over local stars as discussed in the text. The solid line is for a mean age of 12 Gyr, while the dashed line is for 8 Gyr. The region between the dotted lines is the allowed turnoff mass for bulge stars based on current age estimates.

TABLE 1. Photometry in Baade's Window.

TABLE 2. Photometry and heliocentric radial velocities.

TABLE 3. Comparison of radial velocities.

TABLE 4. Coefficients in transformation to FFBG system.

TABLE 5. Residuals after transformation to FFBG system.

TABLE 6. Comparison of CTIO/AAO indices.

TABLE 7. Combined line-strength indices.

TABLE 8. Reddening determinations to Baade's Window.

TABLE 9. Velocity dispersion as a function of magnitude.

TABLE 10. Velocity dispersions of bulge and disk stars.

Table 1. Photometry in Baade's Window.

| Field | Telescope | Field center | | Exposure times (s) | | Seeing ($''$) | |
| | | $\alpha(1950)$ | $\delta(1950)$ | $V$ | $I$ | $V$ | $I$ |
|---|---|---|---|---|---|---|---|
| 1.1 | CTIO 4m | 18:00:12.2 | $-29$:59:55 | $3 \times 10$ | $2 \times 5$ | 1.0 | 1.2 |
| | | | | $2 \times 90$ | $2 \times 30$ | | |
| 2.1 | CTIO 4m | 18:00:33.5 | $-29$:59:33 | $4 \times 10$ | $3 \times 7$ | 1.0 | 1.2 |
| 3.1 | CTIO 0.9m | 18:00:12.2 | $-30$:02:45 | $2 \times 60$ | $2 \times 40$ | 1.1 | 0.9 |
| | | | | $2 \times 500$ | $2 \times 300$ | | |
| 3.2 | CTIO 0.9m | 18:00:33.3 | $-30$:02:38 | $2 \times 60$ | $2 \times 40$ | 1.1 | 0.9 |
| | | | | $2 \times 500$ | $2 \times 300$ | | |
| 3.3 | CTIO 0.9m | 18:00:29.9 | $-30$:04:53 | $2 \times 60$ | $2 \times 30$ | 1.1 | 0.9 |
| | | | | $2 \times 500$ | $2 \times 300$ | | |
| 3.4 | CTIO 0.9m | 18:00:13.4 | $-30$:05:08 | $2 \times 60$ | $2 \times 40$ | 1.1 | 0.9 |
| | | | | $2 \times 500$ | $2 \times 300$ | | |
| 3.5 | CTIO 0.9m | 18:00:41.3 | $-30$:00:14 | $2 \times 120$ | $2 \times 30$ | 2.2 | 2.2 |
| | | | | $2 \times 480$ | $2 \times 360$ | | |
| 3.6 | CTIO 0.9m | 18:00:14.4 | $-30$:00:12 | $2 \times 90$ | $2 \times 30$ | 1.7 | 1.2 |
| | | | | $2 \times 500$ | $2 \times 360$ | | |
| 3.7 | CTIO 0.9m | 18:00:20.4 | $-30$:00:11 | $2 \times 90$ | $2 \times 30$ | 1.8 | 1.5 |
| | | | | $2 \times 500$ | $2 \times 360$ | | |

Table 2. Photometry and heliocentric radial velocities.

| Star | $V$ | Photometry | | | | CTIO velocity | | AAO velocity | | Rich (1990) | | Mean | |
|------|-----|-------|-------|-------|-------|-------|-------|-------|-------|-----------|-----------|-------|-------|
|      |     | error | $V-I$ | error | | $v_r$ | error | $v_r$ | error | $v_r(\text{H})$ | $v_r(\text{L})$ | $v_r$ | error |
| 0−025 | 17.188 | 0.023 | 1.461 | 0.044 | | $\cdots$ | $\cdots$ | 119 | 23 | $\cdots$ | $\cdots$ | 119 | 23 |
| 0−333 | 16.017 | 0.027 | 1.801 | 0.021 | | −73 | 15 | $\cdots$ | $\cdots$ | $\cdots$ | $\cdots$ | −73 | 15 |
| 0−340 | 17.184 | 0.041 | 1.759 | 0.035 | | 80 | 39 | 93 | 22 | $\cdots$ | $\cdots$ | 90 | 19 |
| 1−012 | 15.822 | 0.010 | 1.809 | 0.013 | | $\cdots$ | $\cdots$ | −235 | 25 | $\cdots$ | −271 | −241 | 22 |
| 1−015 | 16.832 | 0.019 | 1.625 | 0.027 | | $\cdots$ | $\cdots$ | $\cdots$ | $\cdots$ | $\cdots$ | $\cdots$ | $\cdots$ | $\cdots$ |
| 1−016 | 14.658 | 0.014 | 2.103 | 0.021 | | −48 | 10 | $\cdots$ | $\cdots$ | $\cdots$ | $\cdots$ | −48 | 10 |
| 1−022 | 16.024 | 0.016 | 1.883 | 0.021 | | $\cdots$ | $\cdots$ | 139 | 23 | $\cdots$ | $\cdots$ | 139 | 23 |
| 1−025 | 16.914 | 0.049 | 1.768 | 0.037 | | $\cdots$ | $\cdots$ | −51 | 16 | −54 | −38 | −53 | 11 |
| 1−027 | 16.818 | 0.021 | 1.607 | 0.023 | | 182 | 28 | 177 | 21 | $\cdots$ | $\cdots$ | 179 | 17 |
| 1−034 | 15.462 | 0.018 | 2.605 | 0.023 | | −106 | 17 | $\cdots$ | $\cdots$ | −111 | $\cdots$ | −111 | 12 |
| 1−039 | 17.032 | 0.041 | 1.817 | 0.023 | | $\cdots$ | $\cdots$ | 45 | 14 | 62 | 81 | 52 | 11 |
| 1−041 | 15.376 | 0.017 | 2.001 | 0.022 | | −104 | 10 | $\cdots$ | $\cdots$ | $\cdots$ | $\cdots$ | −104 | 10 |
| 1−043 | 16.778 | 0.017 | 1.628 | 0.022 | | $\cdots$ | $\cdots$ | 99 | 21 | $\cdots$ | $\cdots$ | 99 | 21 |
| 1−047 | 16.828 | 0.024 | 1.696 | 0.031 | | 85 | 14 | $\cdots$ | $\cdots$ | $\cdots$ | $\cdots$ | 85 | 14 |
| 1−051 | 15.666 | 0.097 | 1.854 | 0.060 | | 156 | 20 | $\cdots$ | $\cdots$ | $\cdots$ | $\cdots$ | 156 | 20 |
| 1−053 | 16.702 | 0.013 | 1.439 | 0.023 | | −62 | 16 | $\cdots$ | $\cdots$ | $\cdots$ | $\cdots$ | −62 | 16 |
| 1−054 | 16.631 | 0.033 | 2.012 | 0.026 | | $\cdots$ | $\cdots$ | −40 | 26 | $\cdots$ | $\cdots$ | −40 | 26 |
| 1−058 | 16.832 | 0.016 | 2.138 | 0.020 | | $\cdots$ | $\cdots$ | 100 | 26 | $\cdots$ | $\cdots$ | 100 | 26 |
| 1−065 | 16.824 | 0.023 | 1.763 | 0.029 | | 192 | 43 | 175 | 34 | $\cdots$ | $\cdots$ | 182 | 27 |
| 1−073 | 15.837 | 0.013 | 2.262 | 0.018 | | 135 | 15 | 84 | 39 | $\cdots$ | $\cdots$ | 128 | 14 |
| 1−076 | 15.753 | 0.024 | 2.006 | 0.053 | | −52 | 25 | $\cdots$ | $\cdots$ | −34 | $\cdots$ | −43 | 14 |
| 1−079 | 16.903 | 0.030 | 1.740 | 0.042 | | $\cdots$ | $\cdots$ | $\cdots$ | $\cdots$ | $\cdots$ | $\cdots$ | $\cdots$ | $\cdots$ |
| 1−083 | 16.758 | 0.019 | 1.717 | 0.026 | | −119 | 36 | −96 | 20 | −131 | −128 | −119 | 12 |
| 1−084 | 16.737 | 0.027 | 1.597 | 0.035 | | −176 | 20 | −179 | 25 | $\cdots$ | $\cdots$ | −177 | 16 |
| 1−089 | 17.102 | 0.063 | 1.436 | 0.035 | | 115 | 44 | $\cdots$ | $\cdots$ | $\cdots$ | $\cdots$ | 115 | 44 |
| 1−090 | 17.082 | 0.081 | 1.636 | 0.040 | | $\cdots$ | $\cdots$ | −62 | 34 | $\cdots$ | $\cdots$ | −62 | 34 |
| 1−093 | 16.411 | 0.036 | 1.861 | 0.028 | | −141 | 13 | −80 | 18 | $\cdots$ | −82 | −118 | 10 |
| 1−102 | 16.181 | 0.012 | 1.720 | 0.019 | | −33 | 28 | 8 | 29 | $\cdots$ | −35 | −15 | 19 |
| 1−105 | 15.471 | 0.016 | 2.127 | 0.017 | | $\cdots$ | $\cdots$ | −5 | 36 | $\cdots$ | $\cdots$ | −5 | 36 |
| 1−108 | 16.719 | 0.030 | 1.606 | 0.025 | | $\cdots$ | $\cdots$ | −198 | 25 | $\cdots$ | $\cdots$ | −198 | 25 |
| 1−109 | 16.792 | 0.054 | 1.701 | 0.028 | | 153 | 41 | $\cdots$ | $\cdots$ | $\cdots$ | $\cdots$ | 153 | 41 |
| 1−116 | 16.886 | 0.051 | 1.780 | 0.030 | | $\cdots$ | $\cdots$ | 62 | 17 | $\cdots$ | $\cdots$ | 62 | 17 |
| 1−118 | 17.020 | 0.014 | 1.588 | 0.027 | | −181 | 40 | $\cdots$ | $\cdots$ | $\cdots$ | $\cdots$ | −181 | 40 |
| 1−129 | 17.038 | 0.049 | 1.844 | 0.034 | | $\cdots$ | $\cdots$ | 194 | 16 | $\cdots$ | 188 | 194 | 15 |
| 1−140 | 15.607 | 0.030 | 1.849 | 0.032 | | −5 | 9 | 32 | 25 | 9 | 1 | 1 | 7 |
| 1−141 | 15.978 | 0.018 | 1.782 | 0.038 | | −26 | 12 | −36 | 36 | −84 | 31 | −43 | 9 |

Table 2—Continued

| Star | V | Photometry error | V − I | error | CTIO velocity $v_r$ | error | AAO velocity $v_r$ | error | Rich (1990) $v_r$(H) | $v_r$(L) | Mean $v_r$ | error |
|------|---|------|-------|-------|------|-------|------|-------|--------|--------|------|-------|
| 1−144 | 16.337 | 0.023 | 2.137 | 0.027 | ⋯ | ⋯ | 55 | 20 | ⋯ | 55 | 56 | 19 |
| 1−148 | 17.262 | 0.110 | 1.735 | 0.066 | ⋯ | ⋯ | −108 | 16 | ⋯ | ⋯ | −108 | 16 |
| 1−151 | 15.942 | 0.020 | 1.728 | 0.025 | 102 | 15 | 54 | 39 | 93 | 81 | 93 | 11 |
| 1−152 | 16.884 | 0.023 | 1.673 | 0.029 | ⋯ | ⋯ | ⋯ | ⋯ | ⋯ | ⋯ | ⋯ | ⋯ |
| 1−155 | 16.669 | 0.246 | 1.708 | 0.139 | ⋯ | ⋯ | −89 | 14 | −70 | −56 | −81 | 11 |
| 1−156 | 16.155 | 0.043 | 1.789 | 0.016 | −164 | 11 | −118 | 39 | ⋯ | −135 | −159 | 10 |
| 1−159 | 16.598 | 0.033 | 2.178 | 0.023 | ⋯ | ⋯ | 59 | 25 | ⋯ | ⋯ | 59 | 25 |
| 1−163 | 16.795 | 0.042 | 1.965 | 0.051 | 61 | 40 | 53 | 16 | ⋯ | ⋯ | 54 | 15 |
| 1−167 | 16.257 | 0.059 | 2.926 | 0.020 | ⋯ | ⋯ | −127 | 43 | ⋯ | ⋯ | −127 | 43 |
| 1−175 | 16.425 | 0.020 | 1.813 | 0.024 | ⋯ | ⋯ | −94 | 13 | ⋯ | ⋯ | −94 | 13 |
| 1−177 | 17.174 | 0.023 | 1.736 | 0.033 | ⋯ | ⋯ | −47 | 14 | ⋯ | ⋯ | −47 | 14 |
| 1−178 | 17.095 | 0.024 | 1.748 | 0.036 | ⋯ | ⋯ | ⋯ | ⋯ | ⋯ | ⋯ | ⋯ | ⋯ |
| 1−179 | 16.514 | 0.018 | 2.352 | 0.016 | −29 | 27 | ⋯ | ⋯ | ⋯ | ⋯ | −29 | 27 |
| 1−180 | 17.134 | 0.022 | 1.881 | 0.025 | ⋯ | ⋯ | −20 | 20 | ⋯ | ⋯ | −20 | 20 |
| 1−181 | 16.877 | 0.035 | 1.588 | 0.021 | ⋯ | ⋯ | ⋯ | ⋯ | ⋯ | 172 | 180 | 49 |
| 1−183 | 16.957 | 0.011 | 1.854 | 0.019 | ⋯ | ⋯ | −133 | 14 | ⋯ | ⋯ | −133 | 14 |
| 1−184 | 16.603 | 0.024 | 1.597 | 0.028 | 3 | 11 | ⋯ | ⋯ | ⋯ | ⋯ | 3 | 11 |
| 1−187 | 16.579 | 0.030 | 1.591 | 0.024 | −105 | 20 | −108 | 14 | ⋯ | ⋯ | −107 | 11 |
| 1−189 | 16.426 | 0.051 | 1.544 | 0.019 | ⋯ | ⋯ | ⋯ | ⋯ | ⋯ | ⋯ | ⋯ | ⋯ |
| 1−191 | 16.512 | 0.022 | 1.718 | 0.027 | −148 | 22 | −130 | 12 | ⋯ | ⋯ | −134 | 11 |
| 1−195 | 15.519 | 0.028 | 2.520 | 0.023 | ⋯ | ⋯ | 33 | 31 | ⋯ | ⋯ | 33 | 31 |
| 1−196 | 16.245 | 0.016 | 2.301 | 0.037 | ⋯ | ⋯ | 16 | 25 | 27 | 10 | 21 | 14 |
| 1−200 | 16.515 | 0.457 | 1.708 | 0.082 | −7 | 14 | ⋯ | ⋯ | ⋯ | ⋯ | −7 | 14 |
| 1−202 | 15.937 | 0.019 | 1.900 | 0.028 | −103 | 12 | −94 | 26 | −97 | ⋯ | −101 | 9 |
| 1−203 | 15.158 | 0.016 | 2.170 | 0.017 | 48 | 15 | 71 | 22 | ⋯ | ⋯ | 55 | 12 |
| 1−218 | 16.956 | 0.024 | 1.655 | 0.026 | 44 | 11 | 47 | 14 | ⋯ | ⋯ | 45 | 9 |
| 1−221 | 17.148 | 0.009 | 1.899 | 0.016 | −82 | 46 | −50 | 20 | ⋯ | ⋯ | −55 | 18 |
| 1−223 | 17.015 | 0.013 | 1.606 | 0.027 | ⋯ | ⋯ | 14 | 26 | ⋯ | ⋯ | 14 | 26 |
| 1−224 | 16.566 | 0.007 | 1.869 | 0.010 | 110 | 32 | ⋯ | ⋯ | ⋯ | ⋯ | 110 | 32 |
| 1−226 | ⋯ | ⋯ | ⋯ | ⋯ | ⋯ | ⋯ | −96 | 27 | ⋯ | ⋯ | −96 | 27 |
| 1−228 | ⋯ | ⋯ | ⋯ | ⋯ | −30 | 39 | −44 | 29 | ⋯ | ⋯ | −39 | 23 |
| 1−232 | 15.938 | 0.008 | 2.194 | 0.011 | −191 | 16 | −177 | 25 | ⋯ | ⋯ | −187 | 13 |
| 1−233 | 16.750 | 0.014 | 1.722 | 0.019 | 38 | 20 | 50 | 20 | ⋯ | ⋯ | 44 | 14 |
| 1−234 | 16.863 | 0.011 | 1.669 | 0.014 | ⋯ | ⋯ | 107 | 27 | ⋯ | ⋯ | 107 | 27 |
| 1−235 | 16.468 | 0.032 | 1.986 | 0.036 | ⋯ | ⋯ | 176 | 18 | ⋯ | ⋯ | 176 | 18 |
| 1−236 | 16.703 | 0.009 | 1.760 | 0.012 | ⋯ | ⋯ | −30 | 31 | ⋯ | ⋯ | −30 | 31 |

Table 2—Continued

| Star | $V$ | Photometry error | $V - I$ | error | CTIO velocity $v_r$ | error | AAO velocity $v_r$ | error | Rich (1990) $v_r(H)$ | $v_r(L)$ | Mean $v_r$ | error |
|------|-----|------------------|---------|-------|---------------------|-------|--------------------|-------|----------------------|----------|-----------|-------|
| 1−239 | 16.801 | 0.032 | 1.601 | 0.035 | 71 | 28 | $\cdots$ | $\cdots$ | $\cdots$ | $\cdots$ | 71 | 28 |
| 1−249 | 16.795 | 0.019 | 1.634 | 0.057 | 22 | 15 | $\cdots$ | $\cdots$ | $\cdots$ | $\cdots$ | 22 | 15 |
| 1−263 | 16.864 | 0.012 | 1.634 | 0.020 | −250 | 37 | $\cdots$ | $\cdots$ | $\cdots$ | $\cdots$ | −250 | 37 |
| 1−264 | 14.414 | 0.004 | 1.871 | 0.008 | −40 | 15 | −18 | 29 | $\cdots$ | $\cdots$ | −35 | 13 |
| 1−285 | 16.584 | 0.010 | 1.826 | 0.014 | $\cdots$ | $\cdots$ | 104 | 17 | $\cdots$ | $\cdots$ | 104 | 17 |
| 1−291 | 17.084 | 0.018 | 1.695 | 0.024 | $\cdots$ | $\cdots$ | 100 | 23 | $\cdots$ | $\cdots$ | 100 | 23 |
| 1−292 | 16.928 | 0.016 | 1.711 | 0.018 | 149 | 31 | $\cdots$ | $\cdots$ | $\cdots$ | $\cdots$ | 149 | 31 |
| 1−293 | 17.131 | 0.033 | 1.763 | 0.039 | −88 | 39 | −60 | 16 | $\cdots$ | $\cdots$ | −64 | 15 |
| 1−298 | 16.073 | 0.009 | 2.391 | 0.011 | $\cdots$ | $\cdots$ | −126 | 26 | $\cdots$ | $\cdots$ | −126 | 26 |
| 1−303 | 16.986 | 0.011 | 1.742 | 0.014 | $\cdots$ | $\cdots$ | −46 | 16 | $\cdots$ | $\cdots$ | −46 | 16 |
| 1−304 | 17.119 | 0.014 | 1.728 | 0.027 | $\cdots$ | $\cdots$ | −182 | 18 | $\cdots$ | −140 | −176 | 17 |
| 1−312 | 16.361 | 0.008 | 2.384 | 0.010 | $\cdots$ | $\cdots$ | $\cdots$ | $\cdots$ | $\cdots$ | $\cdots$ | $\cdots$ | $\cdots$ |
| 1−316 | 16.701 | 0.023 | 2.188 | 0.024 | 162 | 15 | 153 | 25 | $\cdots$ | $\cdots$ | 160 | 13 |
| 1−318 | 15.379 | 0.007 | 2.497 | 0.008 | 82 | 12 | $\cdots$ | $\cdots$ | $\cdots$ | $\cdots$ | 82 | 12 |
| 1−319 | 16.330 | 0.008 | 1.984 | 0.010 | $\cdots$ | $\cdots$ | 9 | 31 | $\cdots$ | 9 | 11 | 26 |
| 1−320 | 15.651 | 0.010 | 2.437 | 0.012 | 140 | 20 | 89 | 31 | $\cdots$ | $\cdots$ | 125 | 17 |
| 1−321 | 16.737 | 0.011 | 1.757 | 0.014 | $\cdots$ | $\cdots$ | −62 | 18 | $\cdots$ | $\cdots$ | −62 | 18 |
| 1−322 | 14.499 | 0.017 | 1.938 | 0.019 | −31 | 18 | −74 | 25 | −74 | $\cdots$ | −59 | 11 |
| 1−324 | 14.634 | 0.008 | 2.616 | 0.011 | −102 | 10 | $\cdots$ | $\cdots$ | $\cdots$ | $\cdots$ | −102 | 10 |
| 1−325 | 16.779 | 0.009 | 1.580 | 0.020 | $\cdots$ | $\cdots$ | $\cdots$ | $\cdots$ | $\cdots$ | $\cdots$ | $\cdots$ | $\cdots$ |
| 1−326 | 16.437 | 0.008 | 1.851 | 0.010 | 49 | 46 | $\cdots$ | $\cdots$ | $\cdots$ | $\cdots$ | 49 | 46 |
| 1−332 | 16.710 | 0.127 | 1.948 | 0.129 | $\cdots$ | $\cdots$ | −52 | 31 | $\cdots$ | $\cdots$ | −52 | 31 |
| 1−335 | 16.277 | 0.020 | 1.894 | 0.020 | $\cdots$ | $\cdots$ | 87 | 25 | $\cdots$ | 257 | 124 | 22 |
| 1−340 | 16.586 | 0.011 | 1.554 | 0.014 | −48 | 50 | −73 | 20 | $\cdots$ | $\cdots$ | −70 | 19 |
| 1−343 | 15.611 | 0.015 | 3.246 | 0.016 | $\cdots$ | $\cdots$ | −116 | 25 | $\cdots$ | $\cdots$ | −116 | 25 |
| 1−344 | 16.910 | 0.016 | 1.704 | 0.025 | $\cdots$ | $\cdots$ | −14 | 39 | $\cdots$ | $\cdots$ | −14 | 39 |
| 1−345 | 16.955 | 0.011 | 1.604 | 0.016 | −20 | 53 | −57 | 16 | $\cdots$ | $\cdots$ | −54 | 15 |
| 1−346 | 17.226 | 0.213 | 1.629 | 0.012 | $\cdots$ | $\cdots$ | $\cdots$ | $\cdots$ | $\cdots$ | $\cdots$ | $\cdots$ | $\cdots$ |
| 1−348 | 16.664 | 0.011 | 1.790 | 0.031 | −14 | 27 | $\cdots$ | $\cdots$ | $\cdots$ | $\cdots$ | −14 | 27 |
| 1−349 | 16.748 | 0.022 | 1.651 | 0.057 | $\cdots$ | $\cdots$ | −13 | 21 | $\cdots$ | $\cdots$ | −13 | 21 |
| 1−353 | 16.262 | 0.008 | 2.151 | 0.043 | $\cdots$ | $\cdots$ | $\cdots$ | $\cdots$ | $\cdots$ | $\cdots$ | $\cdots$ | $\cdots$ |
| 1−357 | 16.944 | 0.015 | 1.588 | 0.034 | $\cdots$ | $\cdots$ | 113 | 22 | $\cdots$ | $\cdots$ | 113 | 22 |
| 1−369 | 16.960 | 0.010 | 1.599 | 0.014 | $\cdots$ | $\cdots$ | 92 | 31 | $\cdots$ | $\cdots$ | 92 | 31 |
| 1−374 | 16.844 | 0.017 | 1.795 | 0.025 | −110 | 27 | −84 | 14 | $\cdots$ | $\cdots$ | −90 | 12 |
| 1−379 | 17.043 | 0.015 | 1.560 | 0.020 | 28 | 16 | −108 | 34 | $\cdots$ | $\cdots$ | −40 | 68[a] |
| 2−014 | 16.979 | 0.050 | 1.247 | 0.064 | −25 | 39 | 61 | 39 | $\cdots$ | $\cdots$ | 18 | 28 |

Table 2—Continued

| Star | $V$ | Photometry | | | | CTIO velocity | | AAO velocity | | Rich (1990) | | Mean | |
|------|-----|-------|-------|-------|-------|-------|-------|-------|-------|-------|-------|-------|-------|
| | | error | $V - I$ | error | | $v_r$ | error | $v_r$ | error | $v_r(H)$ | $v_r(L)$ | $v_r$ | error |
| 2−015 | 16.085 | 0.016 | 2.112 | 0.087 | 201 | 61 | 125 | 34 | $\cdots$ | $\cdots$ | 143 | 30 |
| 2−016 | 16.875 | 0.012 | 1.620 | 0.016 | 222 | 29 | $\cdots$ | $\cdots$ | $\cdots$ | $\cdots$ | 222 | 29 |
| 2−018 | 16.639 | 0.009 | 1.911 | 0.012 | $\cdots$ | $\cdots$ | 191 | 23 | $\cdots$ | $\cdots$ | 191 | 23 |
| 2−019 | 16.783 | 0.009 | 4.936 | 0.017 | $\cdots$ | $\cdots$ | −283 | 20 | $\cdots$ | $\cdots$ | −283 | 20 |
| 2−021 | 16.382 | 0.006 | 3.640 | 0.010 | $\cdots$ | $\cdots$ | −37 | 20 | $\cdots$ | $\cdots$ | −37 | 20 |
| 2−028 | 16.907 | 0.009 | 1.737 | 0.018 | $\cdots$ | $\cdots$ | 81 | 19 | $\cdots$ | $\cdots$ | 81 | 19 |
| 2−031 | 16.916 | 0.011 | 1.617 | 0.015 | $\cdots$ | $\cdots$ | $\cdots$ | $\cdots$ | $\cdots$ | $\cdots$ | $\cdots$ | $\cdots$ |
| 2−033 | 15.514 | 0.008 | 1.838 | 0.010 | 16 | 9 | −21 | 77 | 21 | $\cdots$ | 16 | 8 |
| 2−035 | 17.020 | 0.011 | 1.752 | 0.014 | $\cdots$ | $\cdots$ | −11 | 20 | $\cdots$ | $\cdots$ | −11 | 20 |
| 2−036 | 17.025 | 0.015 | 1.599 | 0.024 | −40 | 23 | −28 | 23 | $\cdots$ | $\cdots$ | −34 | 16 |
| 2−037 | 16.898 | 0.010 | 1.724 | 0.014 | −137 | 39 | $\cdots$ | $\cdots$ | $\cdots$ | $\cdots$ | −137 | 39 |
| 2−040 | 16.630 | 0.012 | 1.533 | 0.015 | $\cdots$ | $\cdots$ | 138 | 20 | 138 | 88 | 133 | 13 |
| 2−042 | 16.546 | 0.008 | 1.875 | 0.011 | $\cdots$ | $\cdots$ | 65 | 39 | −22 | −13 | −11 | 15 |
| 2−043 | 16.533 | 0.007 | 2.164 | 0.009 | $\cdots$ | $\cdots$ | 6 | 25 | $\cdots$ | $\cdots$ | 6 | 25 |
| 2−048 | 16.740 | 0.026 | 1.806 | 0.042 | $\cdots$ | $\cdots$ | 27 | 31 | $\cdots$ | $\cdots$ | 27 | 31 |
| 2−049 | 16.023 | 0.007 | 1.552 | 0.010 | −18 | 14 | −84 | 17 | $\cdots$ | −63 | −45 | 11 |
| 2−050 | 17.154 | 0.022 | 1.478 | 0.025 | $\cdots$ | $\cdots$ | −77 | 59 | $\cdots$ | $\cdots$ | −77 | 59 |
| 2−051 | 17.041 | 0.015 | 1.909 | 0.016 | $\cdots$ | $\cdots$ | 94 | 17 | $\cdots$ | $\cdots$ | 94 | 17 |
| 2−055 | 17.328 | 0.021 | 1.932 | 0.023 | $\cdots$ | $\cdots$ | 112 | 17 | $\cdots$ | $\cdots$ | 112 | 17 |
| 2−059 | 17.364 | 0.012 | 1.597 | 0.026 | $\cdots$ | $\cdots$ | 24 | 48 | $\cdots$ | $\cdots$ | 24 | 48 |
| 2−062 | 16.621 | 0.045 | 1.879 | 0.130 | $\cdots$ | $\cdots$ | −32 | 18 | $\cdots$ | $\cdots$ | −32 | 18 |
| 2−065 | 15.975 | 0.029 | 2.475 | 0.030 | −22 | 15 | −63 | 27 | $\cdots$ | $\cdots$ | −32 | 13 |
| 2−067 | 16.962 | 0.015 | 1.647 | 0.021 | 195 | 30 | 179 | 16 | $\cdots$ | $\cdots$ | 183 | 14 |
| 2−069 | 17.022 | 0.027 | 1.630 | 0.031 | −59 | 41 | $\cdots$ | $\cdots$ | $\cdots$ | $\cdots$ | −59 | 41 |
| 2−075 | 16.328 | 0.010 | 2.074 | 0.013 | $\cdots$ | $\cdots$ | −47 | 20 | $\cdots$ | $\cdots$ | −47 | 20 |
| 2−081 | 16.737 | 0.012 | 1.757 | 0.016 | 54 | 25 | −37 | 21 | $\cdots$ | $\cdots$ | 1 | 16 |
| 2−086 | 16.911 | 0.024 | 1.672 | 0.026 | $\cdots$ | $\cdots$ | −46 | 48 | $\cdots$ | $\cdots$ | −46 | 48 |
| 2−088 | 15.778 | 0.008 | 1.664 | 0.013 | −20 | 17 | −2 | 22 | $\cdots$ | $\cdots$ | −13 | 13 |
| 2−092 | 15.439 | 0.017 | 3.050 | 0.020 | $\cdots$ | $\cdots$ | −234 | 48 | $\cdots$ | $\cdots$ | −234 | 48 |
| 2−096 | 16.975 | 0.011 | 1.610 | 0.015 | $\cdots$ | $\cdots$ | $\cdots$ | $\cdots$ | $\cdots$ | $\cdots$ | $\cdots$ | $\cdots$ |
| 2−097 | 15.332 | 0.005 | 4.134 | 0.040 | $\cdots$ | $\cdots$ | $\cdots$ | $\cdots$ | $\cdots$ | $\cdots$ | $\cdots$ | $\cdots$ |
| 2−100 | 17.018 | 0.024 | 1.839 | 0.027 | $\cdots$ | $\cdots$ | 136 | 25 | $\cdots$ | $\cdots$ | 136 | 25 |
| 2−101 | 15.862 | 0.007 | 1.762 | 0.010 | −20 | 15 | −27 | 27 | $\cdots$ | $\cdots$ | −22 | 13 |
| 2−102 | 16.900 | 0.010 | 1.614 | 0.015 | $\cdots$ | $\cdots$ | −142 | 23 | $\cdots$ | $\cdots$ | −142 | 23 |
| 2−103 | 17.146 | 0.015 | 1.929 | 0.019 | $\cdots$ | $\cdots$ | 65 | 25 | $\cdots$ | $\cdots$ | 65 | 25 |
| 2−105 | 17.158 | 0.016 | 1.731 | 0.019 | −101 | 45 | −48 | 18 | $\cdots$ | $\cdots$ | −55 | 17 |

Table 2—Continued

| Star | $V$ | Photometry | | | CTIO velocity | | AAO velocity | | Rich (1990) | | Mean | |
| | | error | $V-I$ | error | $v_r$ | error | $v_r$ | error | $v_r$(H) | $v_r$(L) | $v_r$ | error |
|------|-----|-------|-------|-------|-------|-------|-------|-------|---------|---------|-------|-------|
| 2−106 | 17.090 | 0.011 | 1.646 | 0.015 | ⋯ | ⋯ | −24 | 29 | ⋯ | ⋯ | −24 | 29 |
| 2−108 | 16.534 | 0.009 | 2.089 | 0.011 | −80 | 51 | −97 | 20 | ⋯ | ⋯ | −95 | 19 |
| 2−109 | 17.104 | 0.014 | 1.634 | 0.021 | −20 | 41 | −6 | 27 | ⋯ | ⋯ | −10 | 23 |
| 2−116 | 17.263 | 0.016 | 1.869 | 0.033 | ⋯ | ⋯ | −8 | 18 | 2 | ⋯ | −5 | 12 |
| 2−117 | 16.581 | 0.008 | 1.662 | 0.011 | −31 | 25 | −84 | 26 | ⋯ | ⋯ | −56 | 18 |
| 2−119 | 15.690 | 0.007 | 1.677 | 0.008 | ⋯ | ⋯ | ⋯ | ⋯ | −218 | −232 | −222 | 16 |
| 2−120 | 16.806 | 0.011 | 1.650 | 0.014 | 26 | 30 | ⋯ | ⋯ | ⋯ | ⋯ | 26 | 30 |
| 2−122 | 14.600 | 0.006 | 2.265 | 0.009 | 100 | 15 | ⋯ | ⋯ | 115 | ⋯ | 105 | 11 |
| 2−126 | 16.791 | 0.013 | 1.728 | 0.019 | ⋯ | ⋯ | 83 | 29 | ⋯ | ⋯ | 83 | 29 |
| 2−131 | 16.611 | 0.009 | 1.582 | 0.012 | 72 | 11 | ⋯ | ⋯ | ⋯ | ⋯ | 72 | 11 |
| 2−133 | 16.231 | 0.008 | 1.682 | 0.011 | −53 | 13 | ⋯ | ⋯ | ⋯ | ⋯ | −53 | 13 |
| 2−135 | ⋯ | ⋯ | ⋯ | ⋯ | ⋯ | ⋯ | ⋯ | ⋯ | ⋯ | ⋯ | ⋯ | ⋯ |
| 2−136 | 15.777 | 0.007 | 1.637 | 0.010 | ⋯ | ⋯ | ⋯ | ⋯ | 84 | 140 | 87 | 16 |
| 2−137 | 16.240 | 0.045 | 1.875 | 0.052 | −125 | 15 | ⋯ | ⋯ | ⋯ | ⋯ | −125 | 15 |
| 2−138 | 16.188 | 0.035 | 1.695 | 0.042 | 32 | 30 | ⋯ | ⋯ | ⋯ | ⋯ | 32 | 30 |
| 2−145 | 17.063 | 0.009 | 1.848 | 0.014 | ⋯ | ⋯ | 40 | 21 | 44 | 86 | 44 | 13 |
| 2−146 | 15.722 | 0.005 | 1.763 | 0.009 | −224 | 17 | ⋯ | ⋯ | ⋯ | ⋯ | −224 | 17 |
| 2−147 | 17.097 | 0.022 | 1.580 | 0.027 | 96 | 44 | ⋯ | ⋯ | 115 | 94 | 102 | 15 |
| 2−149 | 16.350 | 0.006 | 1.982 | 0.009 | 169 | 12 | ⋯ | ⋯ | ⋯ | ⋯ | 169 | 12 |
| 2−152 | 16.427 | 0.007 | 2.524 | 0.009 | ⋯ | ⋯ | 87 | 43 | ⋯ | ⋯ | 87 | 43 |
| 2−155 | 16.838 | 0.009 | 1.688 | 0.028 | ⋯ | ⋯ | −108 | 34 | ⋯ | ⋯ | −108 | 34 |
| 2−158 | 16.734 | 0.013 | 3.159 | 0.017 | 43 | 18 | ⋯ | ⋯ | ⋯ | ⋯ | 43 | 18 |
| 2−159 | 16.867 | 0.014 | 1.905 | 0.021 | ⋯ | ⋯ | −67 | 17 | ⋯ | ⋯ | −67 | 17 |
| 2−167 | 17.211 | 0.009 | 1.903 | 0.028 | ⋯ | ⋯ | −196 | 17 | −235 | −212 | −217 | 12 |
| 2−168 | 17.105 | 0.008 | 1.782 | 0.015 | ⋯ | ⋯ | −34 | 23 | ⋯ | ⋯ | −34 | 23 |
| 2−170 | 16.717 | 0.009 | 1.705 | 0.015 | ⋯ | ⋯ | −76 | 20 | ⋯ | ⋯ | −76 | 20 |
| 2−172 | 17.001 | 0.012 | 1.786 | 0.021 | ⋯ | ⋯ | 4 | 23 | ⋯ | ⋯ | 4 | 23 |
| 2−174 | 16.899 | 0.021 | 1.823 | 0.048 | −139 | 31 | −97 | 18 | −121 | 22 | −116 | 11 |
| 2−175 | 16.728 | 0.021 | 1.996 | 0.024 | 248 | 40 | ⋯ | ⋯ | ⋯ | ⋯ | 248 | 40 |
| 2−178 | 16.858 | 0.012 | 1.918 | 0.036 | 154 | 33 | −2 | 20 | ⋯ | ⋯ | 76 | 78[a] |
| 2−187 | 16.075 | 0.007 | 1.694 | 0.011 | −25 | 12 | −58 | 43 | ⋯ | ⋯ | −27 | 12 |
| 2−188 | 17.072 | 0.031 | 1.730 | 0.034 | −188 | 14 | −178 | 27 | ⋯ | ⋯ | −186 | 12 |
| 2−192 | 16.523 | 0.006 | 1.535 | 0.009 | −44 | 21 | −16 | 25 | ⋯ | ⋯ | −32 | 16 |
| 2−196 | 16.342 | 0.009 | 1.813 | 0.011 | ⋯ | ⋯ | −120 | 14 | ⋯ | ⋯ | −120 | 14 |
| 2−197 | 17.115 | 0.017 | 2.192 | 0.020 | ⋯ | ⋯ | −28 | 31 | ⋯ | ⋯ | −28 | 31 |
| 2−198 | 16.647 | 0.007 | 1.990 | 0.010 | −80 | 29 | −67 | 48 | ⋯ | ⋯ | −77 | 25 |

Table 2—Continued

| Star | $V$ | Photometry error | $V-I$ | error | CTIO velocity $v_r$ | error | AAO velocity $v_r$ | error | Rich (1990) $v_r(\mathrm{H})$ | $v_r(\mathrm{L})$ | Mean $v_r$ | error |
|------|-----|------|------|------|------|------|------|------|------|------|------|------|
| 2−199 | 15.263 | 0.005 | 1.619 | 0.011 | $\cdots$ | $\cdots$ | $\cdots$ | $\cdots$ | −60 | $\cdots$ | −64 | 17 |
| 2−200 | 15.971 | 0.004 | 1.767 | 0.016 | 30 | 25 | 25 | 34 | 16 | $\cdots$ | 19 | 13 |
| 2−201 | 17.103 | 0.011 | 1.942 | 0.015 | $\cdots$ | $\cdots$ | 115 | 27 | $\cdots$ | 39 | 99 | 24 |
| 2−202 | 17.110 | 0.019 | 1.654 | 0.021 | −12 | 50 | $\cdots$ | $\cdots$ | $\cdots$ | $\cdots$ | −12 | 50 |
| 2−203 | 17.270 | 0.010 | 1.799 | 0.014 | $\cdots$ | $\cdots$ | −79 | 27 | $\cdots$ | $\cdots$ | −79 | 27 |
| 2−204 | 16.850 | 0.007 | 1.982 | 0.010 | −277 | 42 | −212 | 23 | $\cdots$ | $\cdots$ | −227 | 20 |
| 2−205 | 17.291 | 0.009 | 1.849 | 0.013 | $\cdots$ | $\cdots$ | 167 | 22 | $\cdots$ | $\cdots$ | 167 | 22 |
| 2−207 | 16.843 | 0.008 | 1.850 | 0.014 | $\cdots$ | $\cdots$ | 40 | 19 | $\cdots$ | $\cdots$ | 40 | 19 |
| 2−213 | 17.315 | 0.038 | 1.930 | 0.041 | $\cdots$ | $\cdots$ | 107 | 17 | $\cdots$ | $\cdots$ | 107 | 17 |
| 2−216 | 16.279 | 0.008 | 1.752 | 0.011 | −68 | 10 | −51 | 48 | −89 | $\cdots$ | −74 | 8 |
| 2−219 | 16.900 | 0.013 | 1.762 | 0.017 | $\cdots$ | $\cdots$ | −24 | 25 | $\cdots$ | $\cdots$ | −24 | 25 |
| 2−220 | 16.949 | 0.009 | 1.773 | 0.012 | 31 | 38 | 53 | 12 | $\cdots$ | $\cdots$ | 51 | 11 |
| 2−228 | 16.157 | 0.013 | 1.659 | 0.014 | −78 | 15 | −81 | 20 | $\cdots$ | $\cdots$ | −79 | 12 |
| 2−232 | 17.086 | 0.008 | 1.873 | 0.011 | $\cdots$ | $\cdots$ | −229 | 17 | $\cdots$ | $\cdots$ | −229 | 17 |
| 2−235 | 17.017 | 0.012 | 1.561 | 0.021 | 130 | 21 | $\cdots$ | $\cdots$ | $\cdots$ | $\cdots$ | 130 | 21 |
| 2−237 | 16.311 | 0.009 | 2.572 | 0.040 | 93 | 38 | 101 | 53 | $\cdots$ | $\cdots$ | 96 | 31 |
| 2−244 | 16.229 | 0.011 | 2.002 | 0.015 | $\cdots$ | $\cdots$ | −163 | 25 | $\cdots$ | $\cdots$ | −163 | 25 |
| 2−247 | 16.811 | 0.015 | 1.615 | 0.025 | 98 | 26 | $\cdots$ | $\cdots$ | $\cdots$ | $\cdots$ | 98 | 26 |
| 2−248 | 15.992 | 0.004 | 1.752 | 0.007 | −108 | 11 | $\cdots$ | $\cdots$ | $\cdots$ | $\cdots$ | −108 | 11 |
| 2−249 | 16.642 | 0.018 | 1.698 | 0.021 | −169 | 40 | −131 | 22 | $\cdots$ | $\cdots$ | −140 | 19 |
| 2−250 | 17.614 | 0.011 | 1.763 | 0.034 | $\cdots$ | $\cdots$ | 137 | 21 | $\cdots$ | $\cdots$ | 137 | 21 |
| 2−252 | 16.610 | 0.006 | 1.823 | 0.009 | 302 | 12 | 258 | 25 | 272 | 283 | 286 | 9 |
| 2−253 | 17.680 | 0.016 | 1.754 | 0.022 | $\cdots$ | $\cdots$ | 48 | 26 | $\cdots$ | $\cdots$ | 48 | 26 |
| 2−254 | 16.983 | 0.012 | 1.818 | 0.030 | $\cdots$ | $\cdots$ | −169 | 17 | $\cdots$ | $\cdots$ | −169 | 17 |
| 2−259 | 17.424 | 0.023 | 1.884 | 0.032 | $\cdots$ | $\cdots$ | −18 | 17 | $\cdots$ | $\cdots$ | −18 | 17 |
| 2−261 | 16.095 | 0.008 | 2.000 | 0.016 | 122 | 16 | 114 | 20 | $\cdots$ | 103 | 118 | 12 |
| 2−264 | 16.880 | 0.024 | 1.651 | 0.098 | −21 | 17 | −12 | 18 | $\cdots$ | $\cdots$ | −17 | 12 |
| 2−272 | 17.093 | 0.008 | 1.924 | 0.012 | 66 | 44 | 81 | 16 | $\cdots$ | $\cdots$ | 79 | 15 |
| 2−276 | 16.981 | 0.008 | 2.104 | 0.017 | $\cdots$ | $\cdots$ | −17 | 26 | $\cdots$ | $\cdots$ | −17 | 26 |
| 2−279 | 16.734 | 0.008 | 1.807 | 0.016 | −141 | 35 | −81 | 20 | $\cdots$ | $\cdots$ | −96 | 17 |
| 2−281 | 16.503 | 0.007 | 1.823 | 0.017 | $\cdots$ | $\cdots$ | 154 | 39 | $\cdots$ | $\cdots$ | 154 | 39 |
| 2−282 | 17.041 | 0.007 | 1.594 | 0.013 | $\cdots$ | $\cdots$ | −146 | 53 | $\cdots$ | $\cdots$ | −146 | 53 |
| 2−284 | 15.994 | 0.006 | 2.329 | 0.009 | 37 | 20 | 37 | 26 | $\cdots$ | $\cdots$ | 37 | 16 |
| 2−285 | 17.053 | 0.009 | 1.686 | 0.014 | 58 | 47 | 79 | 20 | $\cdots$ | $\cdots$ | 76 | 18 |
| 2−286 | 16.183 | 0.017 | 1.797 | 0.018 | 9 | 16 | −53 | 25 | $\cdots$ | $\cdots$ | −9 | 13 |
| 2−288 | 16.976 | 0.026 | 1.507 | 0.035 | $\cdots$ | $\cdots$ | $\cdots$ | $\cdots$ | $\cdots$ | $\cdots$ | $\cdots$ | $\cdots$ |

Table 2—Continued

| Star | $V$ | Photometry error | $V-I$ | error | CTIO velocity $v_r$ | error | AAO velocity $v_r$ | error | Rich (1990) $v_r(H)$ | $v_r(L)$ | Mean $v_r$ | error |
|------|-----|---------|-------|-------|------|-------|------|-------|---------|---------|------|-------|
| 2−297 | 16.571 | 0.010 | 1.632 | 0.014 | $\cdots$ | $\cdots$ | −201 | 20 | $\cdots$ | $\cdots$ | −201 | 20 |
| 2−300 | 15.405 | 0.032 | 2.262 | 0.034 | −46 | 10 | $\cdots$ | $\cdots$ | $\cdots$ | $\cdots$ | −46 | 10 |
| 2−304 | $\cdots$ | $\cdots$ | $\cdots$ | $\cdots$ | −129 | 44 | −84 | 16 | $\cdots$ | $\cdots$ | −89 | 15 |
| 2−306 | 17.208 | 0.021 | 1.816 | 0.038 | $\cdots$ | $\cdots$ | 1 | 20 | $\cdots$ | $\cdots$ | 1 | 20 |
| 2−307 | 16.310 | 0.005 | 2.248 | 0.013 | −39 | 14 | −79 | 22 | $\cdots$ | $\cdots$ | −51 | 12 |
| 2−310 | 16.416 | 0.009 | 1.875 | 0.012 | −108 | 38 | $\cdots$ | $\cdots$ | $\cdots$ | $\cdots$ | −108 | 38 |
| 2−312 | 16.651 | 0.026 | 1.657 | 0.031 | −89 | 17 | −193 | 21 | $\cdots$ | $\cdots$ | −141 | 52[a] |
| 2−316 | 16.874 | 0.008 | 1.580 | 0.015 | $\cdots$ | $\cdots$ | −102 | 29 | $\cdots$ | $\cdots$ | −102 | 29 |
| 2−317 | 17.045 | 0.010 | 1.540 | 0.017 | 50 | 43 | 69 | 22 | $\cdots$ | $\cdots$ | 65 | 20 |
| 2−318 | 17.085 | 0.036 | 1.579 | 0.038 | $\cdots$ | $\cdots$ | −68 | 22 | $\cdots$ | $\cdots$ | −68 | 22 |
| 2−319 | 17.059 | 0.022 | 1.897 | 0.026 | $\cdots$ | $\cdots$ | 157 | 19 | $\cdots$ | $\cdots$ | 157 | 19 |
| 2−321 | 16.883 | 0.012 | 1.818 | 0.017 | $\cdots$ | $\cdots$ | −81 | 18 | $\cdots$ | $\cdots$ | −81 | 18 |
| 2−323 | 17.008 | 0.030 | 1.885 | 0.032 | $\cdots$ | $\cdots$ | −113 | 18 | $\cdots$ | $\cdots$ | −113 | 18 |
| 2−326 | 17.216 | 0.020 | 1.649 | 0.048 | $\cdots$ | $\cdots$ | 121 | 14 | $\cdots$ | $\cdots$ | 121 | 14 |
| 2−327 | 17.300 | 0.014 | 1.768 | 0.018 | 67 | 45 | 27 | 22 | $\cdots$ | $\cdots$ | 35 | 20 |
| 3−018 | $\cdots$ | $\cdots$ | $\cdots$ | $\cdots$ | 93 | 43 | $\cdots$ | $\cdots$ | $\cdots$ | $\cdots$ | 93 | 43 |
| 3−035 | 16.774 | 0.018 | 1.719 | 0.021 | 19 | 43 | $\cdots$ | $\cdots$ | $\cdots$ | $\cdots$ | 19 | 43 |
| 3−036 | 17.183 | 0.025 | 1.560 | 0.028 | 16 | 32 | 16 | 34 | $\cdots$ | $\cdots$ | 16 | 23 |
| 3−043 | 14.166 | 0.017 | 2.024 | 0.019 | 10 | 13 | $\cdots$ | $\cdots$ | $\cdots$ | $\cdots$ | 10 | 13 |
| 3−044 | 16.958 | 0.025 | 1.769 | 0.033 | −148 | 50 | $\cdots$ | $\cdots$ | $\cdots$ | $\cdots$ | −148 | 50 |
| 3−054 | 17.214 | 0.019 | 2.025 | 0.026 | 38 | 47 | 42 | 17 | $\cdots$ | $\cdots$ | 42 | 16 |
| 3−059 | 16.917 | 0.013 | 1.764 | 0.023 | $\cdots$ | $\cdots$ | −110 | 13 | $\cdots$ | $\cdots$ | −110 | 13 |
| 3−060 | 17.056 | 0.012 | 1.598 | 0.016 | −25 | 15 | −58 | 14 | $\cdots$ | $\cdots$ | −43 | 10 |
| 3−069 | 16.628 | 0.011 | 2.549 | 0.016 | $\cdots$ | $\cdots$ | 20 | 36 | $\cdots$ | $\cdots$ | 20 | 36 |
| 3−071 | 17.029 | 0.010 | 1.613 | 0.015 | 183 | 20 | 170 | 12 | $\cdots$ | $\cdots$ | 173 | 10 |
| 3−073 | 16.914 | 0.012 | 1.719 | 0.016 | 247 | 47 | 240 | 17 | $\cdots$ | $\cdots$ | 241 | 16 |
| 3−079 | 16.789 | 0.008 | 1.724 | 0.016 | $\cdots$ | $\cdots$ | 331 | 29 | $\cdots$ | $\cdots$ | 331 | 29 |
| 3−087 | 16.315 | 0.007 | 2.655 | 0.011 | 37 | 8 | 12 | 31 | $\cdots$ | $\cdots$ | 35 | 8 |
| 3−092 | 16.499 | 0.008 | 2.419 | 0.014 | −63 | 26 | −63 | 39 | $\cdots$ | $\cdots$ | −63 | 22 |
| 3−095 | 16.382 | 0.010 | 1.653 | 0.012 | 88 | 13 | 94 | 18 | $\cdots$ | $\cdots$ | 90 | 11 |
| 3−097 | 16.873 | 0.009 | 1.716 | 0.012 | 91 | 14 | 126 | 25 | $\cdots$ | $\cdots$ | 99 | 12 |
| 3−104 | 16.649 | 0.008 | 1.647 | 0.033 | −156 | 22 | 47 | 18 | $\cdots$ | $\cdots$ | −55 | 102[a] |
| 3−110 | 16.981 | 0.012 | 1.558 | 0.015 | $\cdots$ | $\cdots$ | −101 | 36 | $\cdots$ | $\cdots$ | −101 | 36 |
| 3−111 | 15.200 | 0.015 | 2.203 | 0.021 | 12 | 15 | 44 | 25 | $\cdots$ | $\cdots$ | 20 | 13 |
| 3−114 | 15.228 | 0.015 | 2.496 | 0.020 | −140 | 15 | −97 | 39 | $\cdots$ | $\cdots$ | −134 | 14 |
| 3−119 | 16.319 | 0.010 | 1.995 | 0.013 | 117 | 19 | 91 | 26 | $\cdots$ | $\cdots$ | 108 | 15 |

Table 2—Continued

| Star | $V$ | Photometry error | $V - I$ | error | CTIO velocity $v_r$ | error | AAO velocity $v_r$ | error | Rich (1990) $v_r(H)$ | $v_r(L)$ | Mean $v_r$ | error |
|------|-----|------------------|---------|-------|---------------------|-------|---------------------|-------|----------------------|----------|------------|-------|
| 3−122 | 17.066 | 0.011 | 2.163 | 0.014 | $\cdots$ | $\cdots$ | −224 | 18 | $\cdots$ | $\cdots$ | −224 | 18 |
| 3−123 | 16.720 | 0.009 | 1.605 | 0.013 | $\cdots$ | $\cdots$ | −46 | 21 | $\cdots$ | $\cdots$ | −46 | 21 |
| 3−128 | 16.691 | 0.011 | 1.752 | 0.015 | $\cdots$ | $\cdots$ | $\cdots$ | $\cdots$ | $\cdots$ | $\cdots$ | $\cdots$ | $\cdots$ |
| 3−134 | 16.258 | 0.008 | 1.768 | 0.010 | −39 | 12 | $\cdots$ | $\cdots$ | $\cdots$ | $\cdots$ | −39 | 12 |
| 3−135 | 15.498 | 0.005 | 1.901 | 0.007 | $\cdots$ | $\cdots$ | $\cdots$ | $\cdots$ | $\cdots$ | $\cdots$ | $\cdots$ | $\cdots$ |
| 3−143 | 16.587 | 0.010 | 1.730 | 0.012 | 8 | 23 | $\cdots$ | $\cdots$ | $\cdots$ | $\cdots$ | 8 | 23 |
| 3−144 | 15.948 | 0.005 | 1.751 | 0.008 | $\cdots$ | $\cdots$ | −18 | 22 | $\cdots$ | $\cdots$ | −18 | 22 |
| 3−151 | 14.111 | 0.005 | 1.989 | 0.007 | −62 | 13 | $\cdots$ | $\cdots$ | $\cdots$ | $\cdots$ | −62 | 13 |
| 3−152 | 16.144 | 0.006 | 1.913 | 0.009 | −327 | 11 | −305 | 17 | $\cdots$ | −294 | −319 | 9 |
| 3−153 | 16.739 | 0.008 | 1.791 | 0.014 | −137 | 34 | −31 | 34 | $\cdots$ | $\cdots$ | −84 | 24 |
| 3−156 | 16.866 | 0.024 | 1.567 | 0.026 | $\cdots$ | $\cdots$ | −65 | 53 | $\cdots$ | $\cdots$ | −65 | 53 |
| 3−157 | 16.534 | 0.017 | 2.046 | 0.018 | $\cdots$ | $\cdots$ | −35 | 29 | $\cdots$ | −130 | −57 | 25 |
| 3−159 | 16.541 | 0.007 | 1.735 | 0.010 | $\cdots$ | $\cdots$ | 21 | 18 | $\cdots$ | 28 | 23 | 17 |
| 3−160 | 16.495 | 0.006 | 1.646 | 0.009 | $\cdots$ | $\cdots$ | 166 | 53 | 189 | $\cdots$ | 183 | 16 |
| 3−161 | $\cdots$ | $\cdots$ | $\cdots$ | $\cdots$ | $\cdots$ | $\cdots$ | −27 | 36 | $\cdots$ | $\cdots$ | −27 | 36 |
| 3−163 | 17.070 | 0.018 | 4.028 | 0.019 | $\cdots$ | $\cdots$ | $\cdots$ | $\cdots$ | $\cdots$ | $\cdots$ | $\cdots$ | $\cdots$ |
| 3−164 | 15.093 | 0.005 | 1.825 | 0.007 | 50 | 15 | $\cdots$ | $\cdots$ | $\cdots$ | $\cdots$ | 50 | 15 |
| 3−175 | 16.907 | 0.018 | 1.919 | 0.021 | −35 | 41 | −15 | 17 | $\cdots$ | $\cdots$ | −18 | 16 |
| 3−181 | 17.191 | 0.011 | 1.909 | 0.015 | −183 | 43 | −178 | 18 | $\cdots$ | $\cdots$ | −179 | 17 |
| 3−182 | 16.132 | 0.009 | 2.231 | 0.011 | 42 | 16 | −18 | 25 | $\cdots$ | $\cdots$ | 25 | 13 |
| 3−184 | 17.107 | 0.012 | 1.698 | 0.019 | −102 | 32 | −68 | 18 | $\cdots$ | $\cdots$ | −76 | 16 |
| 3−187 | 15.694 | 0.005 | 1.668 | 0.008 | $\cdots$ | $\cdots$ | −209 | 59 | $\cdots$ | $\cdots$ | −209 | 59 |
| 3−188 | 16.516 | 0.011 | 1.910 | 0.021 | −62 | 30 | $\cdots$ | $\cdots$ | $\cdots$ | $\cdots$ | −62 | 30 |
| 3−189 | 17.234 | 0.016 | 1.879 | 0.020 | 192 | 45 | 166 | 14 | $\cdots$ | $\cdots$ | 168 | 13 |
| 3−192 | 16.640 | 0.021 | 1.788 | 0.023 | −210 | 15 | −144 | 59 | $\cdots$ | $\cdots$ | −206 | 15 |
| 3−194 | 16.294 | 0.009 | 2.454 | 0.017 | 59 | 24 | $\cdots$ | $\cdots$ | $\cdots$ | $\cdots$ | 59 | 24 |
| 3−197 | 17.228 | 0.030 | 1.849 | 0.034 | $\cdots$ | $\cdots$ | 12 | 14 | $\cdots$ | −5 | 11 | 13 |
| 3−202 | 16.685 | 0.018 | 2.857 | 0.021 | $\cdots$ | $\cdots$ | 72 | 53 | $\cdots$ | $\cdots$ | 72 | 53 |
| 3−205 | 16.386 | 0.008 | 2.418 | 0.009 | −92 | 15 | −112 | 43 | $\cdots$ | $\cdots$ | −94 | 14 |
| 3−206 | 17.101 | 0.007 | 1.745 | 0.059 | $\cdots$ | $\cdots$ | −3 | 20 | $\cdots$ | $\cdots$ | −3 | 20 |
| 3−209 | 16.674 | 0.019 | 1.991 | 0.012 | −73 | 30 | −71 | 18 | −55 | $\cdots$ | −66 | 11 |
| 3−210 | 16.258 | 0.033 | 3.519 | 0.054 | $\cdots$ | $\cdots$ | 55 | 22 | $\cdots$ | $\cdots$ | 55 | 22 |
| 3−211 | 17.007 | 0.012 | 1.592 | 0.022 | −68 | 15 | $\cdots$ | $\cdots$ | $\cdots$ | $\cdots$ | −68 | 15 |
| 3−220 | 16.719 | 0.009 | 1.736 | 0.013 | $\cdots$ | $\cdots$ | −51 | 39 | $\cdots$ | $\cdots$ | −51 | 39 |
| 3−223 | 13.803 | 0.005 | 1.533 | 0.007 | 63 | 15 | $\cdots$ | $\cdots$ | $\cdots$ | $\cdots$ | 63 | 15 |
| 3−224 | 17.081 | 0.018 | 2.001 | 0.019 | $\cdots$ | $\cdots$ | −13 | 20 | $\cdots$ | 0 | −10 | 19 |

Table 2—Continued

| Star | $V$ | Photometry error | $V - I$ | error | CTIO velocity $v_r$ | error | AAO velocity $v_r$ | error | Rich (1990) $v_r$(H) | $v_r$(L) | Mean $v_r$ | error |
|------|-----|------------------|---------|-------|---------------------|-------|--------------------|-------|----------------------|----------|------------|-------|
| 3−230 | 16.780 | 0.021 | 1.741 | 0.023 | −45 | 29 | −54 | 29 | ⋯ | ⋯ | −50 | 21 |
| 3−231 | 16.521 | 0.072 | 1.519 | 0.074 | 15 | 11 | ⋯ | ⋯ | ⋯ | ⋯ | 15 | 11 |
| 3−234 | 15.806 | 0.007 | 1.673 | 0.009 | −5 | 12 | 20 | 12 | ⋯ | ⋯ | 8 | 8 |
| 3−236 | 16.664 | 0.008 | 2.164 | 0.011 | ⋯ | ⋯ | 26 | 17 | ⋯ | ⋯ | 26 | 17 |
| 3−238 | 16.285 | 0.011 | 2.001 | 0.029 | ⋯ | ⋯ | −20 | 17 | ⋯ | ⋯ | −20 | 17 |
| 3−239 | 17.093 | 0.014 | 1.626 | 0.017 | ⋯ | ⋯ | −86 | 25 | ⋯ | ⋯ | −86 | 25 |
| 3−240 | 17.212 | 0.018 | 1.796 | 0.020 | ⋯ | ⋯ | 134 | 23 | ⋯ | ⋯ | 134 | 23 |
| 3−241 | 16.183 | 0.007 | 2.077 | 0.009 | −20 | 13 | 29 | 19 | ⋯ | ⋯ | −4 | 11 |
| 3−242 | 17.071 | 0.012 | 1.874 | 0.015 | ⋯ | ⋯ | 105 | 15 | ⋯ | ⋯ | 105 | 15 |
| 3−249 | 16.568 | 0.008 | 1.637 | 0.011 | −175 | 17 | −223 | 20 | ⋯ | ⋯ | −195 | 13 |
| 3−254 | 15.282 | 0.011 | 2.135 | 0.019 | 16 | 12 | 28 | 23 | ⋯ | ⋯ | 19 | 11 |
| 3−257 | 16.869 | 0.013 | 1.663 | 0.018 | ⋯ | ⋯ | −56 | 39 | ⋯ | ⋯ | −56 | 39 |
| 3−258 | 16.878 | 0.009 | 1.629 | 0.013 | ⋯ | ⋯ | 58 | 27 | ⋯ | ⋯ | 58 | 27 |
| 3−266 | 16.770 | 0.013 | 1.588 | 0.018 | ⋯ | ⋯ | −56 | 36 | ⋯ | ⋯ | −56 | 36 |
| 3−268 | 15.879 | 0.009 | 1.139 | 0.048 | 11 | 14 | 24 | 21 | ⋯ | ⋯ | 15 | 12 |
| 3−269 | 14.837 | 0.008 | 1.874 | 0.018 | 123 | 14 | ⋯ | ⋯ | ⋯ | ⋯ | 123 | 14 |
| 3−271 | 17.087 | 0.010 | 1.822 | 0.014 | ⋯ | ⋯ | −124 | 13 | ⋯ | ⋯ | −124 | 13 |
| 3−274 | 17.020 | 0.036 | 1.576 | 0.040 | ⋯ | ⋯ | ⋯ | ⋯ | ⋯ | ⋯ | ⋯ | ⋯ |
| 3−275 | 17.125 | 0.010 | 1.805 | 0.014 | ⋯ | ⋯ | 30 | 14 | ⋯ | ⋯ | 30 | 14 |
| 3−278 | 15.879 | 0.007 | 1.677 | 0.010 | 12 | 13 | 90 | 27 | ⋯ | ⋯ | 27 | 12 |
| 3−280 | 17.165 | 0.011 | 1.714 | 0.021 | 54 | 40 | 71 | 14 | ⋯ | ⋯ | 69 | 13 |
| 3−281 | 17.283 | 0.012 | 1.769 | 0.018 | 126 | 47 | ⋯ | ⋯ | ⋯ | ⋯ | 126 | 47 |
| 3−286 | 17.350 | 0.021 | 1.640 | 0.022 | −55 | 40 | −112 | 34 | ⋯ | ⋯ | −88 | 26 |
| 3−291 | 17.043 | 0.019 | 1.664 | 0.023 | 13 | 31 | 19 | 13 | ⋯ | ⋯ | 18 | 12 |
| 3−295 | 17.202 | 0.011 | 3.827 | 0.016 | −97 | 36 | ⋯ | ⋯ | ⋯ | ⋯ | −97 | 36 |
| 4−003 | ⋯ | ⋯ | ⋯ | ⋯ | ⋯ | ⋯ | ⋯ | ⋯ | 208 | 223 | 207 | 16 |
| 4−004 | ⋯ | ⋯ | ⋯ | ⋯ | 0 | 15 | ⋯ | ⋯ | ⋯ | ⋯ | 0 | 15 |
| 4−006 | ⋯ | ⋯ | ⋯ | ⋯ | −165 | 14 | −112 | 53 | ⋯ | ⋯ | −162 | 14 |
| 4−009 | 17.420 | 0.032 | 1.810 | 0.033 | ⋯ | ⋯ | 70 | 20 | ⋯ | ⋯ | 70 | 20 |
| 4−021 | 17.035 | 0.065 | 1.746 | 0.037 | ⋯ | ⋯ | 19 | 19 | ⋯ | ⋯ | 19 | 19 |
| 4−022 | 15.912 | 0.027 | 1.877 | 0.020 | 72 | 6 | ⋯ | ⋯ | 78 | 90 | 72 | 6 |
| 4−026 | 16.895 | 0.025 | 1.620 | 0.022 | ⋯ | ⋯ | −35 | 39 | ⋯ | ⋯ | −35 | 39 |
| 4−027 | 16.924 | 0.012 | 1.624 | 0.020 | ⋯ | ⋯ | −46 | 34 | ⋯ | ⋯ | −46 | 34 |
| 4−031 | 15.340 | 0.018 | 1.862 | 0.016 | ⋯ | ⋯ | −160 | 39 | ⋯ | ⋯ | −160 | 39 |
| 4−033 | 15.013 | 0.032 | 1.746 | 0.015 | 13 | 11 | ⋯ | ⋯ | ⋯ | ⋯ | 13 | 11 |
| 4−036 | 17.249 | 0.027 | 1.510 | 0.034 | ⋯ | ⋯ | ⋯ | ⋯ | ⋯ | ⋯ | ⋯ | ⋯ |

Table 2—Continued

| Star | $V$ | Photometry | | | CTIO velocity | | AAO velocity | | Rich (1990) | | Mean | |
|------|-----|-------|-------|-------|-------|-------|-------|-------|--------|--------|-------|-------|
| | | error | $V-I$ | error | $v_r$ | error | $v_r$ | error | $v_r(H)$ | $v_r(L)$ | $v_r$ | error |
| 4−045 | 16.966 | 0.055 | 1.539 | 0.029 | 57 | 43 | ⋯ | ⋯ | ⋯ | ⋯ | 57 | 43 |
| 4−047 | 16.975 | 0.052 | 1.666 | 0.064 | 200 | 19 | ⋯ | ⋯ | ⋯ | ⋯ | 200 | 19 |
| 4−048 | 16.513 | 0.042 | 3.414 | 0.027 | ⋯ | ⋯ | ⋯ | ⋯ | ⋯ | ⋯ | ⋯ | ⋯ |
| 4−049 | 16.584 | 0.012 | 2.248 | 0.015 | ⋯ | ⋯ | 194 | 39 | ⋯ | ⋯ | 194 | 39 |
| 4−051 | 16.772 | 0.063 | 1.550 | 0.034 | −38 | 18 | −45 | 21 | ⋯ | ⋯ | −41 | 14 |
| 4−052 | 16.430 | 0.026 | 1.881 | 0.014 | −251 | 36 | −248 | 25 | ⋯ | ⋯ | −249 | 21 |
| 4−054 | 15.956 | 0.022 | 2.727 | 0.015 | 173 | 15 | ⋯ | ⋯ | ⋯ | ⋯ | 173 | 15 |
| 4−062 | 17.094 | 0.038 | 1.731 | 0.024 | −139 | 38 | ⋯ | ⋯ | ⋯ | ⋯ | −139 | 38 |
| 4−065 | 16.868 | 0.021 | 1.706 | 0.031 | −107 | 9 | −97 | 14 | ⋯ | ⋯ | −104 | 8 |
| 4−069 | 16.455 | 0.015 | 1.680 | 0.021 | 31 | 13 | −20 | 20 | ⋯ | ⋯ | 16 | 11 |
| 4−070 | 16.044 | 0.006 | 1.840 | 0.012 | −41 | 12 | ⋯ | ⋯ | ⋯ | ⋯ | −41 | 12 |
| 4−071 | 15.815 | 0.012 | 1.667 | 0.017 | −59 | 16 | ⋯ | ⋯ | ⋯ | −80 | −53 | 15 |
| 4−074 | 16.765 | 0.010 | 1.522 | 0.013 | ⋯ | ⋯ | ⋯ | ⋯ | ⋯ | ⋯ | ⋯ | ⋯ |
| 4−075 | 16.280 | 0.020 | 2.245 | 0.009 | ⋯ | ⋯ | 135 | 25 | ⋯ | ⋯ | 135 | 25 |
| 4−086 | 16.450 | 0.026 | 1.914 | 0.017 | 184 | 15 | ⋯ | ⋯ | ⋯ | ⋯ | 184 | 15 |
| 4−093 | 14.357 | 0.015 | 1.979 | 0.012 | −78 | 13 | ⋯ | ⋯ | ⋯ | ⋯ | −78 | 13 |
| 4−102 | 16.846 | 0.041 | 1.580 | 0.020 | ⋯ | ⋯ | 138 | 27 | ⋯ | ⋯ | 138 | 27 |
| 4−111 | 15.478 | 0.033 | 1.975 | 0.015 | 74 | 15 | ⋯ | ⋯ | ⋯ | ⋯ | 74 | 15 |
| 4−114 | 14.596 | 0.066 | 2.437 | 0.040 | 41 | 15 | ⋯ | ⋯ | ⋯ | ⋯ | 41 | 15 |
| 4−121 | 16.654 | 0.025 | 1.979 | 0.025 | 9 | 17 | −3 | 20 | ⋯ | ⋯ | 4 | 13 |
| 4−139 | 17.022 | 0.012 | 1.772 | 0.016 | ⋯ | ⋯ | 73 | 16 | ⋯ | ⋯ | 73 | 16 |
| 4−143 | 16.873 | 0.046 | 1.616 | 0.023 | 113 | 37 | ⋯ | ⋯ | ⋯ | ⋯ | 113 | 37 |
| 4−145 | 16.875 | 0.092 | 1.793 | 0.037 | 87 | 46 | 76 | 18 | ⋯ | ⋯ | 77 | 17 |
| 4−146 | 16.472 | 0.007 | 1.662 | 0.016 | −68 | 12 | −106 | 20 | ⋯ | −70 | −77 | 10 |
| 4−150 | 16.624 | 0.022 | 2.503 | 0.025 | 63 | 50 | 67 | 48 | ⋯ | ⋯ | 65 | 35 |
| 4−155 | 16.688 | 0.017 | 1.693 | 0.053 | ⋯ | ⋯ | 2 | 25 | ⋯ | ⋯ | 2 | 25 |
| 4−160 | 17.004 | 0.023 | 1.605 | 0.021 | 1 | 10 | 97 | 27 | ⋯ | ⋯ | 49 | 48[a] |
| 4−161 | 16.623 | 0.014 | 1.691 | 0.019 | 237 | 40 | 204 | 25 | ⋯ | ⋯ | 213 | 21 |
| 4−165 | 16.311 | 0.009 | 2.030 | 0.028 | −53 | 24 | −109 | 26 | ⋯ | −83 | −78 | 17 |
| 4−167 | 16.998 | 0.040 | 1.840 | 0.012 | ⋯ | ⋯ | −23 | 16 | −16 | −16 | −21 | 11 |
| 4−170 | 17.157 | 0.041 | 1.792 | 0.021 | ⋯ | ⋯ | 131 | 19 | ⋯ | ⋯ | 131 | 19 |
| 4−179 | 16.551 | 0.009 | 1.705 | 0.023 | 102 | 9 | 95 | 22 | ⋯ | ⋯ | 101 | 8 |
| 4−183 | 17.095 | 0.037 | 1.701 | 0.045 | 245 | 34 | 218 | 15 | ⋯ | ⋯ | 222 | 14 |
| 4−184 | 16.758 | 0.059 | 1.916 | 0.019 | 88 | 25 | ⋯ | ⋯ | ⋯ | ⋯ | 88 | 25 |
| 4−185 | 16.468 | 0.010 | 1.600 | 0.015 | −26 | 18 | ⋯ | ⋯ | ⋯ | ⋯ | −26 | 18 |
| 4−186 | 15.633 | 0.005 | 2.236 | 0.009 | 72 | 12 | 74 | 26 | ⋯ | ⋯ | 72 | 11 |

Table 2—Continued

| Star | $V$ | Photometry error | $V - I$ | error | CTIO velocity $v_r$ | error | AAO velocity $v_r$ | error | Rich (1990) $v_r(H)$ | $v_r(L)$ | Mean $v_r$ | error |
|------|-----|------------------|---------|-------|------|-------|------|-------|--------|--------|------|-------|
| 4−187 | 16.202 | 0.087 | 0.981 | 0.016 | $\cdots$ | $\cdots$ | 16 | 14 | $\cdots$ | $\cdots$ | 16 | 14 |
| 4−189 | 16.788 | 0.010 | 1.605 | 0.015 | 128 | 35 | 140 | 21 | $\cdots$ | $\cdots$ | 137 | 18 |
| 4−190 | 16.095 | 0.016 | 1.782 | 0.022 | $\cdots$ | $\cdots$ | −280 | 22 | $\cdots$ | $\cdots$ | −280 | 22 |
| 4−191 | 16.945 | 0.013 | 2.050 | 0.015 | $\cdots$ | $\cdots$ | 32 | 18 | $\cdots$ | $\cdots$ | 32 | 18 |
| 4−198 | 15.002 | 0.010 | 1.766 | 0.015 | −106 | 15 | $\cdots$ | $\cdots$ | $\cdots$ | $\cdots$ | −106 | 15 |
| 4−203 | 14.019 | 0.004 | 2.221 | 0.008 | −32 | 11 | $\cdots$ | $\cdots$ | −8 | $\cdots$ | −26 | 9 |
| 4−206 | 16.306 | 0.041 | 1.771 | 0.044 | −126 | 19 | $\cdots$ | $\cdots$ | $\cdots$ | $\cdots$ | −126 | 19 |
| 4−213 | 16.426 | 0.012 | 2.013 | 0.022 | 106 | 19 | 201 | 20 | $\cdots$ | $\cdots$ | 154 | 48[a] |
| 4−218 | 16.825 | 0.024 | 1.724 | 0.031 | −57 | 17 | $\cdots$ | $\cdots$ | $\cdots$ | $\cdots$ | −57 | 17 |
| 4−243 | 16.253 | 0.030 | 2.079 | 0.022 | −131 | 15 | −119 | 23 | $\cdots$ | $\cdots$ | −127 | 13 |
| 4−250 | 16.063 | 0.022 | 1.977 | 0.032 | $\cdots$ | $\cdots$ | 103 | 31 | $\cdots$ | $\cdots$ | 103 | 31 |
| 4−256 | 16.910 | 0.034 | 1.822 | 0.053 | $\cdots$ | $\cdots$ | 47 | 17 | $\cdots$ | $\cdots$ | 47 | 17 |
| 4−258 | 17.028 | 0.017 | 1.493 | 0.041 | −21 | 14 | $\cdots$ | $\cdots$ | $\cdots$ | $\cdots$ | −21 | 14 |
| 4−262 | 15.108 | 0.180 | 4.330 | 0.049 | $\cdots$ | $\cdots$ | $\cdots$ | $\cdots$ | $\cdots$ | $\cdots$ | $\cdots$ | $\cdots$ |
| 4−270 | 16.541 | 0.018 | 1.576 | 0.023 | $\cdots$ | $\cdots$ | −78 | 23 | $\cdots$ | $\cdots$ | −78 | 23 |
| 4−271 | 16.216 | 0.151 | 4.229 | 0.040 | $\cdots$ | $\cdots$ | $\cdots$ | $\cdots$ | $\cdots$ | $\cdots$ | $\cdots$ | $\cdots$ |
| 4−274 | $\cdots$ | $\cdots$ | $\cdots$ | $\cdots$ | 116 | 11 | $\cdots$ | $\cdots$ | $\cdots$ | $\cdots$ | 116 | 11 |
| 4−275 | 16.164 | 0.016 | 1.620 | 0.024 | −38 | 17 | $\cdots$ | $\cdots$ | $\cdots$ | $\cdots$ | −38 | 17 |
| 4−279 | 15.013 | 0.005 | 1.635 | 0.011 | −24 | 18 | $\cdots$ | $\cdots$ | $\cdots$ | $\cdots$ | −24 | 18 |
| 4−283 | 15.804 | 0.008 | 2.281 | 0.020 | −120 | 31 | $\cdots$ | $\cdots$ | $\cdots$ | $\cdots$ | −120 | 31 |
| 4−285 | 16.670 | 0.009 | 1.801 | 0.013 | 49 | 30 | −5 | 39 | $\cdots$ | 57 | 36 | 21 |
| 4−286 | 16.679 | 0.010 | 1.692 | 0.016 | $\cdots$ | $\cdots$ | 12 | 13 | $\cdots$ | $\cdots$ | 12 | 13 |
| 4−297 | 16.639 | 0.022 | 1.808 | 0.027 | 78 | 47 | $\cdots$ | $\cdots$ | $\cdots$ | 76 | 45 | 34 |
| 4−307 | 15.368 | 0.009 | 1.667 | 0.015 | $\cdots$ | $\cdots$ | 18 | 13 | $\cdots$ | $\cdots$ | 18 | 13 |
| 4−309 | 16.705 | 0.039 | 1.680 | 0.053 | 25 | 26 | −21 | 34 | $\cdots$ | $\cdots$ | 8 | 21 |
| 4−310 | 17.023 | 0.030 | 1.670 | 0.035 | 95 | 34 | 102 | 17 | $\cdots$ | $\cdots$ | 101 | 15 |
| 4−312 | 16.348 | 0.024 | 1.481 | 0.011 | −45 | 57 | $\cdots$ | $\cdots$ | −59 | −47 | −59 | 15 |
| 4−320 | 14.832 | 0.007 | 1.561 | 0.009 | 21 | 15 | 4 | 59 | $\cdots$ | $\cdots$ | 20 | 15 |
| 4−324 | 16.710 | 0.009 | 1.673 | 0.016 | −274 | 43 | −101 | 27 | $\cdots$ | $\cdots$ | −188 | 87[a] |
| 4−325 | 16.830 | 0.007 | $\cdots$ | $\cdots$ | $\cdots$ | $\cdots$ | −21 | 27 | −49 | $\cdots$ | −44 | 14 |
| 4−326 | 17.019 | 0.017 | 1.884 | 0.020 | $\cdots$ | $\cdots$ | 72 | 20 | $\cdots$ | $\cdots$ | 72 | 20 |
| 4−328 | 16.987 | 0.017 | 1.583 | 0.014 | 120 | 45 | $\cdots$ | $\cdots$ | $\cdots$ | $\cdots$ | 120 | 45 |
| 4−335 | 16.922 | 0.014 | 1.518 | 0.019 | 37 | 16 | −28 | 43 | $\cdots$ | $\cdots$ | 29 | 15 |
| vb252 | 17.254 | 0.013 | 1.508 | 0.018 | 14 | 33 | 25 | 27 | $\cdots$ | $\cdots$ | 21 | 21 |
| vb256 | 17.057 | 0.011 | 1.483 | 0.016 | $\cdots$ | $\cdots$ | 100 | 48 | $\cdots$ | $\cdots$ | 100 | 48 |
| vb259 | 17.656 | 0.012 | 1.823 | 0.018 | $\cdots$ | $\cdots$ | 63 | 48 | $\cdots$ | $\cdots$ | 63 | 48 |

Table 2—Continued

| Star | $V$ | Photometry error | $V-I$ | error | CTIO velocity $v_r$ | error | AAO velocity $v_r$ | error | Rich (1990) $v_r$(H) | $v_r$(L) | Mean $v_r$ | error |
|------|-----|------------------|-------|-------|---------------------|-------|--------------------|-------|----------------------|----------|------------|-------|
| vb262 | 17.083 | 0.039 | 1.784 | 0.049 | $\cdots$ | $\cdots$ | $-6$ | 23 | $\cdots$ | $\cdots$ | $-6$ | 23 |
| vb268 | 17.230 | 0.012 | 1.542 | 0.023 | 98 | 38 | 91 | 29 | $\cdots$ | $\cdots$ | 94 | 23 |
| vb271 | 17.283 | 0.035 | 1.513 | 0.045 | 17 | 42 | 16 | 31 | $\cdots$ | $\cdots$ | 16 | 25 |
| vb274 | 17.668 | 0.301 | 1.726 | 0.069 | $\cdots$ | $\cdots$ | 26 | 29 | $\cdots$ | $\cdots$ | 26 | 29 |
| vb282 | 17.349 | 0.027 | 1.589 | 0.036 | $\cdots$ | $\cdots$ | $-44$ | 31 | $\cdots$ | $\cdots$ | $-44$ | 31 |
| vb283 | 17.274 | 0.020 | 1.718 | 0.027 | $\cdots$ | $\cdots$ | $\cdots$ | $\cdots$ | $\cdots$ | $\cdots$ | $\cdots$ | $\cdots$ |
| vb284 | 17.515 | 0.028 | 1.550 | 0.037 | 88 | 45 | 48 | 34 | $\cdots$ | $\cdots$ | 63 | 27 |
| vb315 | 17.716 | 0.031 | 1.653 | 0.031 | $\cdots$ | $\cdots$ | 41 | 26 | $\cdots$ | $\cdots$ | 41 | 26 |
| vb323 | 17.588 | 0.042 | 1.478 | 0.056 | $\cdots$ | $\cdots$ | $-17$ | 36 | $\cdots$ | $\cdots$ | $-17$ | 36 |
| vb326 | 17.130 | 0.026 | 1.711 | 0.024 | $\cdots$ | $\cdots$ | $-36$ | 34 | $\cdots$ | $\cdots$ | $-36$ | 34 |
| vb329 | 17.268 | 0.017 | 1.605 | 0.044 | $\cdots$ | $\cdots$ | 41 | 39 | $\cdots$ | $\cdots$ | 41 | 39 |
| vb330 | 17.283 | 0.028 | 1.764 | 0.034 | $-341$ | 52 | $-249$ | 31 | $\cdots$ | $\cdots$ | $-273$ | 27 |
| vb353 | 17.577 | 0.026 | 1.512 | 0.032 | $-31$ | 47 | $-73$ | 59 | $\cdots$ | $\cdots$ | $-47$ | 37 |
| vb361 | 17.092 | 0.046 | 1.510 | 0.020 | $\cdots$ | $\cdots$ | $\cdots$ | $\cdots$ | $\cdots$ | $\cdots$ | $\cdots$ | $\cdots$ |
| vb362 | 17.296 | 0.031 | 1.777 | 0.040 | $\cdots$ | $\cdots$ | $-25$ | 29 | $\cdots$ | $\cdots$ | $-25$ | 29 |
| vb395 | 17.282 | 0.036 | 1.764 | 0.043 | $\cdots$ | $\cdots$ | $-148$ | 12 | $\cdots$ | $\cdots$ | $-148$ | 12 |
| vb404 | 17.141 | 0.024 | 1.722 | 0.027 | $\cdots$ | $\cdots$ | 81 | 17 | $\cdots$ | $\cdots$ | 81 | 17 |
| vb406 | 17.624 | 0.019 | 1.480 | 0.065 | $\cdots$ | $\cdots$ | 85 | 31 | $\cdots$ | $\cdots$ | 85 | 31 |
| vb411 | 17.928 | 0.067 | 1.419 | 0.072 | 70 | 51 | 5 | 43 | $\cdots$ | $\cdots$ | 32 | 33 |
| vb413 | 17.636 | 0.017 | 1.222 | 0.028 | 29 | 44 | 14 | 27 | $\cdots$ | $\cdots$ | 18 | 23 |
| vb414 | 17.465 | 0.015 | 1.612 | 0.022 | $-93$ | 37 | $-96$ | 19 | $\cdots$ | $\cdots$ | $-95$ | 17 |
| vb417 | 17.116 | 0.012 | 1.860 | 0.015 | $-240$ | 39 | $-196$ | 13 | $\cdots$ | $\cdots$ | $-200$ | 12 |
| vb457 | 17.433 | 0.014 | 1.906 | 0.022 | $\cdots$ | $\cdots$ | $-19$ | 27 | $\cdots$ | $\cdots$ | $-19$ | 27 |
| vb469 | 17.235 | 0.016 | 1.795 | 0.020 | $\cdots$ | $\cdots$ | $-123$ | 18 | $\cdots$ | $\cdots$ | $-123$ | 18 |
| vb470 | 17.486 | 0.012 | 1.838 | 0.019 | $\cdots$ | $\cdots$ | $\cdots$ | $\cdots$ | $\cdots$ | $\cdots$ | $\cdots$ | $\cdots$ |
| vb471 | 17.706 | 0.028 | 1.765 | 0.031 | $\cdots$ | $\cdots$ | 16 | 34 | $\cdots$ | $\cdots$ | 16 | 34 |
| vb472 | 18.774 | 0.032 | 1.193 | 0.073 | $\cdots$ | $\cdots$ | $-50$ | 59 | $\cdots$ | $\cdots$ | $-50$ | 59 |
| vb480 | 17.112 | 0.010 | 1.692 | 0.015 | $\cdots$ | $\cdots$ | 79 | 19 | $\cdots$ | $\cdots$ | 79 | 19 |
| vb483 | 17.284 | 0.015 | 1.991 | 0.019 | $\cdots$ | $\cdots$ | $-41$ | 22 | $\cdots$ | $\cdots$ | $-41$ | 22 |
| vb488 | 17.481 | 0.016 | 1.684 | 0.024 | $\cdots$ | $\cdots$ | 114 | 36 | $\cdots$ | $\cdots$ | 114 | 36 |
| vb491 | 17.536 | 0.043 | 1.855 | 0.074 | $\cdots$ | $\cdots$ | $-154$ | 16 | $\cdots$ | $\cdots$ | $-154$ | 16 |
| vb493 | 17.642 | 0.022 | 1.644 | 0.026 | $\cdots$ | $\cdots$ | 190 | 23 | $\cdots$ | $\cdots$ | 190 | 23 |
| vb497 | 17.181 | 0.031 | 1.835 | 0.045 | $\cdots$ | $\cdots$ | 3 | 17 | $\cdots$ | $\cdots$ | 3 | 17 |
| vb499 | 16.377 | 0.008 | 1.811 | 0.012 | $-24$ | 25 | $\cdots$ | $\cdots$ | $\cdots$ | $\cdots$ | $-24$ | 25 |
| vb500 | 17.416 | 0.013 | 1.924 | 0.016 | 53 | 46 | 52 | 19 | $\cdots$ | $\cdots$ | 52 | 18 |
| vb504 | 17.132 | 0.012 | 1.663 | 0.024 | $\cdots$ | $\cdots$ | 203 | 20 | $\cdots$ | $\cdots$ | 203 | 20 |

[a]CTIO and AAT velocities discrepant.

Table 3. Comparison of radial velocities.

| Difference | Mean | $\sigma$ | N |
|---|---|---|---|
| CTIO − AAO [a] | −0.3 | 35.2 | 130 |
| CTIO − Rich (H)[b] | 3.4 | 22.8 | 19 |
| CTIO − Rich (L)[b] | −9.9 | 44.6 | 20 |
| AAO − Rich (H) | 5.8 | 29.7 | 24 |
| AAO − Rich (L) | −7.0 | 51.3 | 35 |
| Rich (H) − Rich (L) | −8.1 | 51.3 | 16 |

[a]7 stars with discrepant velocities eliminated from comparison. See Table 2.

[b]Velocity from Rich (1990).

Table 4. Coefficients in transformation to FFBG system.

| Index | CTIO observations | | AAT |
|---|---|---|---|
| | 1989 | 1990 | |
| CN | $-0.007 \pm 0.004$ | $-0.105 \pm 0.007$ | $0.023 \pm 0.014$[a] |
| $Mg_1$ | $0.008 \pm 0.004$ | $-0.109 \pm 0.005$ | $0.045 \pm 0.006$ |
| $Mg_2$ | $0.019 \pm 0.003$ | $0.158 \pm 0.002$ | $0.044 \pm 0.010$ |
| G | $0.65 \pm 0.09$ | $-0.22 \pm 0.10$ | $0.00 \pm 0.31$[b] |
| $Mgb$ | $0.49 \pm 0.09$ | $-0.08 \pm 0.03$ | $0.00 \pm 0.16$ |
| Fe 5270 | $0.42 \pm 0.08$ | $-0.55 \pm 0.03$ | $0.00 \pm 0.31$ |
| Fe 5335 | $0.54 \pm 0.07$ | $-0.31 \pm 0.03$ | $0.00 \pm 0.16$ |
| $H\beta$ | $0.00 \pm 0.05$ | $-0.01 \pm 0.03$ | $0.00 \pm 0.21$ |

[a]Before reddening correction. See text.

[b]Zero value indicates that no correction was applied. Errors are $\sigma/N^{1/2}$ for all coefficients.

Table 5.   Residuals after transformation to FFBG system.

| Index | r.m.s. difference | Internal error | FFBG error | Expected error |
|-------|---------|---------|------|--------|
| CN | 0.046 | 0.016 | 0.025 | 0.030 |
| Mg1 | 0.036 | 0.007 | 0.008 | 0.011 |
| Mg2 | 0.019 | 0.009 | 0.007 | 0.011 |
| G | 0.79 | 0.40 | 0.56 | 0.69 |
| Mg$b$ | 0.45 | 0.40 | 0.27 | 0.48 |
| Fe 5270 | 0.40 | 0.43 | 0.28 | 0.51 |
| Fe 5335 | 0.43 | 0.33 | 0.36 | 0.49 |
| H$\beta$ | 0.32 | 0.26 | 0.31 | 0.41 |

Table 6.   Comparison of CTIO/AAO indices.

| index | CTIO−AAO | $\sigma$ | AAO error | CTIO error |
|-------|----------|----------|-----------|------------|
| CN | $-0.065 \pm 0.016$ | 0.111 | 0.038 | 0.101[a] |
| Mg$_1$ | $-0.017 \pm 0.005$ | 0.052 | 0.029 | 0.036 |
| Mg$_2$ | $-0.008 \pm 0.006$ | 0.066 | 0.017 | 0.037 |
| Mgb | $-0.24 \pm 0.08$ | 1.10 | 0.57 | 0.96[b] |
| G | $-0.23 \pm 0.18$ | 2.37 | 0.97 | 1.67 |
| Fe 5270 | $-0.13 \pm 0.10$ | 1.23 | 0.60 | 0.93 |
| Fe 5335 | $-0.17 \pm 0.12$ | 1.39 | 0.73 | 0.76 |
| H$\beta$ | $-0.02 \pm 0.10$ | 1.11 | 0.51 | 0.80 |

[a]Stars with CN ¿ 0.45 mag excluded from statistics.
[b]AAT data transformed to CTIO system before averaging.

Table 7. Combined line-strength indices.

| Star | Fe 5270 | error | Fe 5335 | error | ⟨Fe⟩ | error | Hβ | error | CN | error | Mg₁ | error | Mg₂ | error |
|------|---------|-------|---------|-------|------|-------|-----|-------|-----|-------|-----|-------|-----|-------|
| 0−025 | 1.37 | 0.63 | 0.37 | 0.81 | 0.87 | 0.73 | 1.28 | 0.76 | 0.072 | 0.036 | 0.081 | 0.036 | 0.197 | 0.018 |
| 0−333 | 3.04 | 0.70 | 1.73 | 0.40 | 2.39 | 0.57 | 0.70 | 0.80 | 0.001 | 0.086 | 0.085 | 0.025 | 0.330 | 0.026 |
| 0−340 | 3.68 | 1.42 | 4.74 | 1.08 | 4.21 | 1.26 | -0.37 | 0.71 | 0.216 | 0.282 | 0.246 | 0.039 | 0.409 | 0.219 |
| 1−012 | 1.91 | 0.63 | 4.34 | 0.80 | 3.13 | 0.72 | 2.47 | 0.51 | 0.122 | 0.030 | 0.159 | 0.030 | 0.255 | 0.018 |
| 1−015 | ⋯ | ⋯ | ⋯ | ⋯ | ⋯ | ⋯ | ⋯ | ⋯ | ⋯ | ⋯ | ⋯ | ⋯ | ⋯ | ⋯ |
| 1−016 | 4.06 | 0.38 | 4.41 | 0.30 | 4.23 | 0.34 | 3.86 | 0.80 | 0.137 | 0.050 | 0.276 | 0.013 | 0.411 | 0.013 |
| 1−022 | 2.63 | 0.68 | 3.61 | 0.86 | 3.12 | 0.78 | 1.46 | 0.77 | 0.090 | 0.036 | 0.202 | 0.032 | 0.324 | 0.019 |
| 1−025 | 6.07 | 0.56 | 3.63 | 0.72 | 4.85 | 0.65 | 1.41 | 0.69 | 0.219 | 0.028 | 0.179 | 0.027 | 0.346 | 0.016 |
| 1−027 | 2.29 | 0.72 | 2.07 | 0.76 | 2.18 | 0.74 | 1.40 | 0.64 | 0.252 | 0.099 | 0.090 | 0.029 | 0.217 | 0.102 |
| 1−034 | 4.15 | 0.90 | 3.77 | 0.60 | 3.96 | 0.76 | 3.01 | 0.80 | 0.042 | 0.166 | 0.263 | 0.034 | 0.541 | 0.038 |
| 1−039 | 4.28 | 0.51 | 5.89 | 0.63 | 5.09 | 0.57 | 1.30 | 0.59 | 0.180 | 0.022 | 0.151 | 0.025 | 0.298 | 0.014 |
| 1−041 | 2.38 | 0.38 | 3.16 | 0.30 | 2.77 | 0.34 | 0.16 | 0.80 | 0.044 | 0.081 | 0.235 | 0.021 | 0.329 | 0.020 |
| 1−043 | 2.16 | 0.57 | 1.09 | 0.73 | 1.63 | 0.66 | 0.64 | 0.71 | 0.153 | 0.037 | 0.071 | 0.028 | 0.168 | 0.018 |
| 1−047 | 4.23 | 0.80 | 4.21 | 0.70 | 4.22 | 0.75 | 1.21 | 0.80 | 0.136 | 0.104 | 0.143 | 0.026 | 0.259 | 0.026 |
| 1−051 | 4.81 | 1.30 | 6.11 | 1.40 | 5.46 | 1.35 | 0.82 | 0.80 | 0.290 | 0.222 | 0.239 | 0.040 | 0.377 | 0.041 |
| 1−053 | 1.01 | 0.38 | 1.18 | 0.40 | 1.10 | 0.39 | 1.39 | 0.80 | -0.055 | 0.130 | -0.039 | 0.036 | 0.065 | 0.034 |
| 1−054 | ⋯ | ⋯ | ⋯ | ⋯ | ⋯ | ⋯ | ⋯ | ⋯ | ⋯ | ⋯ | ⋯ | ⋯ | ⋯ | ⋯ |
| 1−058 | 4.90 | 0.71 | 4.01 | 0.87 | 4.46 | 0.79 | -0.62 | 0.94 | 0.220 | 0.044 | 0.253 | 0.034 | 0.457 | 0.019 |
| 1−065 | 3.32 | 1.50 | 3.88 | 1.50 | 3.60 | 1.50 | 0.89 | 0.80 | 0.133 | 0.236 | 0.063 | 0.048 | 0.347 | 0.051 |
| 1−073 | 4.93 | 0.70 | 5.49 | 0.60 | 5.21 | 0.65 | 0.93 | 0.80 | 0.080 | 0.073 | 0.290 | 0.021 | 0.447 | 0.022 |
| 1−076 | 4.66 | 1.90 | 5.95 | 2.10 | 5.30 | 2.00 | -0.89 | 0.80 | -0.025 | 0.256 | 0.340 | 0.065 | 0.519 | 0.069 |
| 1−079 | ⋯ | ⋯ | ⋯ | ⋯ | ⋯ | ⋯ | ⋯ | ⋯ | ⋯ | ⋯ | ⋯ | ⋯ | ⋯ | ⋯ |
| 1−083 | 3.67 | 0.95 | 3.84 | 1.13 | 3.76 | 1.04 | 0.64 | 0.91 | 0.227 | 0.146 | 0.186 | 0.039 | 0.324 | 0.154 |
| 1−084 | 2.79 | 1.88 | 1.91 | 0.61 | 2.35 | 1.40 | 1.90 | 0.66 | 0.026 | 0.078 | 0.060 | 0.029 | 0.201 | 0.148 |
| 1−089 | -0.11 | 0.38 | 1.34 | 0.60 | 0.61 | 0.50 | 1.75 | 0.80 | -0.209 | 0.143 | -0.124 | 0.043 | 0.052 | 0.042 |
| 1−090 | 4.68 | 0.74 | 1.39 | 0.96 | 3.04 | 0.86 | 2.04 | 0.83 | 0.042 | 0.050 | 0.069 | 0.036 | 0.285 | 0.021 |
| 1−093 | 3.36 | 0.62 | 2.80 | 0.66 | 3.08 | 0.64 | 1.56 | 0.62 | 0.134 | 0.062 | 0.214 | 0.025 | 0.356 | 0.187 |
| 1−102 | 2.32 | 0.67 | 3.14 | 0.63 | 2.73 | 0.65 | 0.49 | 0.80 | -0.009 | 0.083 | 0.104 | 0.027 | 0.239 | 0.028 |
| 1−105 | 1.11 | 0.73 | 2.91 | 0.89 | 2.01 | 0.81 | 1.02 | 0.87 | 0.049 | 0.044 | 0.174 | 0.034 | 0.357 | 0.019 |
| 1−108 | 1.89 | 0.47 | 2.53 | 0.58 | 2.21 | 0.53 | 0.81 | 0.60 | 0.036 | 0.050 | 0.098 | 0.023 | 0.217 | 0.014 |
| 1−109 | 3.95 | 1.60 | 3.83 | 1.40 | 3.89 | 1.50 | 2.45 | 0.80 | ⋯ | ⋯ | 0.133 | 0.050 | 0.417 | 0.051 |
| 1−116 | 3.18 | 0.42 | 4.91 | 0.52 | 4.05 | 0.47 | 1.92 | 0.50 | 0.101 | 0.024 | 0.211 | 0.020 | 0.301 | 0.012 |
| 1−118 | 4.04 | 2.00 | 4.68 | 2.10 | 4.36 | 2.05 | -0.28 | 0.80 | 0.315 | 0.207 | -0.045 | 0.061 | ⋯ | ⋯ |
| 1−129 | 4.72 | 0.36 | 4.13 | 0.45 | 4.43 | 0.41 | 0.02 | 0.46 | 0.194 | 0.021 | 0.217 | 0.017 | 0.394 | 0.010 |
| 1−140 | 1.74 | 0.79 | 3.68 | 1.01 | 2.71 | 0.91 | 0.49 | 0.72 | 0.071 | 0.052 | 0.142 | 0.037 | 0.271 | 0.125 |
| 1−141 | 2.99 | 0.40 | 3.06 | 0.40 | 3.02 | 0.40 | 1.09 | 0.80 | 0.097 | 0.093 | 0.196 | 0.024 | 0.278 | 0.023 |

Table 7—Continued

| Star | Fe 5270 | error | Fe 5335 | error | ⟨Fe⟩ | error | Hβ | error | CN | error | Mg$_1$ | error | Mg$_2$ | error |
|---|---|---|---|---|---|---|---|---|---|---|---|---|---|---|
| 1−144 | 4.72 | 0.41 | 3.34 | 0.50 | 4.03 | 0.46 | 0.63 | 0.53 | 0.083 | 0.036 | 0.238 | 0.020 | 0.416 | 0.011 |
| 1−148 | 3.84 | 0.68 | 3.87 | 0.84 | 3.86 | 0.76 | 1.04 | 0.84 | 0.290 | 0.042 | 0.169 | 0.033 | 0.355 | 0.019 |
| 1−151 | 1.93 | 0.40 | 1.78 | 0.30 | 1.85 | 0.35 | 1.05 | 0.80 | -0.039 | 0.066 | -0.042 | 0.020 | 0.163 | 0.020 |
| 1−152 | ⋯ | ⋯ | ⋯ | ⋯ | ⋯ | ⋯ | ⋯ | ⋯ | ⋯ | ⋯ | ⋯ | ⋯ | ⋯ | ⋯ |
| 1−155 | 4.42 | 0.47 | 4.20 | 0.58 | 4.31 | 0.53 | 1.42 | 0.62 | 0.341 | 0.041 | 0.233 | 0.023 | 0.396 | 0.013 |
| 1−156 | 1.98 | 0.38 | 2.83 | 0.30 | 2.40 | 0.34 | 0.88 | 0.80 | 0.043 | 0.065 | 0.115 | 0.020 | 0.180 | 0.019 |
| 1−159 | 4.25 | 0.74 | 6.16 | 0.86 | 5.21 | 0.80 | 0.76 | 0.98 | 0.197 | 0.061 | 0.267 | 0.035 | 0.479 | 0.020 |
| 1−163 | 4.37 | 1.52 | 3.48 | 1.20 | 3.92 | 1.37 | 1.41 | 2.07 | 0.109 | 0.025 | 0.125 | 0.072 | 0.266 | 0.265 |
| 1−167 | 3.55 | 0.51 | 3.93 | 0.61 | 3.74 | 0.56 | 3.03 | 0.66 | 0.021 | 0.143 | 0.169 | 0.024 | 0.481 | 0.013 |
| 1−175 | 4.01 | 0.39 | 4.27 | 0.48 | 4.14 | 0.44 | 1.60 | 0.49 | 0.289 | 0.029 | 0.238 | 0.019 | 0.391 | 0.011 |
| 1−177 | 4.42 | 0.64 | 4.19 | 0.81 | 4.31 | 0.73 | 1.30 | 0.80 | 0.228 | 0.034 | 0.144 | 0.032 | 0.279 | 0.018 |
| 1−178 | ⋯ | ⋯ | ⋯ | ⋯ | ⋯ | ⋯ | ⋯ | ⋯ | ⋯ | ⋯ | ⋯ | ⋯ | ⋯ | ⋯ |
| 1−179 | 3.84 | 1.20 | 3.97 | 1.00 | 3.91 | 1.10 | 1.37 | 0.80 | ⋯ | ⋯ | 0.215 | 0.039 | 0.531 | 0.040 |
| 1−180 | 4.22 | 0.56 | 4.89 | 0.68 | 4.56 | 0.62 | 2.10 | 0.69 | 0.211 | 0.033 | 0.217 | 0.027 | 0.367 | 0.015 |
| 1−181 | ⋯ | ⋯ | ⋯ | ⋯ | ⋯ | ⋯ | ⋯ | ⋯ | ⋯ | ⋯ | ⋯ | ⋯ | ⋯ | ⋯ |
| 1−183 | 4.41 | 0.52 | 4.33 | 0.63 | 4.37 | 0.58 | 1.60 | 0.64 | 0.498 | 0.034 | 0.227 | 0.025 | 0.425 | 0.014 |
| 1−184 | 1.88 | 0.38 | 1.73 | 0.29 | 1.80 | 0.34 | 1.51 | 0.80 | -0.058 | 0.077 | 0.018 | 0.026 | 0.069 | 0.025 |
| 1−187 | 3.08 | 0.71 | 3.00 | 0.52 | 3.04 | 0.62 | 1.18 | 0.62 | 0.059 | 0.077 | 0.130 | 0.028 | 0.259 | 0.143 |
| 1−189 | 0.41 | 0.88 | 1.17 | 1.15 | 0.79 | 1.02 | 0.40 | 0.94 | -0.009 | 0.009 | -0.022 | 0.042 | 0.085 | 0.041 |
| 1−191 | 4.43 | 0.85 | 2.68 | 0.73 | 3.56 | 0.79 | 1.92 | 0.64 | 0.273 | 0.123 | 0.162 | 0.026 | 0.297 | 0.158 |
| 1−195 | 1.02 | 0.74 | 2.59 | 0.91 | 1.81 | 0.83 | 1.05 | 0.91 | 0.075 | 0.046 | 0.176 | 0.034 | 0.427 | 0.019 |
| 1−196 | 5.56 | 0.39 | 5.24 | 0.47 | 5.40 | 0.43 | 1.12 | 0.55 | 0.178 | 0.044 | 0.288 | 0.019 | 0.476 | 0.010 |
| 1−200 | 4.59 | 0.60 | 4.37 | 0.50 | 4.48 | 0.55 | 0.81 | 0.80 | ⋯ | ⋯ | 0.192 | 0.019 | 0.338 | 0.020 |
| 1−202 | 3.86 | 0.86 | 3.54 | 0.67 | 3.70 | 0.77 | 1.26 | 0.80 | 0.183 | 0.141 | 0.207 | 0.030 | 0.392 | 0.029 |
| 1−203 | 3.94 | 0.70 | 4.14 | 0.70 | 4.04 | 0.70 | 1.04 | 0.80 | 0.065 | 0.075 | 0.155 | 0.021 | 0.438 | 0.022 |
| 1−218 | 3.02 | 0.50 | 2.90 | 0.94 | 2.96 | 0.75 | 1.15 | 0.66 | 0.136 | 0.052 | 0.123 | 0.024 | 0.239 | 0.136 |
| 1−221 | 4.43 | 1.36 | 3.76 | 1.19 | 4.09 | 1.28 | 1.42 | 1.10 | 0.530 | 0.285 | 0.281 | 0.050 | 0.423 | 0.241 |
| 1−223 | 2.69 | 0.66 | 3.17 | 0.84 | 2.93 | 0.76 | 2.52 | 0.76 | 0.149 | 0.029 | 0.059 | 0.032 | 0.156 | 0.021 |
| 1−224 | 2.65 | 0.90 | 3.10 | 1.00 | 2.88 | 0.95 | 1.07 | 0.80 | 0.029 | 0.131 | 0.023 | 0.037 | 0.264 | 0.037 |
| 1−226 | 4.17 | 0.60 | 4.67 | 0.72 | 4.42 | 0.66 | 1.27 | 0.78 | 0.383 | 0.043 | 0.288 | 0.028 | 0.435 | 0.016 |
| 1−228 | 4.18 | 1.60 | 2.63 | 0.90 | 3.41 | 1.30 | 2.82 | 0.80 | ⋯ | ⋯ | 0.175 | 0.048 | 0.444 | 0.048 |
| 1−232 | 3.93 | 0.60 | 3.92 | 0.58 | 3.92 | 0.59 | 1.46 | 0.67 | 0.104 | 0.084 | 0.290 | 0.040 | 0.438 | 0.164 |
| 1−233 | 4.39 | 0.80 | 2.96 | 1.01 | 3.68 | 0.91 | 1.58 | 1.02 | 0.316 | 0.050 | 0.118 | 0.040 | 0.280 | 0.023 |
| 1−234 | 2.93 | 0.74 | 2.61 | 0.94 | 2.77 | 0.85 | 1.70 | 0.88 | 0.140 | 0.034 | 0.111 | 0.036 | 0.203 | 0.022 |
| 1−235 | 3.41 | 0.36 | 3.66 | 0.44 | 3.54 | 0.40 | 1.00 | 0.48 | 0.131 | 0.030 | 0.252 | 0.017 | 0.404 | 0.010 |
| 1−236 | ⋯ | ⋯ | ⋯ | ⋯ | ⋯ | ⋯ | ⋯ | ⋯ | ⋯ | ⋯ | ⋯ | ⋯ | ⋯ | ⋯ |

Table 7—Continued

| Star | Fe 5270 | error | Fe 5335 | error | ⟨Fe⟩ | error | Hβ | error | CN | error | Mg$_1$ | error | Mg$_2$ | error |
|------|---------|-------|---------|-------|------|-------|-----|-------|-----|-------|--------|-------|--------|-------|
| 1−239 | 2.22 | 0.70 | 1.45 | 0.50 | 1.84 | 0.61 | 1.43 | 0.80 | -0.109 | 0.097 | -0.040 | 0.032 | 0.173 | 0.032 |
| 1−249 | 3.72 | 0.80 | 3.43 | 0.60 | 3.57 | 0.71 | 0.61 | 0.80 | 0.066 | 0.114 | 0.146 | 0.030 | 0.241 | 0.029 |
| 1−263 | 1.21 | 0.60 | 1.05 | 0.40 | 1.13 | 0.51 | 1.32 | 0.80 | -0.079 | 0.129 | -0.078 | 0.038 | 0.129 | 0.038 |
| 1−264 | 2.00 | 0.44 | 1.86 | 0.50 | 1.93 | 0.47 | 1.65 | 0.65 | 0.105 | 0.061 | 0.085 | 0.020 | 0.154 | 0.070 |
| 1−285 | 3.33 | 0.46 | 4.12 | 0.56 | 3.73 | 0.51 | 1.04 | 0.57 | 0.090 | 0.036 | 0.177 | 0.022 | 0.294 | 0.013 |
| 1−291 | 2.35 | 0.83 | 3.77 | 1.04 | 3.06 | 0.94 | 1.46 | 1.05 | 0.031 | 0.081 | 0.096 | 0.042 | 0.214 | 0.025 |
| 1−292 | 4.14 | 1.30 | 3.55 | 1.00 | 3.85 | 1.16 | 2.80 | 0.80 | 1.374 | 1.310 | 0.018 | 0.035 | 0.283 | 0.036 |
| 1−293 | 4.37 | 1.00 | 3.09 | 1.54 | 3.73 | 1.29 | 0.97 | 0.67 | 0.153 | 0.156 | 0.150 | 0.036 | 0.278 | 0.158 |
| 1−298 | 4.34 | 0.37 | 4.53 | 0.44 | 4.44 | 0.41 | 1.38 | 0.49 | 0.166 | 0.033 | 0.303 | 0.017 | 0.494 | 0.010 |
| 1−303 | 3.35 | 0.41 | 3.29 | 0.51 | 3.32 | 0.46 | 1.43 | 0.50 | 0.151 | 0.022 | 0.198 | 0.019 | 0.374 | 0.011 |
| 1−304 | 4.10 | 0.67 | 2.53 | 0.84 | 3.32 | 0.76 | 2.82 | 0.84 | 0.252 | 0.039 | 0.210 | 0.033 | 0.331 | 0.019 |
| 1−312 | ⋯ | ⋯ | ⋯ | ⋯ | ⋯ | ⋯ | ⋯ | ⋯ | ⋯ | ⋯ | ⋯ | ⋯ | ⋯ | ⋯ |
| 1−316 | 3.90 | 0.98 | 4.55 | 0.77 | 4.23 | 0.88 | 0.98 | 0.61 | 0.099 | 0.093 | 0.290 | 0.025 | 0.455 | 0.225 |
| 1−318 | 3.65 | 0.40 | 4.40 | 0.40 | 4.02 | 0.40 | 0.82 | 0.80 | 0.070 | 0.078 | 0.305 | 0.019 | 0.510 | 0.020 |
| 1−319 | 3.72 | 0.85 | 4.60 | 1.04 | 4.16 | 0.95 | 1.70 | 1.04 | -0.038 | 0.028 | 0.222 | 0.040 | 0.343 | 0.023 |
| 1−320 | ⋯ | ⋯ | 4.95 | 1.10 | ⋯ | ⋯ | 0.38 | 0.80 | 0.021 | 0.198 | 0.342 | 0.045 | 0.545 | 0.048 |
| 1−321 | ⋯ | ⋯ | ⋯ | ⋯ | ⋯ | ⋯ | ⋯ | ⋯ | ⋯ | ⋯ | ⋯ | ⋯ | ⋯ | ⋯ |
| 1−322 | 4.45 | 1.00 | 3.85 | 0.70 | 4.15 | 0.86 | 1.66 | 0.80 | 0.074 | 0.174 | 0.244 | 0.036 | 0.359 | 0.035 |
| 1−324 | 1.56 | 0.38 | 2.50 | 0.29 | 2.03 | 0.34 | 1.60 | 0.80 | 0.009 | 0.042 | 0.243 | 0.011 | 0.457 | 0.012 |
| 1−325 | ⋯ | ⋯ | ⋯ | ⋯ | ⋯ | ⋯ | ⋯ | ⋯ | ⋯ | ⋯ | ⋯ | ⋯ | ⋯ | ⋯ |
| 1−326 | 4.00 | 1.90 | 3.83 | 1.70 | 3.92 | 1.80 | 0.91 | 0.80 | 0.451 | 0.355 | ⋯ | ⋯ | 0.302 | 0.055 |
| 1−332 | ⋯ | ⋯ | ⋯ | ⋯ | ⋯ | ⋯ | ⋯ | ⋯ | ⋯ | ⋯ | ⋯ | ⋯ | ⋯ | ⋯ |
| 1−335 | ⋯ | ⋯ | ⋯ | ⋯ | ⋯ | ⋯ | ⋯ | ⋯ | ⋯ | ⋯ | ⋯ | ⋯ | ⋯ | ⋯ |
| 1−340 | 2.43 | 0.43 | 2.43 | 0.41 | 2.43 | 0.42 | 0.62 | 0.62 | 0.013 | 0.073 | 0.112 | 0.022 | 0.216 | 0.108 |
| 1−343 | 1.17 | 0.38 | -0.53 | 0.29 | 0.32 | 0.34 | 3.46 | 0.80 | ⋯ | ⋯ | 0.093 | 0.011 | 0.571 | 0.015 |
| 1−344 | 1.93 | 0.85 | 1.09 | 1.11 | 1.51 | 0.99 | -1.55 | 1.09 | 0.090 | 0.049 | 0.092 | 0.042 | 0.196 | 0.026 |
| 1−345 | 2.47 | 1.52 | 2.16 | 0.75 | 2.32 | 1.20 | 1.11 | 1.09 | 0.141 | 0.022 | 0.108 | 0.056 | 0.236 | 0.133 |
| 1−346 | ⋯ | ⋯ | ⋯ | ⋯ | ⋯ | ⋯ | ⋯ | ⋯ | ⋯ | ⋯ | ⋯ | ⋯ | ⋯ | ⋯ |
| 1−348 | 1.69 | 0.60 | 0.79 | 0.30 | 1.24 | 0.47 | 1.34 | 0.80 | -0.117 | 0.102 | -0.025 | 0.032 | 0.189 | 0.032 |
| 1−349 | 3.77 | 0.55 | 2.96 | 0.70 | 3.37 | 0.63 | 1.60 | 0.63 | 0.219 | 0.032 | 0.131 | 0.026 | 0.241 | 0.016 |
| 1−353 | ⋯ | ⋯ | ⋯ | ⋯ | ⋯ | ⋯ | ⋯ | ⋯ | ⋯ | ⋯ | ⋯ | ⋯ | ⋯ | ⋯ |
| 1−357 | 1.71 | 0.49 | 2.35 | 0.61 | 2.03 | 0.55 | 1.31 | 0.59 | 0.001 | 0.002 | 0.066 | 0.024 | 0.154 | 0.015 |
| 1−369 | ⋯ | ⋯ | ⋯ | ⋯ | ⋯ | ⋯ | ⋯ | ⋯ | ⋯ | ⋯ | ⋯ | ⋯ | ⋯ | ⋯ |
| 1−374 | 4.74 | 1.19 | 3.70 | 0.85 | 4.22 | 1.03 | 1.26 | 0.69 | 0.323 | 0.121 | 0.211 | 0.034 | 0.344 | 0.188 |
| 1−379 | 1.69 | 0.76 | 0.71 | 0.78 | 1.20 | 0.77 | 0.99 | 1.62 | -0.001 | 0.130 | 0.076 | 0.068 | 0.268 | 0.310 |
| 2−014 | 2.57 | 0.92 | 1.65 | 1.11 | 2.11 | 1.01 | 0.45 | 0.92 | -0.040 | 0.101 | 0.084 | 0.042 | 0.183 | 0.191 |

Table 7—Continued

| Star | Fe 5270 | error | Fe 5335 | error | ⟨Fe⟩ | error | Hβ | error | CN | error | Mg$_1$ | error | Mg$_2$ | error |
|------|---------|-------|---------|-------|------|-------|-----|-------|-----|-------|--------|-------|--------|-------|
| 2−015 | 4.55 | 1.20 | 4.61 | 1.00 | 4.58 | 1.11 | 0.98 | 0.80 | 0.151 | 0.191 | 0.309 | 0.039 | 0.453 | 0.038 |
| 2−016 | 1.57 | 0.55 | 0.76 | 0.30 | 1.17 | 0.44 | 1.55 | 0.80 | -0.160 | 0.097 | ⋯ | ⋯ | 0.080 | 0.031 |
| 2−018 | 3.03 | 0.75 | 1.55 | 0.95 | 2.29 | 0.86 | -0.38 | 0.99 | 0.059 | 0.056 | 0.212 | 0.036 | 0.369 | 0.021 |
| 2−019 | 1.75 | 0.39 | -1.76 | 0.47 | -0.01 | 0.43 | 3.46 | 0.64 | -0.143 | 0.015 | -0.091 | 0.019 | 0.410 | 0.010 |
| 2−021 | 3.06 | 0.59 | 0.75 | 0.72 | 1.91 | 0.66 | 4.16 | 0.84 | -0.117 | 0.032 | 0.069 | 0.029 | 0.546 | 0.015 |
| 2−028 | 3.89 | 0.58 | 2.13 | 0.74 | 3.01 | 0.66 | 0.63 | 0.70 | 0.062 | 0.036 | 0.125 | 0.028 | 0.237 | 0.017 |
| 2−031 | ⋯ | ⋯ | ⋯ | ⋯ | ⋯ | ⋯ | ⋯ | ⋯ | ⋯ | ⋯ | ⋯ | ⋯ | ⋯ | ⋯ |
| 2−033 | 2.56 | 0.38 | 2.39 | 0.30 | 2.48 | 0.34 | 1.30 | 0.80 | -0.031 | 0.052 | 0.023 | 0.016 | 0.198 | 0.017 |
| 2−035 | 4.21 | 0.65 | 3.99 | 0.81 | 4.10 | 0.73 | 1.54 | 0.78 | 0.286 | 0.036 | 0.167 | 0.032 | 0.321 | 0.018 |
| 2−036 | 2.78 | 1.68 | 1.30 | 0.84 | 2.04 | 1.33 | 1.45 | 0.67 | 0.041 | 0.122 | 0.048 | 0.048 | 0.141 | 0.081 |
| 2−037 | 1.62 | 0.70 | ⋯ | ⋯ | ⋯ | ⋯ | 1.44 | 0.80 | ⋯ | ⋯ | 0.121 | 0.046 | 0.337 | 0.044 |
| 2−040 | ⋯ | ⋯ | ⋯ | ⋯ | ⋯ | ⋯ | ⋯ | ⋯ | ⋯ | ⋯ | ⋯ | ⋯ | ⋯ | ⋯ |
| 2−042 | ⋯ | ⋯ | ⋯ | ⋯ | ⋯ | ⋯ | ⋯ | ⋯ | ⋯ | ⋯ | ⋯ | ⋯ | ⋯ | ⋯ |
| 2−043 | 4.13 | 0.63 | 3.18 | 0.78 | 3.66 | 0.71 | 1.17 | 0.83 | 0.102 | 0.053 | 0.285 | 0.030 | 0.353 | 0.017 |
| 2−048 | 3.13 | 0.57 | 2.84 | 0.73 | 2.99 | 0.66 | 2.48 | 0.69 | 0.043 | 0.038 | 0.138 | 0.028 | 0.129 | 0.020 |
| 2−049 | 2.69 | 0.61 | 2.93 | 0.71 | 2.81 | 0.66 | 1.72 | 0.72 | 0.173 | 0.051 | 0.084 | 0.029 | 0.178 | 0.108 |
| 2−050 | 3.14 | 0.67 | -1.73 | 0.93 | 0.71 | 0.81 | 2.16 | 0.80 | -0.019 | 0.016 | 0.030 | 0.034 | 0.087 | 0.032 |
| 2−051 | 3.65 | 0.39 | 3.32 | 0.50 | 3.49 | 0.45 | 0.83 | 0.52 | 0.119 | 0.024 | 0.146 | 0.020 | 0.245 | 0.012 |
| 2−055 | 3.98 | 0.37 | 3.32 | 0.47 | 3.65 | 0.42 | 1.19 | 0.48 | 0.149 | 0.022 | 0.192 | 0.018 | 0.317 | 0.010 |
| 2−059 | 2.19 | 0.65 | -0.68 | 0.86 | 0.76 | 0.76 | 1.11 | 0.80 | -0.077 | 0.023 | 0.017 | 0.033 | 0.076 | 0.035 |
| 2−062 | 4.05 | 0.65 | 4.93 | 0.79 | 4.49 | 0.72 | 0.54 | 0.84 | 0.301 | 0.044 | 0.247 | 0.031 | 0.385 | 0.018 |
| 2−065 | 3.95 | 2.19 | 4.60 | 0.94 | 4.28 | 1.69 | 0.07 | 0.85 | -0.040 | 0.163 | 0.279 | 0.032 | 0.453 | 0.189 |
| 2−067 | 3.14 | 0.89 | 1.93 | 1.08 | 2.53 | 0.99 | 1.68 | 0.75 | 0.194 | 0.134 | 0.092 | 0.031 | 0.214 | 0.112 |
| 2−069 | 1.90 | 0.80 | 1.95 | 0.80 | 1.93 | 0.80 | 2.71 | 0.80 | 0.030 | 0.252 | -0.031 | 0.043 | 0.159 | 0.042 |
| 2−075 | 3.95 | 0.38 | 3.93 | 0.46 | 3.94 | 0.42 | -0.17 | 0.49 | 0.087 | 0.034 | 0.271 | 0.018 | 0.404 | 0.010 |
| 2−081 | 3.77 | 0.98 | 2.96 | 0.97 | 3.36 | 0.97 | 1.87 | 0.71 | 0.175 | 0.090 | 0.142 | 0.035 | 0.249 | 0.112 |
| 2−086 | 2.81 | 0.66 | 1.05 | 0.84 | 1.93 | 0.76 | 1.44 | 0.78 | 0.131 | 0.033 | 0.014 | 0.033 | 0.075 | 0.036 |
| 2−088 | 2.94 | 0.75 | 3.11 | 0.92 | 3.02 | 0.84 | 1.84 | 0.72 | 0.144 | 0.080 | 0.101 | 0.048 | 0.192 | 0.077 |
| 2−092 | 2.40 | 1.43 | 2.51 | 1.09 | 2.45 | 1.27 | 2.17 | 0.72 | -0.010 | 0.047 | 0.162 | 0.024 | 0.509 | 0.190 |
| 2−096 | ⋯ | ⋯ | ⋯ | ⋯ | ⋯ | ⋯ | ⋯ | ⋯ | ⋯ | ⋯ | ⋯ | ⋯ | ⋯ | ⋯ |
| 2−097 | 2.44 | 0.63 | -1.24 | 0.77 | 0.60 | 0.70 | 2.80 | 1.02 | -0.089 | 0.033 | -0.071 | 0.030 | 0.465 | 0.016 |
| 2−100 | 5.11 | 0.70 | 3.13 | 0.90 | 4.12 | 0.81 | 1.90 | 0.91 | 0.091 | 0.044 | 0.179 | 0.035 | 0.338 | 0.020 |
| 2−101 | 1.29 | 0.54 | 1.32 | 0.97 | 1.31 | 0.78 | 0.74 | 0.70 | 0.110 | 0.091 | 0.038 | 0.026 | 0.088 | 0.075 |
| 2−102 | 2.75 | 0.75 | 0.52 | 0.97 | 1.64 | 0.87 | 1.11 | 0.91 | 0.109 | 0.042 | 0.099 | 0.036 | 0.155 | 0.024 |
| 2−103 | 2.55 | 0.81 | 4.26 | 0.99 | 3.41 | 0.90 | 2.17 | 1.09 | -0.075 | 0.039 | 0.181 | 0.040 | 0.321 | 0.023 |
| 2−105 | 3.32 | 1.01 | 3.00 | 0.95 | 3.16 | 0.98 | 1.70 | 0.69 | 0.070 | 0.042 | 0.168 | 0.042 | 0.262 | 0.139 |

Table 7—Continued

| Star | Fe 5270 | error | Fe 5335 | error | ⟨Fe⟩ | error | Hβ | error | CN | error | Mg$_1$ | error | Mg$_2$ | error |
|------|---------|-------|---------|-------|------|-------|------|-------|------|-------|------|-------|------|-------|
| 2−106 | 3.28 | 0.60 | 1.09 | 0.78 | 2.19 | 0.70 | 2.35 | 0.70 | 0.086 | 0.032 | 0.105 | 0.029 | 0.292 | 0.017 |
| 2−108 | 3.59 | 0.69 | 5.68 | 0.84 | 4.63 | 0.76 | 1.48 | 0.88 | 0.409 | 0.051 | 0.280 | 0.018 | 0.397 | 0.019 |
| 2−109 | 2.36 | 0.90 | 2.09 | 0.77 | 2.22 | 0.84 | 1.89 | 1.14 | 0.042 | 0.104 | 0.067 | 0.038 | 0.201 | 0.156 |
| 2−116 | 4.99 | 0.53 | 4.46 | 0.66 | 4.73 | 0.60 | 1.99 | 0.70 | 0.268 | 0.036 | 0.224 | 0.026 | 0.334 | 0.015 |
| 2−117 | 1.78 | 0.47 | 1.10 | 0.54 | 1.44 | 0.50 | 0.98 | 0.64 | -0.008 | 0.081 | 0.052 | 0.028 | 0.083 | 0.087 |
| 2−119 | ⋯ | ⋯ | ⋯ | ⋯ | ⋯ | ⋯ | ⋯ | ⋯ | ⋯ | ⋯ | ⋯ | ⋯ | ⋯ | ⋯ |
| 2−120 | 2.14 | 0.70 | 3.11 | 0.90 | 2.63 | 0.81 | 2.03 | 0.80 | 0.022 | 0.114 | -0.027 | 0.034 | 0.216 | 0.035 |
| 2−122 | 2.64 | 0.38 | 3.00 | 0.40 | 2.82 | 0.39 | 0.85 | 0.80 | 0.017 | 0.072 | 0.103 | 0.016 | 0.358 | 0.016 |
| 2−126 | 3.63 | 0.80 | 1.04 | 1.03 | 2.34 | 0.92 | 1.34 | 0.97 | 0.043 | 0.067 | 0.080 | 0.039 | 0.209 | 0.023 |
| 2−131 | 2.58 | 0.40 | 1.80 | 0.29 | 2.19 | 0.35 | 0.83 | 0.80 | 0.020 | 0.075 | 0.049 | 0.024 | 0.081 | 0.022 |
| 2−133 | 4.13 | 0.80 | ⋯ | ⋯ | ⋯ | ⋯ | 1.02 | 0.80 | 0.177 | 0.115 | ⋯ | ⋯ | ⋯ | ⋯ |
| 2−135 | ⋯ | ⋯ | ⋯ | ⋯ | ⋯ | ⋯ | ⋯ | ⋯ | ⋯ | ⋯ | ⋯ | ⋯ | ⋯ | ⋯ |
| 2−136 | ⋯ | ⋯ | ⋯ | ⋯ | ⋯ | ⋯ | ⋯ | ⋯ | ⋯ | ⋯ | ⋯ | ⋯ | ⋯ | ⋯ |
| 2−137 | 1.39 | 0.40 | 1.61 | 0.40 | 1.50 | 0.40 | 1.97 | 0.80 | -0.075 | 0.062 | -0.067 | 0.022 | 0.160 | 0.022 |
| 2−138 | ⋯ | ⋯ | ⋯ | ⋯ | ⋯ | ⋯ | ⋯ | ⋯ | ⋯ | ⋯ | ⋯ | ⋯ | ⋯ | ⋯ |
| 2−145 | 4.91 | 0.61 | 3.55 | 0.77 | 4.23 | 0.69 | 2.64 | 0.80 | 0.233 | 0.041 | 0.244 | 0.030 | 0.377 | 0.017 |
| 2−146 | 3.76 | 0.70 | 3.09 | 0.50 | 3.42 | 0.61 | 1.25 | 0.80 | -0.198 | 0.129 | 0.126 | 0.027 | 0.245 | 0.027 |
| 2−147 | 1.04 | 1.11 | 2.55 | 0.86 | 1.80 | 0.99 | 2.51 | 0.88 | -0.060 | 0.121 | 0.079 | 0.057 | 0.129 | 0.118 |
| 2−149 | 5.18 | 0.90 | 5.42 | 0.80 | 5.30 | 0.85 | -0.03 | 0.80 | 0.090 | 0.096 | 0.227 | 0.025 | 0.345 | 0.025 |
| 2−152 | ⋯ | ⋯ | ⋯ | ⋯ | ⋯ | ⋯ | ⋯ | ⋯ | ⋯ | ⋯ | ⋯ | ⋯ | ⋯ | ⋯ |
| 2−155 | ⋯ | ⋯ | ⋯ | ⋯ | ⋯ | ⋯ | ⋯ | ⋯ | ⋯ | ⋯ | ⋯ | ⋯ | ⋯ | ⋯ |
| 2−158 | 3.77 | 2.20 | 2.89 | 1.60 | 3.33 | 1.92 | 1.74 | 0.80 | 0.011 | 0.238 | -0.045 | 0.059 | 0.258 | 0.064 |
| 2−159 | 3.81 | 0.52 | 4.35 | 0.63 | 4.08 | 0.58 | 0.95 | 0.67 | 0.262 | 0.031 | 0.191 | 0.025 | 0.363 | 0.014 |
| 2−167 | 4.45 | 0.45 | 4.69 | 0.55 | 4.57 | 0.50 | 1.87 | 0.59 | 0.188 | 0.029 | 0.193 | 0.022 | 0.339 | 0.013 |
| 2−168 | 2.53 | 0.66 | 4.04 | 0.81 | 3.29 | 0.74 | 2.28 | 0.78 | 0.119 | 0.035 | 0.074 | 0.032 | 0.189 | 0.020 |
| 2−170 | 3.66 | 0.63 | 2.03 | 0.82 | 2.85 | 0.73 | 0.08 | 0.83 | 0.175 | 0.036 | 0.159 | 0.031 | 0.241 | 0.019 |
| 2−172 | 2.45 | 0.67 | 3.08 | 0.85 | 2.77 | 0.77 | 0.89 | 0.80 | 0.120 | 0.035 | 0.126 | 0.033 | 0.234 | 0.019 |
| 2−174 | 3.75 | 1.26 | 2.75 | 0.86 | 3.25 | 1.08 | 1.49 | 0.68 | 0.211 | 0.122 | 0.147 | 0.033 | 0.288 | 0.140 |
| 2−175 | 4.11 | 1.40 | 3.16 | 0.80 | 3.63 | 1.14 | 0.09 | 0.80 | ⋯ | ⋯ | 0.371 | 0.059 | 0.436 | 0.055 |
| 2−178 | 4.33 | 0.39 | 3.75 | 0.48 | 4.04 | 0.44 | 1.20 | 0.49 | 0.122 | 0.025 | 0.221 | 0.018 | 0.354 | 0.011 |
| 2−187 | 1.83 | 0.64 | 1.47 | 0.68 | 1.65 | 0.66 | 0.76 | 0.71 | 0.197 | 0.039 | 0.043 | 0.028 | 0.072 | 0.047 |
| 2−188 | 1.57 | 0.47 | 1.53 | 0.52 | 1.55 | 0.50 | 0.75 | 0.64 | -0.035 | 0.040 | 0.079 | 0.024 | 0.149 | 0.107 |
| 2−192 | 3.22 | 0.90 | 2.65 | 0.70 | 2.94 | 0.81 | 1.72 | 0.80 | 0.126 | 0.142 | 0.003 | 0.030 | 0.243 | 0.030 |
| 2−196 | 4.74 | 0.45 | 4.27 | 0.56 | 4.51 | 0.51 | 1.15 | 0.57 | 0.257 | 0.033 | 0.212 | 0.022 | 0.407 | 0.012 |
| 2−197 | 4.12 | 0.67 | 3.11 | 0.79 | 3.62 | 0.73 | 0.44 | 0.88 | -0.011 | 0.017 | 0.305 | 0.031 | 0.515 | 0.018 |
| 2−198 | 2.84 | 0.85 | 1.43 | 1.81 | 2.14 | 1.41 | -0.15 | 0.67 | 0.128 | 0.126 | 0.203 | 0.034 | 0.345 | 0.018 |

Table 7—Continued

| Star | Fe 5270 | error | Fe 5335 | error | ⟨Fe⟩ | error | Hβ | error | CN | error | Mg₁ | error | Mg₂ | error |
|---|---|---|---|---|---|---|---|---|---|---|---|---|---|---|
| 2−199 | ... | ... | ... | ... | ... | ... | ... | ... | ... | ... | ... | ... | ... | ... |
| 2−200 | 2.58 | 0.55 | 2.72 | 0.52 | 2.65 | 0.53 | 1.35 | 0.80 | -0.144 | 0.088 | 0.045 | 0.025 | 0.211 | 0.025 |
| 2−201 | 3.47 | 0.74 | 3.41 | 0.92 | 3.44 | 0.83 | 1.34 | 0.97 | 0.093 | 0.049 | 0.159 | 0.037 | 0.243 | 0.022 |
| 2−202 | 2.67 | 1.00 | 2.45 | 0.90 | 2.56 | 0.95 | 1.77 | 0.80 | 0.070 | 0.431 | -0.016 | 0.040 | 0.196 | 0.040 |
| 2−203 | 3.67 | 0.76 | 4.15 | 0.94 | 3.91 | 0.85 | 0.93 | 0.99 | 0.026 | 0.115 | 0.121 | 0.038 | 0.298 | 0.021 |
| 2−204 | 3.06 | 1.27 | 1.70 | 0.83 | 2.38 | 1.07 | 1.43 | 0.67 | 0.102 | 0.036 | 0.161 | 0.040 | 0.280 | 0.147 |
| 2−205 | 4.01 | 0.52 | 3.61 | 0.64 | 3.81 | 0.58 | 2.23 | 0.64 | 0.099 | 0.031 | 0.168 | 0.026 | 0.272 | 0.015 |
| 2−207 | 4.12 | 0.74 | 0.79 | 0.93 | 2.46 | 0.84 | 1.92 | 0.90 | 0.154 | 0.038 | 0.150 | 0.035 | 0.290 | 0.020 |
| 2−213 | 4.23 | 0.45 | 3.18 | 0.56 | 3.71 | 0.51 | 1.50 | 0.56 | 0.095 | 0.027 | 0.190 | 0.022 | 0.341 | 0.013 |
| 2−216 | 2.00 | 0.67 | 2.09 | 1.16 | 2.04 | 0.95 | 1.20 | 0.79 | -0.031 | 0.044 | 0.060 | 0.034 | 0.085 | 0.036 |
| 2−219 | 2.47 | 0.87 | 4.90 | 1.04 | 3.69 | 0.96 | 1.82 | 1.01 | 0.343 | 0.051 | 0.245 | 0.040 | 0.339 | 0.023 |
| 2−220 | 3.67 | 1.01 | 3.78 | 0.94 | 3.72 | 0.98 | 1.87 | 0.67 | 0.199 | 0.117 | 0.179 | 0.037 | 0.327 | 0.186 |
| 2−228 | 1.72 | 0.58 | 1.50 | 0.68 | 1.61 | 0.63 | 0.86 | 0.72 | 0.040 | 0.062 | 0.052 | 0.035 | 0.155 | 0.082 |
| 2−232 | 3.79 | 0.64 | 3.16 | 0.79 | 3.48 | 0.72 | 0.31 | 0.85 | 0.434 | 0.043 | 0.218 | 0.031 | 0.342 | 0.017 |
| 2−235 | 2.82 | 0.82 | 1.83 | 0.84 | 2.32 | 0.83 | 2.60 | 1.37 | -0.025 | 0.086 | 0.040 | 0.043 | 0.149 | 0.149 |
| 2−237 | 3.56 | 1.02 | 2.21 | 1.97 | 2.89 | 1.57 | 0.59 | 1.10 | 0.062 | 0.214 | 0.308 | 0.039 | 0.513 | 0.252 |
| 2−244 | 2.95 | 0.65 | 2.84 | 0.80 | 2.90 | 0.73 | 2.47 | 0.83 | 0.019 | 0.271 | 0.200 | 0.032 | 0.296 | 0.018 |
| 2−247 | ... | ... | ... | ... | ... | ... | ... | ... | ... | ... | ... | ... | ... | ... |
| 2−248 | 2.04 | 0.38 | 2.22 | 0.29 | 2.13 | 0.34 | 1.49 | 0.80 | ... | ... | 0.067 | 0.018 | 0.108 | 0.017 |
| 2−249 | 3.50 | 0.74 | 2.25 | 0.97 | 2.88 | 0.86 | 0.83 | 0.68 | 0.051 | 0.110 | 0.121 | 0.034 | 0.223 | 0.161 |
| 2−250 | 3.32 | 0.78 | 3.89 | 0.95 | 3.61 | 0.87 | 1.32 | 1.01 | 0.278 | 0.049 | 0.169 | 0.038 | 0.293 | 0.022 |
| 2−252 | 2.38 | 0.56 | 3.10 | 0.57 | 2.74 | 0.57 | 0.97 | 0.66 | 0.051 | 0.086 | 0.123 | 0.024 | 0.210 | 0.118 |
| 2−253 | ... | ... | ... | ... | ... | ... | ... | ... | ... | ... | ... | ... | ... | ... |
| 2−254 | 4.82 | 0.57 | 3.75 | 0.71 | 4.29 | 0.64 | 2.03 | 0.76 | 0.311 | 0.038 | 0.163 | 0.029 | 0.339 | 0.016 |
| 2−259 | 3.74 | 0.56 | 3.12 | 0.70 | 3.43 | 0.63 | 1.22 | 0.74 | 0.185 | 0.035 | 0.223 | 0.027 | 0.340 | 0.016 |
| 2−261 | 3.03 | 0.75 | 3.11 | 0.72 | 3.07 | 0.74 | 0.70 | 0.72 | 0.111 | 0.074 | 0.199 | 0.032 | 0.338 | 0.153 |
| 2−264 | 4.02 | 0.46 | 3.52 | 0.50 | 3.77 | 0.48 | 1.38 | 0.64 | 0.114 | 0.068 | 0.122 | 0.023 | 0.248 | 0.137 |
| 2−272 | 4.72 | 1.70 | 5.01 | 1.48 | 4.87 | 1.60 | 1.54 | 0.79 | 0.355 | 0.479 | 0.252 | 0.042 | 0.395 | 0.147 |
| 2−276 | 2.41 | 0.64 | 4.74 | 0.76 | 3.58 | 0.70 | 1.99 | 0.83 | 0.247 | 0.046 | 0.230 | 0.030 | 0.373 | 0.017 |
| 2−279 | 3.65 | 0.79 | 2.76 | 1.29 | 3.21 | 1.07 | 0.74 | 0.72 | 0.176 | 0.028 | 0.170 | 0.034 | 0.280 | 0.105 |
| 2−281 | ... | ... | ... | ... | ... | ... | ... | ... | ... | ... | ... | ... | ... | ... |
| 2−282 | 1.93 | 0.92 | 1.30 | 1.23 | 1.62 | 1.09 | -0.10 | 1.16 | 0.083 | 0.048 | 0.053 | 0.046 | 0.104 | 0.036 |
| 2−284 | 4.39 | 0.76 | 4.84 | 0.87 | 4.61 | 0.82 | 1.87 | 0.76 | 0.000 | 0.105 | 0.269 | 0.033 | 0.452 | 0.205 |
| 2−285 | 2.77 | 1.41 | 2.85 | 1.37 | 2.81 | 1.39 | 2.23 | 1.66 | 0.289 | 0.743 | 0.186 | 0.040 | 0.308 | 0.177 |
| 2−286 | 3.17 | 0.64 | 2.79 | 1.00 | 2.98 | 0.84 | 1.09 | 0.71 | 0.064 | 0.043 | 0.125 | 0.030 | 0.209 | 0.126 |
| 2−288 | 0.18 | 0.50 | 0.19 | 0.30 | 0.19 | 0.41 | 2.17 | 0.80 | -0.173 | 0.232 | -0.121 | 0.078 | 0.137 | 0.083 |

Table 7—Continued

| Star | Fe 5270 | error | Fe 5335 | error | $\langle$Fe$\rangle$ | error | H$\beta$ | error | CN | error | Mg$_1$ | error | Mg$_2$ | error |
|------|---------|-------|---------|-------|---------|-------|------|-------|-----|-------|--------|-------|--------|-------|
| 2−297 | 2.63 | 0.51 | 2.84 | 0.64 | 2.74 | 0.58 | -0.36 | 0.60 | 0.208 | 0.032 | 0.122 | 0.024 | 0.212 | 0.015 |
| 2−300 | 2.92 | 0.38 | 3.80 | 0.30 | 3.36 | 0.34 | 0.95 | 0.80 | 0.114 | 0.060 | 0.259 | 0.016 | 0.385 | 0.016 |
| 2−304 | 3.13 | 0.94 | 2.42 | 0.86 | 2.77 | 0.90 | 1.61 | 0.92 | 0.210 | 0.032 | 0.167 | 0.042 | 0.282 | 0.016 |
| 2−306 | 4.98 | 0.51 | 4.35 | 0.63 | 4.67 | 0.57 | 1.90 | 0.69 | 0.244 | 0.034 | 0.201 | 0.026 | 0.347 | 0.014 |
| 2−307 | 3.80 | 0.66 | 4.89 | 0.71 | 4.35 | 0.68 | 0.35 | 0.67 | 0.169 | 0.121 | 0.327 | 0.029 | 0.513 | 0.301 |
| 2−310 | 3.63 | 1.40 | 3.12 | 1.10 | 3.38 | 1.26 | 0.78 | 0.80 | 0.116 | 0.210 | 0.090 | 0.044 | 0.336 | 0.045 |
| 2−312 | 2.72 | 0.50 | 2.94 | 0.52 | 2.83 | 0.51 | 1.60 | 0.63 | 0.055 | 0.064 | 0.073 | 0.024 | 0.169 | 0.119 |
| 2−316 | 2.46 | 0.86 | 2.40 | 1.08 | 2.43 | 0.98 | 1.64 | 1.01 | 0.097 | 0.046 | 0.093 | 0.041 | 0.248 | 0.024 |
| 2−317 | 3.24 | 1.07 | 2.33 | 0.96 | 2.79 | 1.02 | 0.26 | 0.59 | 0.102 | 0.130 | 0.094 | 0.041 | 0.204 | 0.131 |
| 2−318 | 2.63 | 0.55 | 2.30 | 0.70 | 2.47 | 0.63 | 0.49 | 0.66 | -0.008 | 0.007 | 0.069 | 0.027 | 0.153 | 0.018 |
| 2−319 | 4.44 | 0.32 | 3.93 | 0.39 | 4.19 | 0.36 | 0.78 | 0.42 | 0.119 | 0.020 | 0.199 | 0.016 | 0.353 | 0.009 |
| 2−321 | 2.65 | 0.66 | 3.59 | 0.84 | 3.12 | 0.76 | 1.03 | 0.80 | 0.124 | 0.035 | 0.127 | 0.033 | 0.235 | 0.019 |
| 2−323 | 4.77 | 0.67 | 3.72 | 0.84 | 4.25 | 0.76 | 0.84 | 0.83 | 0.270 | 0.036 | 0.149 | 0.033 | 0.344 | 0.019 |
| 2−326 | 3.19 | 0.44 | 3.59 | 0.55 | 3.39 | 0.50 | 2.19 | 0.53 | 0.240 | 0.024 | 0.120 | 0.022 | 0.258 | 0.012 |
| 2−327 | 4.19 | 0.67 | 1.45 | 0.74 | 2.82 | 0.71 | 1.03 | 0.99 | -0.011 | 0.117 | 0.102 | 0.043 | 0.198 | 0.101 |
| 3−018 | 1.69 | 0.90 | 2.43 | 1.00 | 2.06 | 0.95 | 1.64 | 0.80 | -0.005 | 0.168 | $\cdots$ | $\cdots$ | 0.194 | 0.047 |
| 3−035 | $\cdots$ | $\cdots$ | 2.09 | 0.80 | $\cdots$ | $\cdots$ | 1.88 | 0.90 | $\cdots$ | $\cdots$ | 0.098 | 0.051 | 0.286 | 0.047 |
| 3−036 | 0.99 | 1.33 | 2.06 | 0.97 | 1.52 | 1.16 | 1.61 | 1.39 | -0.077 | 0.132 | 0.037 | 0.054 | 0.078 | 0.069 |
| 3−043 | 4.45 | 0.40 | 4.65 | 0.30 | 4.55 | 0.35 | 0.49 | 0.80 | 0.197 | 0.057 | 0.292 | 0.014 | 0.408 | 0.014 |
| 3−044 | 2.65 | 0.90 | 2.31 | 0.70 | 2.48 | 0.81 | 1.01 | 0.80 | 1.148 | 1.222 | 0.142 | 0.039 | 0.359 | 0.038 |
| 3−054 | 4.50 | 1.31 | 4.56 | 1.47 | 4.53 | 1.39 | 0.54 | 0.97 | 0.192 | 0.030 | 0.278 | 0.022 | 0.425 | 0.195 |
| 3−059 | 5.11 | 0.55 | 3.47 | 0.68 | 4.29 | 0.62 | 1.13 | 0.70 | 0.248 | 0.035 | 0.240 | 0.027 | 0.382 | 0.015 |
| 3−060 | 3.95 | 0.92 | 3.78 | 0.88 | 3.87 | 0.90 | 2.24 | 0.77 | 0.263 | 0.170 | 0.138 | 0.035 | 0.283 | 0.151 |
| 3−069 | 3.54 | 0.64 | 4.23 | 0.76 | 3.89 | 0.70 | 1.34 | 0.85 | 0.060 | 0.053 | 0.231 | 0.030 | 0.504 | 0.017 |
| 3−071 | 3.52 | 0.82 | 3.61 | 0.82 | 3.56 | 0.82 | 0.97 | 0.69 | 0.257 | 0.093 | 0.112 | 0.029 | 0.226 | 0.096 |
| 3−073 | 4.93 | 0.51 | 4.50 | 1.48 | 4.71 | 1.11 | 1.27 | 1.11 | 0.316 | 0.031 | 0.149 | 0.045 | 0.326 | 0.165 |
| 3−079 | 4.36 | 0.82 | 3.23 | 1.07 | 3.80 | 0.95 | 1.94 | 0.99 | 0.018 | 0.344 | 0.070 | 0.041 | 0.209 | 0.025 |
| 3−087 | 2.81 | 0.63 | 3.76 | 0.55 | 3.29 | 0.59 | 2.61 | 0.65 | 0.220 | 0.096 | 0.280 | 0.034 | 0.492 | 0.186 |
| 3−092 | 3.34 | 1.00 | 3.92 | 1.00 | 3.63 | 1.00 | 1.25 | 0.80 | -0.090 | 0.163 | 0.167 | 0.037 | 0.479 | 0.039 |
| 3−095 | 3.78 | 0.69 | 4.16 | 0.77 | 3.97 | 0.73 | 0.54 | 0.76 | 0.383 | 0.071 | 0.166 | 0.031 | 0.260 | 0.088 |
| 3−097 | 4.81 | 0.70 | 4.30 | 0.50 | 4.55 | 0.61 | 0.82 | 0.80 | 0.348 | 0.096 | 0.211 | 0.022 | 0.340 | 0.022 |
| 3−104 | 3.84 | 1.18 | 2.67 | 0.76 | 3.26 | 0.99 | 2.44 | 0.86 | 1.221 | 0.544 | 0.166 | 0.047 | 0.308 | 0.173 |
| 3−110 | $\cdots$ | $\cdots$ | $\cdots$ | $\cdots$ | $\cdots$ | $\cdots$ | $\cdots$ | $\cdots$ | $\cdots$ | $\cdots$ | $\cdots$ | $\cdots$ | $\cdots$ | $\cdots$ |
| 3−111 | 3.25 | 0.48 | 3.15 | 0.53 | 3.20 | 0.50 | 0.43 | 0.64 | 0.032 | 0.075 | 0.262 | 0.033 | 0.403 | 0.196 |
| 3−114 | 1.97 | 0.77 | 2.55 | 0.40 | 2.26 | 0.61 | 0.65 | 0.60 | -0.010 | 0.085 | 0.241 | 0.020 | 0.467 | 0.213 |
| 3−119 | 3.70 | 0.82 | 3.14 | 0.76 | 3.42 | 0.79 | 2.34 | 0.88 | 0.202 | 0.136 | 0.256 | 0.031 | 0.364 | 0.154 |

Table 7—Continued

| Star | Fe 5270 | error | Fe 5335 | error | ⟨Fe⟩ | error | Hβ | error | CN | error | Mg₁ | error | Mg₂ | error |
|------|---------|-------|---------|-------|------|-------|-----|-------|-----|-------|------|-------|------|-------|
| 3−122 | 4.86 | 0.38 | 5.03 | 0.46 | 4.95 | 0.42 | 0.87 | 0.50 | 0.184 | 0.029 | 0.271 | 0.018 | 0.477 | 0.010 |
| 3−123 | 4.54 | 0.57 | 4.04 | 0.71 | 4.29 | 0.64 | 1.30 | 0.73 | 0.313 | 0.042 | 0.128 | 0.028 | 0.254 | 0.016 |
| 3−128 | ⋯ | ⋯ | ⋯ | ⋯ | ⋯ | ⋯ | ⋯ | ⋯ | ⋯ | ⋯ | ⋯ | ⋯ | ⋯ | ⋯ |
| 3−134 | 2.93 | 0.50 | 2.04 | 0.29 | 2.48 | 0.41 | 0.89 | 0.80 | 0.018 | 0.071 | 0.112 | 0.024 | 0.209 | 0.023 |
| 3−135 | ⋯ | ⋯ | ⋯ | ⋯ | ⋯ | ⋯ | ⋯ | ⋯ | ⋯ | ⋯ | ⋯ | ⋯ | ⋯ | ⋯ |
| 3−143 | 0.52 | 0.80 | 3.02 | 2.40 | 1.77 | 1.79 | -1.46 | 1.70 | ⋯ | ⋯ | 0.077 | 0.103 | 0.329 | 0.105 |
| 3−144 | 1.91 | 0.51 | 2.52 | 0.64 | 2.22 | 0.58 | 0.28 | 0.67 | 0.098 | 0.034 | 0.118 | 0.025 | 0.249 | 0.015 |
| 3−151 | 3.70 | 0.38 | 3.85 | 0.29 | 3.77 | 0.34 | 0.68 | 0.80 | 0.137 | 0.039 | 0.238 | 0.010 | 0.403 | 0.010 |
| 3−152 | 2.67 | 1.13 | 2.05 | 0.84 | 2.36 | 1.00 | 0.69 | 0.64 | 0.140 | 0.043 | 0.219 | 0.020 | 0.330 | 0.160 |
| 3−153 | 1.74 | 0.70 | 2.20 | 0.70 | 1.97 | 0.70 | 1.38 | 0.80 | 0.068 | 0.145 | 0.059 | 0.039 | 0.273 | 0.039 |
| 3−156 | 0.33 | 0.73 | 3.25 | 0.91 | 1.79 | 0.83 | 0.68 | 0.59 | 0.124 | 0.042 | 0.062 | 0.036 | 0.115 | 0.018 |
| 3−157 | ⋯ | ⋯ | ⋯ | ⋯ | ⋯ | ⋯ | ⋯ | ⋯ | ⋯ | ⋯ | ⋯ | ⋯ | ⋯ | ⋯ |
| 3−159 | 4.24 | 0.50 | 4.07 | 0.62 | 4.16 | 0.56 | 1.46 | 0.66 | 0.281 | 0.040 | 0.203 | 0.025 | 0.402 | 0.014 |
| 3−160 | ⋯ | ⋯ | ⋯ | ⋯ | ⋯ | ⋯ | ⋯ | ⋯ | ⋯ | ⋯ | ⋯ | ⋯ | ⋯ | ⋯ |
| 3−161 | ⋯ | ⋯ | ⋯ | ⋯ | ⋯ | ⋯ | ⋯ | ⋯ | ⋯ | ⋯ | ⋯ | ⋯ | ⋯ | ⋯ |
| 3−163 | 1.05 | 0.44 | 0.08 | 0.68 | 0.57 | 0.57 | 5.36 | 0.66 | -0.026 | 0.015 | -0.011 | 0.021 | 0.483 | 0.011 |
| 3−164 | 2.90 | 0.40 | 2.95 | 0.40 | 2.93 | 0.40 | 1.35 | 0.80 | 0.049 | 0.056 | 0.107 | 0.016 | 0.328 | 0.016 |
| 3−175 | 4.83 | 1.10 | 5.04 | 1.04 | 4.93 | 1.07 | 1.05 | 0.62 | 0.264 | 0.228 | 0.263 | 0.038 | 0.385 | 0.161 |
| 3−181 | 4.46 | 1.40 | 3.85 | 0.88 | 4.15 | 1.17 | 1.82 | 0.77 | 0.409 | 0.203 | 0.250 | 0.040 | 0.406 | 0.195 |
| 3−182 | 3.19 | 0.46 | 4.10 | 0.50 | 3.64 | 0.48 | 0.90 | 0.63 | 0.078 | 0.032 | 0.258 | 0.029 | 0.408 | 0.162 |
| 3−184 | 2.33 | 0.62 | 1.94 | 0.55 | 2.14 | 0.59 | 1.86 | 0.61 | 0.002 | 0.087 | 0.085 | 0.028 | 0.208 | 0.098 |
| 3−187 | 1.36 | 0.54 | 1.65 | 0.71 | 1.51 | 0.63 | 1.26 | 0.64 | -0.042 | 0.015 | 0.044 | 0.027 | 0.033 | 0.036 |
| 3−188 | ⋯ | ⋯ | ⋯ | ⋯ | ⋯ | ⋯ | ⋯ | ⋯ | ⋯ | ⋯ | ⋯ | ⋯ | ⋯ | ⋯ |
| 3−189 | 5.37 | 1.46 | 4.11 | 1.14 | 4.74 | 1.31 | 1.88 | 1.19 | 0.283 | 0.253 | 0.231 | 0.020 | 0.421 | 0.202 |
| 3−192 | 1.68 | 0.63 | 2.94 | 1.75 | 2.31 | 1.32 | 0.40 | 0.68 | 0.104 | 0.059 | 0.135 | 0.030 | 0.234 | 0.094 |
| 3−194 | ⋯ | ⋯ | 5.41 | 1.00 | ⋯ | ⋯ | 1.20 | 0.80 | 0.153 | 0.194 | 0.352 | 0.037 | 0.545 | 0.038 |
| 3−197 | 4.64 | 0.45 | 4.45 | 0.56 | 4.55 | 0.51 | 1.15 | 0.60 | 0.133 | 0.027 | 0.213 | 0.023 | 0.342 | 0.013 |
| 3−202 | 3.90 | 0.58 | 1.82 | 0.71 | 2.86 | 0.65 | 1.09 | 0.80 | -0.024 | 0.020 | 0.215 | 0.027 | 0.450 | 0.015 |
| 3−205 | 2.20 | 0.73 | 3.12 | 0.68 | 2.66 | 0.71 | 0.86 | 0.70 | 0.206 | 0.235 | 0.269 | 0.030 | 0.506 | 0.248 |
| 3−206 | 5.74 | 0.58 | 4.77 | 0.73 | 5.26 | 0.66 | 1.20 | 0.76 | 0.236 | 0.034 | 0.191 | 0.029 | 0.294 | 0.017 |
| 3−209 | 4.97 | 0.96 | 3.80 | 0.96 | 4.38 | 0.96 | 1.50 | 0.88 | 0.248 | 0.132 | 0.279 | 0.036 | 0.423 | 0.249 |
| 3−210 | 2.89 | 0.48 | 1.86 | 0.46 | 2.38 | 0.47 | 4.49 | 0.98 | -0.009 | 0.110 | 0.108 | 0.022 | 0.496 | 0.229 |
| 3−211 | ⋯ | ⋯ | 3.52 | 0.70 | ⋯ | ⋯ | 2.64 | 0.80 | ⋯ | ⋯ | 0.246 | 0.037 | 0.365 | 0.037 |
| 3−220 | 1.29 | 2.00 | -0.39 | 0.29 | 0.45 | 1.43 | 3.00 | 0.80 | 0.173 | 0.795 | ⋯ | ⋯ | ⋯ | ⋯ |
| 3−223 | 2.44 | 0.38 | 2.35 | 0.29 | 2.40 | 0.34 | 1.10 | 0.80 | -0.018 | 0.024 | 0.005 | 0.009 | 0.260 | 0.009 |
| 3−224 | 4.59 | 0.55 | 3.15 | 0.71 | 3.87 | 0.64 | 0.58 | 0.70 | 0.137 | 0.034 | 0.157 | 0.027 | 0.301 | 0.016 |

Table 7—Continued

| Star | Fe 5270 | error | Fe 5335 | error | ⟨Fe⟩ | error | Hβ | error | CN | error | Mg$_1$ | error | Mg$_2$ | error |
|------|---------|-------|---------|-------|------|-------|-----|-------|-----|-------|--------|-------|--------|-------|
| 3−230 | 3.87 | 0.93 | 2.64 | 0.87 | 3.25 | 0.90 | 1.53 | 0.72 | 0.398 | 0.114 | 0.167 | 0.036 | 0.253 | 0.071 |
| 3−231 | 4.06 | 0.90 | 3.63 | 0.60 | 3.84 | 0.76 | 2.56 | 0.80 | 0.461 | 0.157 | 0.159 | 0.031 | 0.248 | 0.030 |
| 3−234 | 4.12 | 0.48 | 3.64 | 0.74 | 3.88 | 0.62 | 1.63 | 0.63 | 0.350 | 0.053 | 0.153 | 0.019 | 0.288 | 0.151 |
| 3−236 | 4.26 | 0.35 | 4.25 | 0.43 | 4.26 | 0.39 | 0.29 | 0.48 | 0.084 | 0.029 | 0.228 | 0.017 | 0.391 | 0.010 |
| 3−238 | 4.69 | 0.62 | 3.79 | 0.77 | 4.24 | 0.70 | 1.13 | 0.80 | 0.185 | 0.038 | 0.159 | 0.031 | 0.295 | 0.018 |
| 3−239 | 3.43 | 0.54 | 1.78 | 0.69 | 2.61 | 0.62 | 1.71 | 0.64 | -0.029 | 0.014 | 0.090 | 0.026 | 0.181 | 0.016 |
| 3−240 | 3.68 | 0.62 | 4.31 | 0.77 | 4.00 | 0.70 | 1.27 | 0.78 | 0.266 | 0.033 | 0.183 | 0.030 | 0.311 | 0.018 |
| 3−241 | 4.04 | 0.40 | 4.34 | 0.40 | 4.19 | 0.40 | 1.08 | 0.80 | 0.214 | 0.065 | 0.280 | 0.018 | 0.414 | 0.018 |
| 3−242 | 5.06 | 0.56 | 4.04 | 0.70 | 4.55 | 0.63 | 0.82 | 0.74 | 0.190 | 0.039 | 0.189 | 0.028 | 0.329 | 0.016 |
| 3−249 | 3.50 | 0.67 | 2.12 | 0.58 | 2.81 | 0.63 | 1.62 | 0.66 | 0.197 | 0.036 | 0.084 | 0.027 | 0.223 | 0.147 |
| 3−254 | 3.46 | 0.53 | 4.12 | 0.60 | 3.79 | 0.56 | 1.25 | 0.90 | 0.124 | 0.047 | 0.257 | 0.052 | 0.394 | 0.145 |
| 3−257 | ⋯ | ⋯ | ⋯ | ⋯ | ⋯ | ⋯ | ⋯ | ⋯ | ⋯ | ⋯ | ⋯ | ⋯ | ⋯ | ⋯ |
| 3−258 | ⋯ | ⋯ | ⋯ | ⋯ | ⋯ | ⋯ | ⋯ | ⋯ | ⋯ | ⋯ | ⋯ | ⋯ | ⋯ | ⋯ |
| 3−266 | 0.90 | 0.46 | 1.44 | 0.59 | 1.17 | 0.53 | 0.88 | 0.55 | 0.101 | 0.024 | 0.025 | 0.023 | 0.071 | 0.027 |
| 3−268 | 3.33 | 0.68 | 3.34 | 0.79 | 3.33 | 0.74 | 0.50 | 0.75 | 0.071 | 0.068 | 0.158 | 0.032 | 0.277 | 0.123 |
| 3−269 | 3.64 | 0.38 | 4.08 | 0.30 | 3.86 | 0.34 | 0.66 | 0.80 | 0.150 | 0.039 | 0.182 | 0.011 | 0.311 | 0.011 |
| 3−271 | 4.71 | 0.36 | 3.97 | 0.44 | 4.34 | 0.40 | 1.13 | 0.46 | 0.265 | 0.022 | 0.191 | 0.017 | 0.372 | 0.010 |
| 3−274 | ⋯ | ⋯ | ⋯ | ⋯ | ⋯ | ⋯ | ⋯ | ⋯ | ⋯ | ⋯ | ⋯ | ⋯ | ⋯ | ⋯ |
| 3−275 | 3.83 | 0.50 | 3.41 | 0.63 | 3.62 | 0.57 | 0.45 | 0.67 | 0.186 | 0.029 | 0.228 | 0.024 | 0.346 | 0.014 |
| 3−278 | 2.42 | 0.53 | 1.97 | 0.61 | 2.19 | 0.57 | 1.62 | 0.70 | 0.195 | 0.068 | 0.081 | 0.028 | 0.126 | 0.025 |
| 3−280 | 3.91 | 1.31 | 3.94 | 1.20 | 3.93 | 1.25 | 0.86 | 0.82 | 0.182 | 0.208 | 0.161 | 0.046 | 0.291 | 0.116 |
| 3−281 | 2.83 | 1.40 | 3.12 | 1.30 | 2.98 | 1.35 | 0.48 | 0.80 | ⋯ | ⋯ | 0.170 | 0.062 | 0.389 | 0.060 |
| 3−286 | 2.46 | 1.00 | 2.14 | 0.89 | 2.30 | 0.95 | 0.76 | 0.64 | -0.021 | 0.185 | 0.162 | 0.075 | 0.217 | 0.041 |
| 3−291 | 3.99 | 1.04 | 2.54 | 0.64 | 3.27 | 0.87 | 1.23 | 0.62 | 0.176 | 0.106 | 0.118 | 0.029 | 0.208 | 0.093 |
| 3−295 | 3.07 | 0.57 | -0.95 | 0.69 | 1.06 | 0.63 | 6.08 | 0.84 | 0.003 | 0.008 | 0.032 | 0.025 | 0.579 | 0.014 |
| 4−003 | ⋯ | ⋯ | ⋯ | ⋯ | ⋯ | ⋯ | ⋯ | ⋯ | ⋯ | ⋯ | ⋯ | ⋯ | ⋯ | ⋯ |
| 4−004 | 1.59 | 0.50 | 2.98 | 0.70 | 2.29 | 0.61 | 0.73 | 0.80 | 0.097 | 0.118 | 0.062 | 0.030 | 0.300 | 0.031 |
| 4−006 | 2.38 | 0.54 | 2.70 | 0.74 | 2.54 | 0.65 | 0.95 | 0.80 | 0.171 | 0.165 | 0.217 | 0.038 | 0.348 | 0.039 |
| 4−009 | 2.90 | 0.76 | 5.03 | 0.92 | 3.97 | 0.84 | 1.96 | 0.88 | 0.130 | 0.043 | 0.229 | 0.036 | 0.331 | 0.021 |
| 4−021 | 2.63 | 0.56 | 1.67 | 0.72 | 2.15 | 0.65 | 2.93 | 0.64 | 0.087 | 0.028 | 0.105 | 0.027 | 0.213 | 0.017 |
| 4−022 | 4.31 | 0.88 | 5.08 | 0.96 | 4.69 | 0.92 | 0.88 | 0.80 | 0.254 | 0.164 | 0.193 | 0.027 | 0.418 | 0.028 |
| 4−026 | 3.52 | 0.87 | -0.80 | 1.15 | 1.36 | 1.02 | 2.28 | 1.04 | 0.083 | 0.051 | 0.174 | 0.042 | 0.256 | 0.025 |
| 4−027 | ⋯ | ⋯ | ⋯ | ⋯ | ⋯ | ⋯ | ⋯ | ⋯ | ⋯ | ⋯ | ⋯ | ⋯ | ⋯ | ⋯ |
| 4−031 | 2.19 | 0.85 | 4.16 | 1.03 | 3.18 | 0.94 | 0.80 | 1.01 | 0.126 | 0.046 | 0.178 | 0.040 | 0.291 | 0.023 |
| 4−033 | 3.21 | 0.41 | 3.08 | 0.30 | 3.14 | 0.36 | 0.88 | 0.80 | 0.124 | 0.055 | 0.117 | 0.015 | 0.277 | 0.015 |
| 4−036 | ⋯ | ⋯ | ⋯ | ⋯ | ⋯ | ⋯ | ⋯ | ⋯ | ⋯ | ⋯ | ⋯ | ⋯ | ⋯ | ⋯ |

Table 7—Continued

| Star | Fe 5270 | error | Fe 5335 | error | ⟨Fe⟩ | error | Hβ | error | CN | error | Mg₁ | error | Mg₂ | error |
|------|---------|-------|---------|-------|------|-------|------|-------|------|-------|------|-------|------|-------|
| 4−045 | 2.72 | 1.20 | 3.00 | 1.20 | 2.86 | 1.20 | 0.95 | 0.80 | 0.571 | 0.525 | -0.035 | 0.045 | 0.160 | 0.044 |
| 4−047 | 5.21 | 0.80 | 4.95 | 0.70 | 5.08 | 0.75 | 1.62 | 0.80 | 0.186 | 0.091 | 0.224 | 0.023 | ⋯ | ⋯ |
| 4−048 | 2.62 | 0.50 | -0.02 | 0.29 | 1.30 | 0.41 | 3.40 | 0.80 | ⋯ | ⋯ | ⋯ | ⋯ | 0.512 | 0.033 |
| 4−049 | ⋯ | ⋯ | ⋯ | ⋯ | ⋯ | ⋯ | ⋯ | ⋯ | ⋯ | ⋯ | ⋯ | ⋯ | ⋯ | ⋯ |
| 4−051 | 3.31 | 0.66 | 3.87 | 0.72 | 3.59 | 0.69 | 1.34 | 0.64 | 0.157 | 0.091 | 0.123 | 0.029 | 0.238 | 0.128 |
| 4−052 | 3.70 | 0.82 | 3.61 | 1.69 | 3.66 | 1.32 | 2.02 | 1.21 | 0.070 | 0.131 | 0.270 | 0.032 | 0.418 | 0.216 |
| 4−054 | 5.32 | 0.80 | 5.64 | 0.70 | 5.48 | 0.75 | 2.66 | 0.80 | 0.034 | 0.079 | 0.241 | 0.020 | 0.497 | 0.022 |
| 4−062 | 4.03 | 1.50 | 2.49 | 0.80 | 3.26 | 1.20 | 0.99 | 0.80 | ⋯ | ⋯ | 0.127 | 0.045 | 0.412 | 0.046 |
| 4−065 | 3.59 | 0.38 | 3.98 | 0.48 | 3.79 | 0.43 | 0.75 | 0.61 | 0.232 | 0.061 | 0.179 | 0.020 | 0.310 | 0.187 |
| 4−069 | 5.15 | 0.78 | 4.45 | 0.64 | 4.80 | 0.71 | 2.40 | 0.64 | 0.355 | 0.112 | 0.170 | 0.033 | 0.340 | 0.159 |
| 4−070 | 3.62 | 0.80 | 4.57 | 0.90 | 4.09 | 0.85 | 0.57 | 0.80 | 0.082 | 0.132 | 0.180 | 0.035 | 0.305 | 0.035 |
| 4−071 | 3.47 | 0.57 | 2.68 | 0.29 | 3.07 | 0.45 | 1.69 | 0.80 | 0.097 | 0.071 | 0.133 | 0.023 | 0.242 | 0.023 |
| 4−074 | ⋯ | ⋯ | ⋯ | ⋯ | ⋯ | ⋯ | ⋯ | ⋯ | ⋯ | ⋯ | ⋯ | ⋯ | ⋯ | ⋯ |
| 4−075 | 3.75 | 0.37 | 5.42 | 0.44 | 4.59 | 0.41 | 0.55 | 0.46 | 0.159 | 0.025 | 0.283 | 0.017 | 0.504 | 0.010 |
| 4−086 | 3.37 | 1.10 | 3.87 | 1.10 | 3.62 | 1.10 | 0.78 | 0.80 | 0.020 | 0.151 | ⋯ | ⋯ | 0.396 | 0.040 |
| 4−093 | 2.88 | 0.38 | 3.97 | 0.30 | 3.42 | 0.34 | 0.66 | 0.80 | 0.110 | 0.045 | 0.253 | 0.012 | 0.376 | 0.012 |
| 4−102 | 4.16 | 0.46 | 1.96 | 1.06 | 3.06 | 0.82 | 0.88 | 0.92 | 0.044 | 0.051 | 0.064 | 0.040 | 0.172 | 0.025 |
| 4−111 | 4.61 | 1.00 | 3.87 | 0.70 | 4.24 | 0.86 | 0.16 | 0.80 | 0.127 | 0.109 | ⋯ | ⋯ | 0.373 | 0.026 |
| 4−114 | 2.39 | 0.38 | 2.99 | 0.30 | 2.69 | 0.34 | 0.83 | 0.80 | 0.022 | 0.064 | 0.123 | 0.013 | 0.445 | 0.014 |
| 4−121 | 4.21 | 0.91 | 4.24 | 0.87 | 4.23 | 0.89 | 0.87 | 0.63 | 0.161 | 0.124 | 0.229 | 0.030 | 0.380 | 0.171 |
| 4−139 | 4.28 | 0.65 | 2.79 | 0.81 | 3.54 | 0.73 | 2.38 | 0.81 | 0.125 | 0.041 | 0.171 | 0.032 | 0.327 | 0.018 |
| 4−143 | 3.15 | 1.20 | 2.78 | 1.00 | 2.97 | 1.10 | 2.07 | 0.80 | -0.098 | 0.124 | -0.025 | 0.039 | 0.178 | 0.039 |
| 4−145 | 3.37 | 1.28 | 2.87 | 1.40 | 3.12 | 1.34 | 1.42 | 0.96 | 0.084 | 0.162 | 0.183 | 0.045 | 0.319 | 0.186 |
| 4−146 | 3.85 | 0.65 | 4.58 | 0.88 | 4.22 | 0.77 | 2.12 | 0.65 | 0.238 | 0.083 | 0.194 | 0.036 | 0.301 | 0.151 |
| 4−150 | 2.62 | 0.85 | 2.50 | 0.79 | 2.56 | 0.83 | 1.59 | 0.74 | 0.002 | 0.101 | 0.163 | 0.035 | 0.327 | 0.256 |
| 4−155 | 5.04 | 0.73 | 4.20 | 0.91 | 4.62 | 0.83 | 1.49 | 0.87 | 0.181 | 0.039 | 0.166 | 0.035 | 0.325 | 0.020 |
| 4−160 | 3.13 | 0.67 | 3.01 | 1.20 | 3.07 | 0.97 | 2.02 | 0.64 | 0.078 | 0.082 | 0.116 | 0.045 | 0.191 | 0.060 |
| 4−161 | 3.23 | 0.55 | 1.88 | 0.69 | 2.56 | 0.62 | 1.91 | 0.74 | 0.015 | 0.743 | 0.115 | 0.028 | 0.204 | 0.017 |
| 4−165 | 2.36 | 0.69 | 4.08 | 0.72 | 3.22 | 0.71 | 2.02 | 0.93 | 0.133 | 0.132 | 0.253 | 0.033 | 0.398 | 0.201 |
| 4−167 | 4.22 | 0.43 | 3.62 | 0.53 | 3.92 | 0.48 | 1.47 | 0.52 | 0.096 | 0.027 | 0.208 | 0.020 | 0.362 | 0.011 |
| 4−170 | 5.28 | 0.60 | 3.98 | 0.75 | 4.63 | 0.68 | 2.71 | 0.73 | 0.164 | 0.035 | 0.197 | 0.029 | 0.391 | 0.016 |
| 4−179 | 3.52 | 0.75 | 4.26 | 0.53 | 3.89 | 0.65 | 1.25 | 0.64 | 0.138 | 0.054 | 0.155 | 0.021 | 0.276 | 0.160 |
| 4−183 | 3.63 | 1.07 | 3.37 | 0.99 | 3.50 | 1.03 | 1.86 | 0.85 | 0.186 | 0.142 | 0.176 | 0.034 | 0.337 | 0.175 |
| 4−184 | ⋯ | ⋯ | 4.68 | 1.20 | ⋯ | ⋯ | -0.78 | 0.80 | ⋯ | ⋯ | 0.140 | 0.036 | 0.377 | 0.036 |
| 4−185 | 2.05 | 0.40 | 1.60 | 0.29 | 1.82 | 0.35 | 1.77 | 0.80 | 0.096 | 0.142 | 0.056 | 0.031 | 0.105 | 0.029 |
| 4−186 | 4.93 | 0.87 | 3.96 | 0.94 | 4.44 | 0.91 | 0.12 | 0.73 | 0.124 | 0.092 | 0.270 | 0.032 | 0.470 | 0.261 |

Table 7—Continued

| Star | Fe 5270 | error | Fe 5335 | error | ⟨Fe⟩ | error | Hβ | error | CN | error | Mg₁ | error | Mg₂ | error |
|------|---------|-------|---------|-------|------|-------|-----|-------|-----|-------|------|-------|------|-------|
| 4−187 | 2.20 | 0.46 | 1.60 | 0.60 | 1.90 | 0.53 | 2.17 | 0.49 | 0.012 | 0.052 | 0.035 | 0.022 | 0.185 | 0.014 |
| 4−189 | 3.41 | 0.83 | 2.73 | 0.61 | 3.07 | 0.73 | 1.64 | 0.62 | 0.094 | 0.099 | 0.113 | 0.030 | 0.227 | 0.142 |
| 4−190 | 2.04 | 0.71 | 3.79 | 0.88 | 2.92 | 0.80 | 0.79 | 0.87 | 0.083 | 0.052 | 0.113 | 0.035 | 0.208 | 0.021 |
| 4−191 | 4.13 | 0.57 | 4.04 | 0.70 | 4.09 | 0.64 | 1.57 | 0.71 | 0.379 | 0.045 | 0.295 | 0.027 | 0.466 | 0.015 |
| 4−198 | 1.62 | 0.38 | 1.80 | 0.29 | 1.71 | 0.34 | 1.36 | 0.80 | -0.014 | 0.046 | 0.002 | 0.014 | 0.195 | 0.013 |
| 4−203 | 2.51 | 0.38 | 2.94 | 0.29 | 2.72 | 0.34 | 0.96 | 0.80 | 0.069 | 0.050 | 0.080 | 0.008 | 0.315 | 0.011 |
| 4−206 | 3.28 | 0.50 | 3.42 | 0.50 | 3.35 | 0.50 | 1.66 | 0.80 | 0.261 | 0.117 | 0.226 | 0.027 | 0.325 | 0.022 |
| 4−213 | 4.47 | 1.00 | 3.50 | 0.86 | 3.98 | 0.94 | 0.20 | 0.70 | 0.221 | 0.120 | 0.254 | 0.030 | 0.456 | 0.267 |
| 4−218 | 1.91 | 1.30 | 3.68 | 2.10 | 2.80 | 1.75 | 0.81 | 0.80 | -0.125 | 0.159 | -0.077 | 0.061 | 0.178 | 0.064 |
| 4−243 | 2.71 | 0.49 | 3.51 | 0.82 | 3.11 | 0.67 | 0.67 | 0.62 | 0.079 | 0.084 | 0.270 | 0.024 | 0.441 | 0.285 |
| 4−250 | 3.89 | 0.74 | 3.66 | 0.93 | 3.78 | 0.84 | -1.03 | 0.95 | 0.108 | 0.043 | 0.211 | 0.035 | 0.425 | 0.020 |
| 4−256 | 2.38 | 0.60 | 1.22 | 0.77 | 1.80 | 0.69 | 1.54 | 0.73 | 0.132 | 0.031 | 0.158 | 0.029 | 0.309 | 0.016 |
| 4−258 | 1.35 | 0.38 | ⋯ | ⋯ | ⋯ | ⋯ | 1.95 | 0.80 | -0.012 | 0.049 | 0.056 | 0.018 | 0.069 | 0.016 |
| 4−262 | 1.02 | 0.72 | -1.86 | 0.83 | -0.42 | 0.78 | 4.77 | 0.99 | -0.151 | 0.027 | -0.151 | 0.033 | 0.376 | 0.017 |
| 4−270 | 1.24 | 0.61 | 2.67 | 0.77 | 1.96 | 0.69 | 0.81 | 0.69 | 0.123 | 0.030 | 0.041 | 0.029 | 0.184 | 0.018 |
| 4−271 | 0.00 | 0.24 | -2.06 | 0.82 | -1.03 | 0.60 | 5.08 | 1.01 | -0.157 | 0.026 | -0.166 | 0.035 | 0.368 | 0.016 |
| 4−274 | 4.70 | 0.80 | 5.25 | 0.70 | 4.97 | 0.75 | 1.99 | 0.80 | -0.002 | 0.103 | 0.293 | 0.026 | 0.542 | 0.028 |
| 4−275 | 3.79 | 0.80 | 3.99 | 0.70 | 3.89 | 0.75 | 1.50 | 0.80 | 0.230 | 0.152 | 0.151 | 0.030 | 0.227 | 0.029 |
| 4−279 | 3.16 | 1.50 | 3.89 | 1.70 | 3.53 | 1.60 | 1.53 | 0.80 | 0.661 | 0.502 | 0.104 | 0.055 | 0.361 | 0.057 |
| 4−283 | 1.95 | 0.45 | 3.47 | 0.44 | 2.71 | 0.45 | 0.79 | 0.80 | 0.201 | 0.159 | 0.297 | 0.023 | 0.512 | 0.024 |
| 4−285 | 4.61 | 1.50 | 4.06 | 1.20 | 4.34 | 1.36 | 1.52 | 0.80 | 0.401 | 0.203 | 0.112 | 0.037 | 0.361 | 0.038 |
| 4−286 | 4.30 | 0.73 | 2.23 | 0.94 | 3.27 | 0.84 | -0.19 | 0.88 | 0.266 | 0.043 | 0.088 | 0.030 | 0.287 | 0.020 |
| 4−297 | 4.09 | 1.30 | 3.74 | 1.10 | 3.92 | 1.20 | 1.03 | 0.80 | 0.082 | 0.161 | 0.043 | 0.034 | 0.321 | 0.036 |
| 4−307 | 3.87 | 0.68 | 2.29 | 0.86 | 3.08 | 0.78 | 1.96 | 0.70 | 0.094 | 0.030 | 0.104 | 0.030 | 0.272 | 0.018 |
| 4−309 | 3.55 | 1.10 | 3.54 | 1.00 | 3.55 | 1.05 | 0.91 | 0.80 | 0.751 | 0.404 | ⋯ | ⋯ | 0.358 | 0.036 |
| 4−310 | 5.12 | 1.28 | 3.81 | 1.02 | 4.46 | 1.16 | 1.69 | 0.75 | 0.152 | 0.152 | 0.144 | 0.038 | 0.314 | 0.162 |
| 4−312 | 1.54 | 0.38 | 0.41 | 0.29 | 0.98 | 0.34 | 1.58 | 0.80 | -0.099 | 0.085 | -0.052 | 0.028 | 0.069 | 0.027 |
| 4−320 | 1.79 | 0.38 | 1.08 | 0.29 | 1.43 | 0.34 | 1.36 | 0.80 | -0.099 | 0.033 | -0.087 | 0.012 | 0.089 | 0.012 |
| 4−324 | 2.39 | 1.10 | 1.66 | 0.70 | 2.03 | 0.92 | 1.41 | 0.80 | 0.334 | 0.368 | 0.076 | 0.050 | 0.325 | 0.051 |
| 4−325 | 4.80 | 0.78 | 2.26 | 0.98 | 3.53 | 0.89 | 2.13 | 0.88 | 0.216 | 0.036 | 0.195 | 0.030 | 0.370 | 0.021 |
| 4−326 | 4.64 | 0.52 | 2.99 | 0.66 | 3.82 | 0.59 | 1.19 | 0.64 | 0.026 | 0.062 | 0.176 | 0.035 | 0.334 | 0.014 |
| 4−328 | 4.29 | 2.60 | ⋯ | ⋯ | ⋯ | ⋯ | 2.48 | 0.80 | 0.036 | 0.325 | 0.054 | 0.080 | 0.197 | 0.082 |
| 4−335 | 2.17 | 0.77 | 1.66 | 0.69 | 1.91 | 0.73 | 2.49 | 0.87 | -0.064 | 0.095 | 0.059 | 0.042 | 0.155 | 0.124 |
| vb252 | 1.54 | 0.53 | 0.97 | 0.54 | 1.26 | 0.54 | 1.42 | 0.69 | -0.027 | 0.100 | 0.026 | 0.028 | 0.111 | 0.120 |
| vb256 | ⋯ | ⋯ | ⋯ | ⋯ | ⋯ | ⋯ | ⋯ | ⋯ | ⋯ | ⋯ | ⋯ | ⋯ | ⋯ | ⋯ |
| vb259 | ⋯ | ⋯ | ⋯ | ⋯ | ⋯ | ⋯ | ⋯ | ⋯ | ⋯ | ⋯ | ⋯ | ⋯ | ⋯ | ⋯ |

Table 7—Continued

| Star | Fe 5270 | error | Fe 5335 | error | ⟨Fe⟩ | error | Hβ | error | CN | error | Mg$_1$ | error | Mg$_2$ | error |
|---|---|---|---|---|---|---|---|---|---|---|---|---|---|---|
| vb262 | 4.17 | 0.65 | 2.54 | 0.82 | 3.36 | 0.74 | 2.27 | 0.81 | 0.037 | 0.064 | 0.225 | 0.031 | 0.337 | 0.018 |
| vb268 | 3.02 | 1.07 | 4.07 | 1.04 | 3.55 | 1.06 | 1.77 | 0.72 | 0.019 | 0.091 | 0.101 | 0.039 | 0.225 | 0.100 |
| vb271 | 2.38 | 0.93 | 1.90 | 1.04 | 2.14 | 0.99 | 2.52 | 0.80 | -0.096 | 0.094 | 0.037 | 0.039 | 0.091 | 0.031 |
| vb274 | 3.93 | 0.77 | 1.63 | 0.98 | 2.78 | 0.88 | 0.34 | 1.04 | 0.318 | 0.063 | 0.201 | 0.038 | 0.318 | 0.022 |
| vb282 | 1.34 | 0.69 | 1.37 | 0.87 | 1.36 | 0.79 | 1.44 | 0.80 | 0.087 | 0.037 | 0.053 | 0.033 | 0.108 | 0.026 |
| vb283 | ⋯ | ⋯ | ⋯ | ⋯ | ⋯ | ⋯ | ⋯ | ⋯ | ⋯ | ⋯ | ⋯ | ⋯ | ⋯ | ⋯ |
| vb284 | 2.49 | 0.91 | 1.15 | 0.78 | 1.82 | 0.85 | 2.25 | 0.98 | 0.023 | 0.105 | 0.086 | 0.052 | 0.184 | 0.178 |
| vb315 | 3.60 | 0.82 | 1.72 | 1.06 | 2.66 | 0.95 | 2.97 | 0.97 | 0.138 | 0.043 | 0.173 | 0.040 | 0.231 | 0.024 |
| vb323 | 1.71 | 0.84 | 1.91 | 1.06 | 1.81 | 0.96 | 2.05 | 0.99 | 0.052 | 0.056 | 0.152 | 0.040 | 0.239 | 0.024 |
| vb326 | 1.60 | 0.87 | 3.01 | 1.04 | 2.31 | 0.96 | 0.66 | 1.02 | 0.087 | 0.048 | 0.148 | 0.040 | 0.315 | 0.023 |
| vb329 | 1.97 | 0.81 | 3.55 | 1.02 | 2.76 | 0.92 | 1.89 | 0.88 | 0.075 | 0.037 | 0.098 | 0.039 | 0.167 | 0.025 |
| vb330 | 2.40 | 1.89 | 1.42 | 0.95 | 1.91 | 1.49 | 1.08 | 1.92 | 0.182 | 0.328 | 0.185 | 0.062 | 0.297 | 0.114 |
| vb353 | 2.97 | 1.60 | 1.34 | 0.70 | 2.16 | 1.24 | 1.53 | 0.80 | -0.180 | 0.175 | -0.078 | 0.051 | 0.134 | 0.051 |
| vb361 | 3.50 | 0.94 | -5.52 | 1.33 | -1.01 | 1.15 | 0.02 | 1.09 | 0.025 | 0.114 | 0.087 | 0.046 | 0.185 | 0.028 |
| vb362 | 3.07 | 0.72 | 3.84 | 0.89 | 3.46 | 0.81 | 2.06 | 0.88 | 0.207 | 0.045 | 0.112 | 0.035 | 0.256 | 0.020 |
| vb395 | 4.50 | 0.35 | 3.97 | 0.44 | 4.24 | 0.40 | 1.82 | 0.43 | 0.350 | 0.020 | 0.152 | 0.017 | 0.299 | 0.010 |
| vb404 | 3.95 | 0.58 | 4.68 | 0.71 | 4.32 | 0.65 | 1.73 | 0.69 | 0.184 | 0.033 | 0.158 | 0.036 | 0.335 | 0.016 |
| vb406 | 1.85 | 0.53 | 1.08 | 0.68 | 1.47 | 0.61 | 1.62 | 0.63 | 0.003 | 0.006 | 0.027 | 0.036 | 0.118 | 0.019 |
| vb411 | 3.40 | 1.92 | 1.64 | 1.10 | 2.52 | 1.56 | 0.45 | 0.75 | -0.076 | 0.139 | 0.051 | 0.054 | 0.138 | 0.062 |
| vb413 | 2.70 | 0.91 | 1.65 | 0.77 | 2.17 | 0.84 | 2.71 | 1.67 | -0.045 | 0.077 | 0.041 | 0.042 | 0.075 | 0.048 |
| vb414 | 2.78 | 0.84 | 1.78 | 0.72 | 2.28 | 0.78 | 1.50 | 0.69 | 0.223 | 0.115 | 0.144 | 0.039 | 0.242 | 0.104 |
| vb417 | 2.93 | 0.75 | 2.25 | 1.25 | 2.59 | 1.03 | 0.79 | 0.61 | 0.235 | 0.194 | 0.228 | 0.043 | 0.343 | 0.160 |
| vb457 | 5.83 | 0.70 | 2.24 | 0.92 | 4.04 | 0.82 | 1.00 | 0.94 | -0.141 | 0.032 | 0.186 | 0.036 | 0.265 | 0.021 |
| vb469 | 4.46 | 0.45 | 3.08 | 0.57 | 3.77 | 0.51 | 2.26 | 0.59 | 0.170 | 0.026 | 0.120 | 0.036 | 0.256 | 0.013 |
| vb470 | ⋯ | ⋯ | ⋯ | ⋯ | ⋯ | ⋯ | ⋯ | ⋯ | ⋯ | ⋯ | ⋯ | ⋯ | ⋯ | ⋯ |
| vb471 | 3.13 | 0.77 | 0.91 | 0.98 | 2.02 | 0.88 | 1.55 | 0.97 | 0.121 | 0.048 | 0.145 | 0.038 | 0.249 | 0.022 |
| vb472 | 2.02 | 0.77 | 2.42 | 0.97 | 2.22 | 0.88 | 3.00 | 0.90 | -0.082 | 0.022 | 0.057 | 0.038 | 0.085 | 0.036 |
| vb480 | 3.81 | 0.57 | 2.12 | 0.73 | 2.97 | 0.66 | 1.59 | 0.69 | 0.092 | 0.029 | 0.099 | 0.036 | 0.162 | 0.018 |
| vb483 | 1.58 | 0.59 | 3.45 | 0.74 | 2.52 | 0.67 | 2.10 | 0.71 | 0.111 | 0.035 | 0.162 | 0.029 | 0.266 | 0.016 |
| vb488 | 4.60 | 0.78 | 0.51 | 0.94 | 2.56 | 0.86 | 3.03 | 0.95 | -0.008 | 0.012 | 0.070 | 0.040 | 0.163 | 0.026 |
| vb491 | 4.58 | 0.48 | 4.46 | 0.59 | 4.52 | 0.54 | 1.80 | 0.63 | 0.119 | 0.032 | 0.228 | 0.023 | 0.329 | 0.013 |
| vb493 | 4.34 | 0.56 | 2.39 | 0.73 | 3.37 | 0.65 | 1.31 | 0.71 | 0.194 | 0.033 | 0.061 | 0.029 | 0.183 | 0.018 |
| vb497 | 2.51 | 0.52 | 5.21 | 0.61 | 3.86 | 0.57 | 1.35 | 0.67 | 0.231 | 0.033 | 0.211 | 0.025 | 0.373 | 0.014 |
| vb499 | 2.08 | 2.10 | 3.62 | 3.20 | 2.85 | 2.71 | 3.22 | 0.80 | -0.088 | 0.330 | -0.127 | 0.092 | 0.124 | 0.096 |
| vb500 | 4.70 | 1.85 | 4.70 | 1.68 | 4.70 | 1.77 | 0.67 | 0.76 | 0.549 | 0.549 | 0.260 | 0.048 | 0.416 | 0.201 |
| vb504 | 3.89 | 0.61 | 3.29 | 0.78 | 3.59 | 0.70 | 2.13 | 0.74 | 0.168 | 0.032 | 0.100 | 0.030 | 0.200 | 0.018 |

Table 8.   Reddening determinations to Baade's Window.

| Reference | $E(B-V)_0$ | $E(V-I)$ | Note |
|---|---|---|---|
| Arp (1965) | $0.50 \pm 0.03$ | $\cdots$ | adjusted to zero color |
| van den Bergh (1971) | $0.49 \pm 0.03$ | $\cdots$ | adjusted to zero olor |
| Walker & Mack (1986) | $0.56 \pm 0.03$ | $\cdots$ | |
| Terndrup & Walker (1994) | $0.52 \pm 0.05$ | $\cdots$ | |
| This paper | $0.51 \pm 0.04$ | $0.64 \pm 0.08$ | full sample |
| This paper | $0.47 \pm 0.04$ | $0.59 \pm 0.08$ | region A |
| This paper | $0.54 \pm 0.06$ | $0.67 \pm 0.09$ | regions B/C |

Table 9.   Velocity dispersion as a function of magnitude.

| | $V < 15.5$ | $V < 16.0$ | $V \geq 16.0$ |
|---|---|---|---|
| $\langle v_r \rangle$ | $-17 \pm 13$ | $-7 \pm 10$ | $-8 \pm 6$ |
| $\langle \mu_l \rangle$ | $2.75 \pm 0.44$ | $1.95 \pm 0.36$ | $-0.17 \pm 0.16$ |
| $\langle \mu_b \rangle$ | $-0.32 \pm 0.44$ | $-0.04 \pm 0.33$ | $-0.27 \pm 0.15$ |
| $\sigma(v_r)$ | $69 \pm 21$ | $79 \pm 16$ | $110 \pm 10$ |
| $\sigma(\mu_l)$ | $2.51 \pm 0.61$ | $2.94 \pm 0.51$ | $3.05 \pm 0.23$ |
| $\sigma(\mu_b)$ | $2.52 \pm 0.61$ | $2.64 \pm 0.46$ | $2.78 \pm 0.21$ |

Note. — Heliocentric radial velocities are in units of km s$^{-1}$. Proper motions are in units of mas yr$^{-1}$.

Table 10.   Velocity dispersions of bulge and disk stars.

| | Stars with $V < 16.0$ | | $V > 16.0$ |
|---|---|---|---|
| | $R = 4$ kpc | $R = 8$ kpc | $R = 8$ kpc |
| $\sigma(v_r)$ | $79 \pm 16$ | $79 \pm 16$ | $110 \pm 10$ |
| $\sigma(v_l)$ | $56 \pm 10$ | $112 \pm 19$ | $116 \pm 9$ |
| $\sigma(v_b)$ | $50 \pm 9$ | $100 \pm 17$ | $105 \pm 8$ |